%% file: main.tex
\documentclass[twocolumn,superscriptaddress,showpacs,prd,aps,amsmath,amssymb,nofootinbib]{revtex4-1}

\usepackage{graphicx}
\usepackage{longtable}

\input kigou.tex

\usepackage[normalem]{ulem}
\usepackage{epstopdf}
\usepackage{color}
\usepackage{soul}
\usepackage{bm}

\newcommand{\cocal}{\textsc{cocal}}

\usepackage[]{amsmath}
\usepackage[]{amsfonts}
\usepackage[]{amssymb}
\usepackage[]{mathrsfs}
\usepackage{times}
\usepackage{hyperref}
\hypersetup{
  colorlinks=true,        
  linkcolor=blue,         
  citecolor=cyan,         
}
\newcommand{\cf}{cf.,~}
\newcommand{\ie}{i.e.,~}
\newcommand{\eg}{e.g.,~}

\def\QEQ{{%
			\setbox0\hbox{$I$}%
			\rlap{\hbox to \wd0{\hss--\hss}}\box0
		}}

\begin{document}

\title{Uniformly rotating, axisymmetric and triaxial quark stars in general relativity}

\author{Enping Zhou}
\affiliation{State Key Laboratory of Nuclear Science and Technology and
  School of Physics, Peking University, Beijing 100871, People's Republic
  of China} \affiliation{Institute for Theoretical Physics, Frankfurt am
  Main 60438, Germany}

\author{Antonios Tsokaros}
\affiliation{Institute for Theoretical Physics, Frankfurt am Main 60438,
  Germany} \affiliation{Department of Physics, University of Illinois at
  Urbana-Champaign, Urbana, IL 61801, USA}

\author{Luciano Rezzolla}
\affiliation{Institute for Theoretical Physics, Frankfurt am Main 60438,
  Germany} \affiliation{Frankfurt Institute of Advanced Studies,
  Frankfurt am Main 60438, Germany}

\author{Renxin Xu}
\affiliation{State Key Laboratory of Nuclear Science and Technology and
  School of Physics, Peking University, Beijing 100871, People's Republic
  of China} \affiliation{Kavli Institute for Astronomy and Astrophysics,
  Peking University, Beijing, 100871, People's Republic of China}

\author{K\=oji Ury\=u}
\affiliation{Department of Physics, University of the Ryukyus, Senbaru,
  Nishihara, Okinawa 903-0213, Japan}

\date{\today}

\begin{abstract}
Quasi-equilibrium models of uniformly rotating axisymmetric and triaxial
quark stars are computed in general relativistic gravity scenario. The
Isenberg-Wilson-Mathews (IWM) formulation is employed and the Compact
Object CALculator (\cocal{}) code is extended to treat rotating stars
with finite surface density and new equations of state (EOSs). Besides
the MIT bag model for quark matter which is composed of de-confined
quarks, we examine a new EOS proposed by Lai and Xu that is based on
quark clustering and results in a stiff EOS that can support masses up to
$3.3M_\odot$ in the case we considered. We perform convergence tests for
our new code to evaluate the effect of finite surface density in the
accuracy of our solutions and construct sequences of solutions for both
small and high compactness. The onset of secular instability due to
viscous dissipation is identified and possible implications are
discussed. An estimate of the gravitational wave amplitude and luminosity
based on quadrupole formulas is presented and comparison with neutron
stars is discussed.
\end{abstract}

\maketitle

\section{Introduction}

The recent gravitational-wave (GW) event GW170817 together with
accompanying electromagnetic emission observations
\cite{Abbott2017,Abbott2017b} from a binary neutron star (BNS) merger has
opened a brand new multi-messenger observation era for us to explore the
Universe. Apart from enriching our knowledge on origins of short
gamma-ray bursts \cite{Abbott2017d} and nucleosynthesis
associated with BNS mergers \cite{Abbott2017c, Baiotti2016}, it also
provides an effective way for us to constrain the equation of state (EOS)
of neutron stars (NS).  In addition to systems such as binary black-hole
mergers and BNS mergers, rapidly rotating compact stars have also been
considered as important candidates of GW sources
\cite{Andersson:2009yt}, which could be detected by ground-based GW
observatories \cite{Abramovici92, Punturo:2010, Accadia2011_etal,
  Kuroda2010, Aso:2013} and help us understand the nature of strong
interaction of dense matter.

It has been long since the equilibrium models of self-gravitating,
uniformly rotating, incompressible fluid stars were
systematically studied in a Newtonian gravity scheme
\cite{Chandrasekhar1969book}. Depending on the rotational kinetic energy,
the configuration could be axisymmetric Maclaurin ellipsoids as well as
nonaxisymmetric ellipsoids, such as Jacobian (triaxial) ellipsoid. For
compact stars that we are interested in for GW astronomy, however,
general relativity is required to replace Newtonian gravity. The field of
relativistic rotating stars has been studied for many years
\cite{Meinel:2008,Friedman2012}.

A rotating NS will spontaneously break its axial symmetry if the
rotational kinetic energy to gravitational binding energy ratio, $T/|W|$
exceeds a critical value. This instability can either be of secular type
\cite{shapiro98b, Bonazzola1996b, rosinska2002, Bonazzola1998c} or
dynamical \cite{Houser96, Pickett96, brown2000, New2000, Liu02,
  Watts:2003nn, Baiotti06b, Manca07, Corvino:2010}, depending on the
process driving the instability and with only small modifications if a
magnetisation is present \cite{Camarda:2009mk, Franci2013,
  Muhlberger2014} (see \cite{Andersson03} for a review). A high $T/|W|$
ratio can also be reached for a newly born rotating compact star during a
core collapse supernova or for a NS which is spun up by accretion
\cite{Lai95, Bildsten98, woosley2005, watts2008, piro2012b}.

Quasi-equilibrium figures of triaxially rotating NSs have also been
created and studied in full general relativity \cite{Huang08, Uryu2016a}.
In this case, the bifurcation from an axisymmetric to triaxial
configuration happens very close to the mass shedding limit, and, for
soft NS EOSs or for NSs with large compactness, the triaxial sequence
could totally vanish \cite{Jame64, Bonazzola1996b, Bonazzola1998c,
  hachisu1982,Lai93}. As a result, it is presently unclear whether
triaxial configurations of NSs can actually be realized
in practice.

On the other hand, it is worth noting that the EOS of compact stars is
still a matter of lively debate since astronomical observations are not
sufficient to rule out many of the nuclear-physics EOSs that are
compatible with the observations. As a result, besides the popular idea
of NSs, other models for compact stars are possible and have been
considered in the past. A particularly well developed literature is the
one concerned with strange quark stars (QSs), since it was long
conjectured that strange quark matter composed of de-confined up, down
and strange quarks could be absolutely stable
\cite{Bodmer1971,Witten84}. There is also possible observational evidence
indicating the existence of QSs (for a recent example, see
\cite{Dai_ZG:2016}). Additionally, the small tidal deformability of QSs
is favoured by the observation of GW170817 \cite{Lai2017b} and possible
models with QS merger or QS formations are also suggested to explain the
electromagnetic counterparts for a short gamma-ray burst
(c.f. \cite{Li2016,Lai2017b}).

Following this possibility, a large effort has been developed to
calculate equilibrium configurations of QSs, starting 
from the first attempts \cite{Itoh70,Alcock86,Haensel1986}. At present,
both uniformly rotating \cite{rosinska2000b, gourgoulhon1999,
  Stergioulas99a} and differentially rotating QSs \cite{szkudlarek2012}
have been studied in full general relativity. Unlike NSs, which are bound
by self-gravity, QSs are self-bound by strong interaction. Consequently,
rotating QSs can reach a much larger $T/|W|$ ratio compared with NSs and
the triaxial instability could play a more important role
\cite{rosinska2000a, rosinska2000b, rosinska2001}. The triaxial bar mode
(Jacobi-like) instability for MIT bag-model EOS has been investigated in
a general relativistic framework \cite{rosinska2003}.

We here use the Compact Object CALculator code, \cocal{}, to build
general-relativistic triaxial QS solution sequences using different EOS
models. \cocal{} is a code to calculate general-relativistic equilibrium
and quasi-equilibrium solutions for binary compact stars (black hole and
NS) as well as rotating (uniformly or differentially)
NSs \cite{Uryu2012,Uryu:2012b,Tsokaros2012,Tsokaros2015,
  Huang08,Uryu2016a}.  The part of the \cocal{} code handling the
calculation of the EOS was originally designed for piecewise polytropic
EOSs. We have here extended the code to include polynomial type of EOSs,
as those that can be used to describe QSs. In doing so the trivial
relationship between the thermodynamic quantities for a piecewise
polytrope (\eg see Eqs.  (64)--(68) in \cite{Tsokaros2015}), is lost and
now one has to apply root finding methods. Another issue is related to
the surface fitted coordinates that are used in \cocal{} to track the
surface of the star. For NSs, the surface was identified as the place
where the rest-mass density goes to zero or where the specific enthalpy
becomes one. This is no longer generally true for a
self-bound QS and a different approach needs to be developed. The
nonlinear algebraic system that determines the angular velocity, the
constant from the Euler equation, and the renormalization constant of the
spherical grid has to be modified in order to accommodate
the arbitrary surface enthalpy.

We here compute solutions for both axisymmetric and triaxial rotating QSs
with the new code, as well as sequences with various QS EOS and different
compactnesses. We checked our new implementation for those cases were
previous studies have been possible \cite{Huang08}, and we confirm the
accuracy of our new code. We discuss the astrophysical implications of
the quantities of rotating QSs at the bifurcation point. For instance,
the spin frequency at the bifurcation point could be a more realistic
spin up limit for compact stars rather than the mass shedding limit,
which relates to the fastest spinning pulsar we might be able to
observe. The GW strain and luminosity estimates for our models are given,
while full numerical simulations are left for the future (see
\cite{Tsokaros2017} for recent simulations involving triaxial NSs).

The structure of this paper is organized as follows: In
Sec. \ref{sec:formandmethod} we discuss the formulation we used and the
field equations (Sec. \ref{sec:IWMformulation}), the hydrodynamics
(Sec. \ref{sec:fluidformulation}) and the EOS part
(Sec. \ref{sec:eos}). In order to test the behavior of the modified code,
we have performed convergence tests with five resolutions and compared
with rotating NS solutions built by the original \cocal{} code. These
tests can be found in Sec. \ref{sec:codetest}. Triaxially deformed
rotating QS sequences for different compactnesses, are presented in
Sec. \ref{sec:triasolution}, while the implications for the astrophysical
observations of this work are presented in Sec. \ref{sec:disandconclu}.
Hereafter we use units with $G=c=M_\odot=1$ unless otherwise stated; a
conversion table to the standard cgs units can be found, for instance, in
\cite{Rezzolla_book:2013}.

\section{Formulation and numerical method}
\label{sec:formandmethod}
\subsection{Field equations}
\label{sec:IWMformulation}

In order to solve the field equations numerically, the
Isenberg-Wilson-Mathews (IWM) formulation
\cite{Isenberg1980,Isenberg08,Wilson89} is employed. In a coordinate
chart $\{t,x^i\}$, the $\mathrm{3+1}$ decomposition of the spacetime
metric gives
\begin{equation} 
ds^2 = -\alpha^2 dt^2 + \psi^4 \delta_{ij}
(dx^{i}+\beta^{i}dt) (dx^{j}+\beta^{j}dt)\,, 
\end{equation} 
where $\alpha,\ \beta^i$ are the lapse and shift vector (the kinematical
quantities), while $\gamma_{ij}=\psi^4\delta_{ij}$ is the IWM
approximation for the three-metric.

The extrinsic curvature of the foliation is defined by 
\begin{equation}
\Kabd 
\,:=\,-\frac1{2\alpha}\pa_t \gmabd +\frac1{2\alpha}\Lie_\beta \gmabd\,. 
\end{equation}
and a maximal slicing condition $K=0$ is assumed.

Decomposing the Einstein equations with respect to the normal $n^\alpha$
of foliation, we get the following 5 equations in terms of the five
metric coefficients $\{\psi, \beta^a, \alpha\}$ on the initial slice
$\Sigma_0$: 
\begin{eqnarray} 
&&(\Gabd-8\pi\Tabd)\,n^\alpha n^\beta \ \,=\, 0,
\label{eq:Ham}\\
&&(\Gabd-8\pi\Tabd)\,\gamma^{i\alpha} n^\beta \,=\, 0,
\label{eq:Mom}\\
&&(\Gabd-8\pi\Tabd)\,\Big(\gamma^{\albe}+\frac12 n^\alpha n^\beta\Big)
\,=\, 0, 
\label{eq:trG}
\end{eqnarray} 
where the first and second equations are the Hamiltonian and momentum
constraints, respectively. Here $\gamma_{\alpha\beta} =
g_{\alpha\beta}+n_\alpha n_\beta$ is the projection tensor onto the
spatial slices. These equations can be written in the form of elliptic
equations with the non-linear source terms, respectively,
\begin{eqnarray} 
&&
\!\!\!\!\!\!\!  \nabla^2\psi \,=\, - \frac{\psi^5}{8}A_{ab}A^{ab} -
2\pi\psi^5\rhoH,
\label{eq:HaC_elip2}   \\
&& \!\!\!\!\!\!\!
\nabla^2\beta^a + \frac13 \partial^a\partial_b\beta^b \,=\,
-2\alpha A^{ab}\partial_b\ln\frac{\psi^6}{\alpha} + 16\pi\alpha j^a,
\label{eq:MoC_elip2} \\
&& \!\!\!\!\!\!\!
\nabla^2(\alpha\psi) \,=\, \frac{7}{8}\alpha\psi^5A_{ab}A^{ab}
+ 2\pi\alpha\psi^5(\rhoH+2S).  \label{eq:trG_elip2}
\end{eqnarray}
where $A^{ij}=K^{ij}={\psi^{-4}}(\partial^i \beta^j + \partial^j\beta^i
-\frac{2}{3}\delta^{ij}\partial_k\beta^k)/{2\alpha}$, and the source
terms of matter are defined by $\rhoH:=\Tabd n^\alpha n^\beta$,
$j^i:=-\Tabd \gamma^{i\alpha} n^\beta$, and $S:=\Tabd \gamma^{\alpha
  \beta}$.

The above set of equations must be supplied with 
boundary conditions at infinity. Since we are working
in the inertial frame and we impose asymptotic flatness, we must have
\begin{equation}
\lim_{r\rightarrow\infty} \psi = 1\,,\qquad 
\lim_{r\rightarrow\infty} \alpha = 1\,,\qquad 
\lim_{r\rightarrow\infty} \beta^i = 0\,.
\label{eq:bcpsal}
\end{equation}

\subsection{Hydrostatic equilibrium}
\label{sec:fluidformulation}

The hydrostatic equation for a perfect fluid in quasi-equilibrium can be
derived from the relativistic Euler equation \cite{Rezzolla_book:2013}
\begin{equation}
u^\beta \na_\beta(hu_\alpha) + \na_\alpha h\,=0\,, 
\end{equation}
where $u^\alpha=u^t(1,v^i)=u^t(1,\Omega\phi^i)$ is the 4-velocity of the
fluid, $\phi^i=(-y,x,0)$, and $h$ is the specific enthalpy defined by
$h:=(\epsilon+p)/\rho$ ($\rho$ is the rest-mass density and
$\epsilon$ the total energy density).

When the symmetry along a helical Killing vector $k^\alpha=t^\alpha +
\Omega \phi^\alpha$ is imposed for the fluid variables, which is
approximately true also in the case for a rotating nonaxisymmetric star
in quasi-equilibrium, the integral of the Euler equation becomes
\begin{equation}
\frac{h}{u^t}\,=\, \mathcal{E}\, ,
\label{eq:firstint}
\end{equation}
where $\mathcal{E}$ is a constant. From the normalization of the four
velocity $u_\alpha u^\alpha=-1$, one obtains
\begin{equation}
u^t = \frac1{\sqrt{\alpha^2 - \omega_a \omega^a}}
= \frac1{\sqrt{\alpha^2 - \psi^4 \delta_{ab}\,\omega^a \omega^b}}\,, 
\label{eq:ut}
\end{equation}
where $\omega^a = \beta^a + \Omega \phi^a$.  The fluid sources of
Eqs.~\eqref{eq:HaC_elip2}--\ref{eq:trG_elip2}, \ie $\rho_{\mathrm{H}}$,
$j_a$ and $S$, are defined in terms of the energy momentum tensor in the
previous section.  In terms of the fluid and field variables they can be
written as \cite{Rezzolla_book:2013}
\begin{eqnarray}
&&\rho_{\mathrm{H}}=\rho[h(\alpha u^t)^2-q],   \label{eq:rhoh}\\
&& j^i=\rho h\alpha(u^t)^2\gamma^{i\alpha} u_\alpha,  \label{eq:ja}\\
&& S=\rho h (\alpha u^t)^2 - \rho h +3\rho q,  \label{eq:s}
\end{eqnarray}
in which $q:=p/\rho$ is the relativistic analogue of the Emden function.
Here $u^t$ is related to $h$ through
Eq.~\eqref{eq:firstint}. Therefore in order to close the
system, an additional relationship is needed between the specific
enthalpy, the pressure and the rest-mass density of the fluid, \ie an
EOS. Once such a relation is available, to solve the field equations
[Eqs.~\eqref{eq:HaC_elip2}--\ref{eq:trG_elip2}] and the hydrostatic
equation [Eq.~\eqref{eq:firstint}] one has to find the two constants
$\{\Omega,\mathcal{E}\}$ that appear in all of them. This procedure is
described in detail for example in Ref. \cite{Tsokaros2015}.

\subsection{Equation of State}
\label{sec:eos}

In this work, we have considered two types of EOS for QSs. One of them is
the MIT bag-model EOS \cite{chodos1974}, since it is the most widely used
EOS for QSs. In the case when strange quark mass is neglected, the
pressure is related to total energy density according to
\begin{equation}
p=\sigma (\epsilon-\epsilon_s) \, ,
\label{eq:mit_gen}
\end{equation}
where $\sigma,\ \epsilon_s$ are two constants, the second being the total
energy density at the surface. Related to $\epsilon_s$ is the so called
bag constant, $B=\epsilon_s/4$. In this work, and following
\cite{limousin2005}, the simplest MIT bag-model EOS has been employed,
where $\sigma=1/3$ and $B^{1/4}=138\,\mathrm{MeV}$.

Besides the MIT bag-model EOS, we have also considered another QS EOS
suggested by Lai and Xu \cite{lai2009}, which we will refer to as the LX
EOS hereafter. Unlike the conventional QS models (\eg the MIT bag-model
EOS) which are composed of de-confined quarks, Lai and Xu \cite{lai2009}
suggested that quark clustering is possible at the density of a cold
compact star since the coupling of strong interaction is still decent at
such energy scale. Due to the non-perturbative effect of strong
interaction at low energy scales and the many-body problem, it is very
difficult to derive the EOS of such a quark-cluster star\footnote{Such a
  quark-cluster star has also been named a {\em strangeon} star in
  Ref. \cite{lai2017}.} from first principles.

Lai and Xu attempted to approach the EOS of such quark-cluster star with
phenomenological models, i.e., to compare the intercluster potential with
the interaction between inert molecules (a similar approach has also been
discussed in \cite{Guo2014}). They also take the lattice effects into
account as the potential could be deep enough to trap the quark
clusters. Combining the inter-cluster potential and the lattice
thermodynamics, they have derived an EOS in the following form:
\begin{equation}
p=4U_0(12.4r_0^{12}n^5-8.4r_0^6n^3)+\frac{1}{8}(6\pi^2)^{\frac{1}{3}}\hbar cn^{\frac{4}{3}}\,,
\label{eq:lx09eos}
\end{equation}
where $\hbar$ is the reduced Planck constant.  The parameters in this
expression, $U_0$ and $r_0$, are the depth of the potential and
characteristic range of the interaction, respectively. The EOS is also
dependent on the number of quarks in each cluster ($N_\mathrm{q}$) since
it relates the energy density ($\epsilon$) and rest-mass density ($\rho$)
to the number-density of quark clusters ($n$ in
Eq. (\ref{eq:lx09eos})). Similarly to the MIT bag-model EOS case, we use
the rest-mass density parameter, which is
\begin{equation}
\rho=m_\mathrm{u}\frac{N_\mathrm{q}}{3}n\,,
\end{equation}
where $m_\mathrm{u}=931\mathrm{MeV}/c^2$ is the atomic mass unit. While
several different choices of parameters are considered in
Ref. \cite{lai2009}, in our work we restrict our attention to
$U_0=50\,\mathrm{MeV}$ and for $N_\mathrm{q}=18$. We also note that
although it is not as obvious as for the MIT bag-model EOS, the LX
EOS also has a nonzero surface density since expression
Eq.~\eqref{eq:lx09eos} has a unique zero root when the number density is
positive.

\begin{figure}
\begin{center}
\includegraphics[width=0.95\columnwidth]{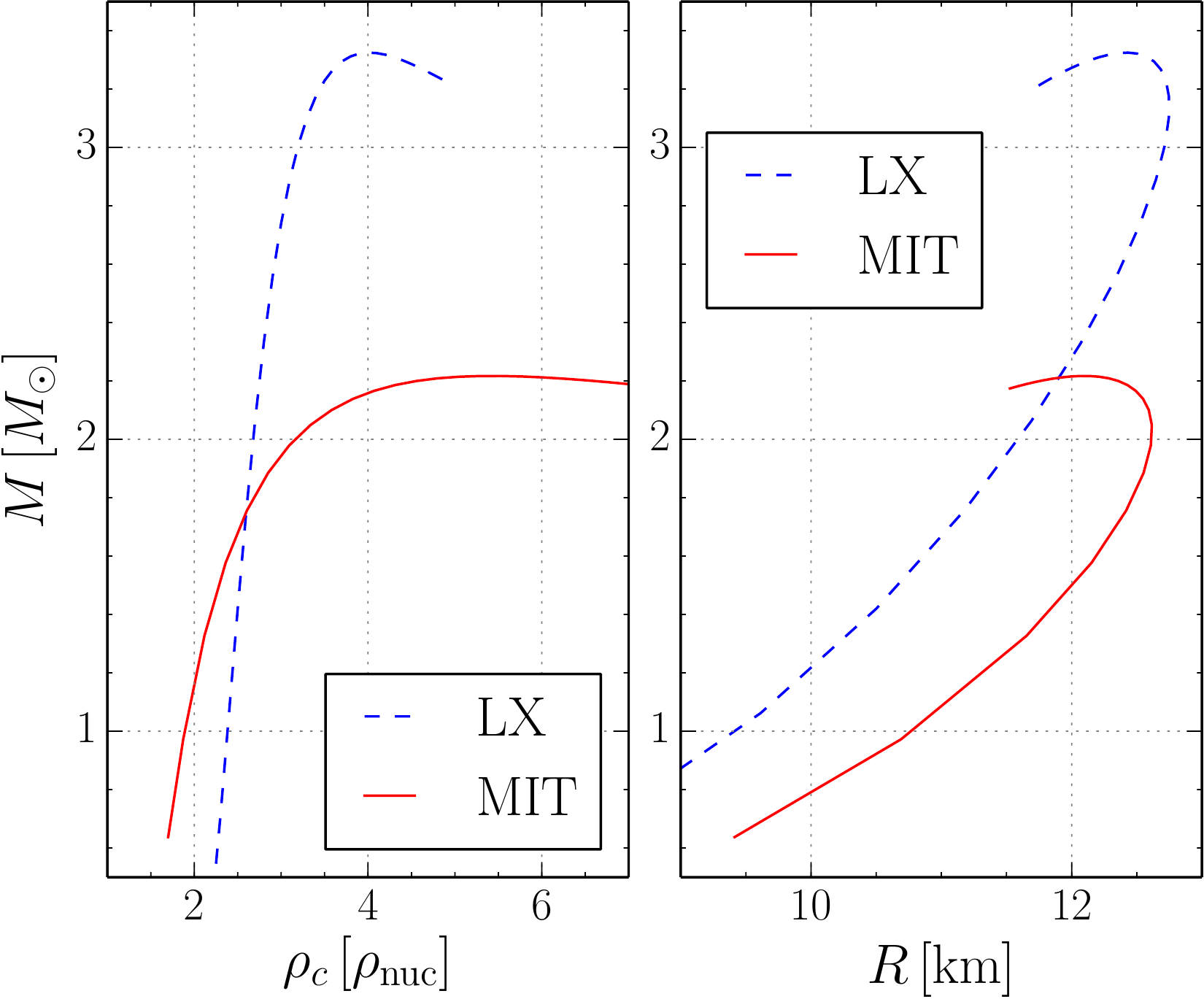}
\end{center}
\caption{The TOV solution sequences for MIT bag-model EOS (red solid line)
  and the LX EOS (blue dashed line) respectively. The left panel shows the
  mass-central density relationship for each model and the right panel is
  the mass-radius diagram. The bag constant we apply in this work for the
  MIT bag-model EOS satisfies the 2 solar mass constraint from observations.}
\label{fig:plot_tov}
\end{figure}

Being a stiff EOS, the LX EOS is favored by the discovery of massive
pulsars \cite{lai2011, Demorest2010,Antoniadis2013}. The rest-mass
density and mass-radius relationships for spherical models can be seen in
Fig.~\ref{fig:plot_tov}, and the characteristics of the maximum mass
models are reported in Table \ref{tab:tovmax}. The LX EOS has also been
discussed in relation with the possibility of understanding some puzzling
observations related to compact stars, such as the energy release during
pulsar glitches \cite{zhou2014}, the peculiar X-ray flares \cite{xu2009}
and the optical/UV excess of X-ray-dim isolated NSs
\cite{wang2017}. Particularly, a solid QS model has been suggested in
order to understand those observations \cite{Xu2003}. However, as pointed
out in \cite{zhou2014}, the critical strain of such a star is very
small. A starquake will be induced when the relative difference in
ellipticity is $10^{-6}$, for most, between the actual configuration of
the star and the configuration as if the star is a perfect fluid. This is
consistent with the pulsar glitch observations on Vela. Therefore, we
find it a good approximation to calculate the quasi-equilibrium
configuration of such a star with perfect fluid assumption.

Understanding the models and properties of the QS EOSs that we want to
consider, we can modify the EOS part of the simulation code, which was
originally designed for NS models, accordingly. 

As mentioned above, in the case of NSs, a piecewise-polytropic EOS is
usually assumed to describe the EOS \cite{Read:2009a,
  Rezzolla_book:2013}. In each piece, the pressure and rest-mass density
are related as
\begin{equation}
p_i = \kappa_i \rho^{\Gamma_i}= \kappa_i \rho^{1+1/n_i}\,, 
\qquad   i=1,2,\ldots,N\,.
\label{eq:polyEOS}
\end{equation}

For QSs, due to the nonzero surface density and nonzero energy density
integration constant, we will assume that the EOS is generally a
polynomial
\begin{equation}
p = \sum\limits_{i=1}^{N}\kappa_i \rho^{\Gamma_i}\,.   
\label{eq:qsEOS}
\end{equation}

Given the relationship between $p$ and $\rho$, one can apply the first
law of thermodynamics to obtain other quantities such as the energy
density and the specific enthalpy. In the zero-temperature case, the
first law of thermodynamics can be expressed as \cite{Rezzolla_book:2013}
\begin{equation}
d\epsilon =\frac{\epsilon+p}{\rho} d\rho, \quad\mbox{or}\quad 
d\left(\frac{\epsilon}{\rho}\right)=\frac{p}{\rho^2}d\rho\,,
\label{eq:1stlaw}
\end{equation}
which can be integrated to obtain the total energy density. The integral
constant is usually chosen to be 1, since when there is no internal
energy, the energy density and the rest-mass density coincide (apart from
the square of the speed of light). However, for QS EOSs
(like in \cite{Guo2014} and the MIT bag-model EOS), the integral constant
is different from unity and needs to be properly taken into account.

\begin{table}
\begin{tabular}{ccc}
\multicolumn{3}{c}{MIT bag-model EOS}  \\
\hline
$C=-c^2$ & $\rho_s=1.4\rho_\mathrm{nuc}$ & $h_s=0.89697478c^2$ \\
\hline
$N=2$ & $\kappa_i$ & $\Gamma_i$  \\
\hline
$i=1$ & $2.7977907\times10^{15}$  & $\frac{4}{3}$ \\
$i=2$ & $-7.5279768\times10^{34}$ & $0$  \\
\hline
\hline  \\[-5pt]
\multicolumn{3}{c}{LX EOS}  \\
\hline
$C=0$ & $\rho_s=2.0\rho_\mathrm{nuc}$ & $h_s=0.96828675c^2$ \\
\hline
$N=3$ & $\kappa_i$ & $\Gamma_i$  \\
\hline
$i=1$ & $ 2.0824706\times10^{-39}$  & $5$  \\
$i=2$ & $ -6.1559375\times10^{-10}$ & $3$  \\
$i=3$ & $ 7.1307226\times10^{13}$   & $\frac{4}{3}$  \\
\hline
\hline
\end{tabular}
\caption{Quark EOS parameters in cgs units for MIT and the LX EOSs
  respectively. $\rho_\mathrm{nuc}$ is the nuclear saturation density
  which is chosen to be $2.67\times10^{14}\,\mathrm{g\, cm^{-3}}$. Unlike
  NSs which have a surface enthalpy $h_s=1$, for QSs this value is an
  input parameter for a specific model. In practice one inputs the
  rest-mass density at the surface $\rho_s$ and then $h_s$ is computed
  from the EOS. The constant $C$ is the integration constant of the first
  law and determines the limit of the thermodynamic variables at the
  surface of the star. }
\label{tab:EOSparfile}
\end{table}

In this case, the energy density and specific enthalpy are related to the
rest-mass density by 
\begin{equation}
\epsilon=\sum\limits_{i=1}^{N}\frac{\kappa_{i}}{\Gamma_{i}-1}\rho^{\Gamma_{i}}+\rho(1+C)\,,
\label{eq:ene}
\end{equation}
\begin{equation}
h=\frac{\epsilon+p}{\rho}=\sum\limits_{i=1}^{N}\frac{\Gamma_{i}\kappa_{i}}{\Gamma_{i}-1}\rho^{\Gamma_{i}-1}+1+C\,.
\label{eq:enthalpy}
\end{equation}
Here $C$ is the integral constant we mentioned above. It is usually taken
to be zero for NS models. Here it is introduced again in order to
accommodate stars that have different surface limit for
the thermodynamic variables.

The nonzero surface density of QSs requires a different boundary
condition in our simulation. For typical NSs when we adjust the position
of the surface, we are actually locating the points where the specific
enthalpy is $1$. For QSs, the surface identification will be at values of
the specific enthalpy different from unity and
consistent with Eq. (\ref{eq:enthalpy}). What we
typically use as input parameter is the surface rest-mass density
$\rho_s$, from which we then calculate $h_s$ using
Eq. (\ref{eq:enthalpy}).

As an example, for MIT bag-model EOS the first law at zero temperature
implies
\begin{eqnarray}
\epsilon & = & \frac{1}{1+\sigma}\left( \bar{C} \rho^{1+\sigma} + \sigma \epsilon_s \right) , \label{eq:mit_gen_eps} \\
p & = & \frac{\sigma}{1+\sigma}\left( \bar{C} \rho^{1+\sigma} -\epsilon_s \right) , \label{eq:mit_gen_p} \\
h & = & \bar{C} \rho^\sigma,
\end{eqnarray}
where $\bar{C}$ is a constant of integration. The above EOS is of the
form (\ref{eq:qsEOS}) with
\begin{equation}
\kappa_1 = \frac{\sigma \bar{C}}{1+\sigma},\quad \Gamma_1 = 1+\sigma,\quad 
    \kappa_2 =-\frac{\sigma \epsilon_s}{1+\sigma},\quad \Gamma_2 = 0\,, 
\end{equation}
and $C=-1$. Having all thermodynamical variables in terms of the
rest-mass density is convenient from the computational point of view since
this is one of the fundamental variables used in the \cocal{} code,
therefore the modifications with respect to the EOS will be minimal.

Given a fixed choice of $\sigma$ and $B$ for the MIT bag-model
EOS, one can obtain a unique solution of the field
equations under hydrostatic equilibrium (see Secs.
\ref{sec:IWMformulation} and \ref{sec:fluidformulation}). In other words,
the relationship between the gravitational mass versus the central energy
density, the mass-radius relationship and the spacetime metric will not
depend on the coefficient $\bar{C}$, as it is eliminated out from
Eq.~\eqref{eq:mit_gen} (similar argument can be found in
\cite{Li2017,Bhattacharyya2016}). At the same time, it will indeed affect
the rest mass hence the binding energy of the QS since
it relates the rest-mass density and the number density of the
components. Hence a reasonable choice for $\bar{C}$ will still be helpful
although it will not affect anything that we are interested in for this
work. Here we chose $\bar{C}$ such that the EOS corresponds to the $a_4 =
0.8$ model as in \cite{Alford2005} and we assume a rest mass of
$931\,{\rm MeV}/c^2$ for each baryon number ($n_b=n_q/3$, where $n_q$ is
the number density of quarks). Any other choices for $\bar{C}$ are in
principle possible and they will not affect our solution except for the
rest mass of the star. Actual values of these constants can be found in
Table \ref{tab:EOSparfile} in cgs units.

In Eqs.~\eqref{eq:mit_gen_eps} and \eqref{eq:mit_gen_p} we have employed
the relationship which is quite similar to the explicit form of MIT bag
model. By factoring out the rest mass, those two equations can be rewritten
as a function of number density instead of rest-mass density. However,
one clarification we want to discuss at this point is that, the choice of
using Eq.~\eqref{eq:mit_gen_eps} and \eqref{eq:mit_gen_p} is not
essential. Moreover, Eqs.~\eqref{eq:mit_gen_eps} and
  \eqref{eq:mit_gen_p} are related by the first law of thermodynamics [see
  Eq.~\eqref{eq:1stlaw}], which is not essential as well. We can describe
the MIT EOS Eq. (\ref{eq:mit_gen}), in a parametric form
$(p(\rho),\epsilon(\rho))$ with an arbitrary parameter $\rho$, as long as
the functions $p(\rho)$ and $\epsilon(\rho)$ satisfy
Eq. (\ref{eq:mit_gen}). In doing so we choose to satisfy
Eq. (\ref{eq:1stlaw}) and therefore arrive at Eq. (\ref{eq:mit_gen_eps}),
(\ref{eq:mit_gen_p}). As one can see from
Eqs. (\ref{eq:rhoh})--(\ref{eq:s}), the fluid terms that appear are
$q\rho=p$ and $h\rho=\epsilon+p$. Thus, for the MIT bag-model EOS, the
only thermodynamic variable that appears in the field equations is
$\epsilon$. Every model thus calculated will be uniquely defined by a
deformation parameter and the central total energy density. The scaling
constant which is analogous to scaling as $\kappa^{n/2}$ for NSs
\cite{Rezzolla_book:2013}, is here $\epsilon_s^{-1/2}$.

\begin{table}
\begin{tabular}{cccccccccccc}
\hline
EOS & $(p/\rho)_c$ & $\epsilon_c$ & $\rho_c$ & $M$ & $\compa$ \\
\hline
MIT& 0.2940 & $2.609\times10^{-3}$ & 	$2.342\times10^{-3}$ & 	2.217 & 	0.2706 & 	 \\
LX& 2.326 & $2.451\times10^{-3}$	& $1.744\times10^{-3}$ & 3.325	 & 	0.3956 & 	 \\
\hline
\end{tabular}
\caption{Pressure, energy density, rest-mass density, gravitational mass,
  and compactness at the maximum mass of spherical solutions for the two
  EOSs in this work.}
\label{tab:tovmax}
\end{table}

Before concluding this discussion on the EOSs we should mention that
although a polynomial-type EOS is implemented for the calculation of QSs
with a finite surface density, the developments in the new version of
\cocal{} allow us to calculate any compact star with an EOS that can be
described by a polynomial function, including NSs and hybrid stars. For
instance, some phenomenological approaches suggested recently in
Ref. \cite{Baym2017} also result in a polynomial-type EOS, which can be
computed straightforwardly with the new code.

\section{Code tests}
\label{sec:codetest}

The working properties of the \cocal{} code for single rotating stars is
presented in detail in previous works, \eg
\cite{Huang08,Uryu2016a}\footnote{See
  \cite{Uryu2012,Uryu:2012b,Tsokaros2012,Tsokaros2015} for the general
  binary case.}, so that here we will only mention the most important
quantities that are used in our simulations. The method has its origins
in the works of Ostriker and Marck \cite{Ostriker1968}, who used it to
compute Newtonian stars, and the works of Komatsu, Eriguchi, and Hachisu
\cite{Komatsu89}, who devised a stable numerical algorithm and obtained
first axisymmetric general relativistic rotating stars. From this latest
work the method is commonly referred as the KEH method
and consists of an integral representation of the Poisson equation
commonly referred as the representation formula. Since we have only one
computational domain with trivial boundary conditions at infinity, the
Green's function is $G(x,x')=1/|x-x'|$ and is expanded as a series of the
associated Legendre polynomials and trigonometric functions. The maximum
number of the terms included in this expansion is given as $L$ in Table
\ref{tab:grid}.

This approach is used to compute the gravitational fields
$\{\alpha,\psi,\beta^i\}$ while hydrostatic equilibrium is achieved
through Eq.~(\ref{eq:firstint}). At every step in the iteration to reach
the solution at the desired accuracy, three constants need to be
computed. The first one is the angular velocity of the star, $\Omega$,
while the second one is the constant from the Euler integral,
$\mathcal{E}$, in Eq.~(\ref{eq:firstint}), and, finally, the third
constant is $R_0$, a normalization factor for the whole domain where the
equations are solved.\footnote{We recall that \cocal{} uses normalized
  variables $\hat{x}^i:=x^i/R_0$ and the quantities listed in Table
  \ref{tab:reso} refer to those and should be denoted
  by a hat. For simplicity, however, we have omitted these hats in the
  Table.} At every step during the iteration the nonlinear equation with
respect to these three constants is solved typically by evaluating
Eq.~(\ref{eq:firstint}) at three points in the star. For axisymmetric
configurations, we use the center of the star and two points on the
surface, one on the positive $x$-axis and one at the North pole of the
stellar model. An axisymmetric equilibrium is achieved by setting the
ratio between the polar axis over the equatorial radius along the
$x$-axis. For triaxial configurations, on the other hand, the three
points are the center of the star together with two points again on the
surface, one at the positive $x$-axis, and one on the positive
$y$-axis. Each triaxial solution has a fixed ratio of the radius on the
$y$-axis over the radius of the $x$-axis.

From a numerical point of view, \cocal{} is a finite-difference code
that uses spherical coordinates $(r,\theta,\phi)\in [0,r_b]\times
[0,\pi]\times [0,2\pi]$\footnote{Note that the field equations for the
  shift vector Eq.~(\ref{eq:MoC_elip2}) are expressed in Cartesian
  coordinates.}  and the basic parameters are summarized in Table
\ref{tab:grid}.  In the angular directions $\theta,\phi$ the
discretization is uniform, \ie $\Delta\theta=\pi/N_\theta$, and
$\Delta_\phi=2\pi/N_\phi$. In the radial direction the grid is uniform
until point $r_c$ with $\Delta r_i = r_c/\Nrm$, and in the interval
$[r_c,r_b]$ the radial grid in non-uniform and follows a geometric series
law \cite{Huang08}. While field variables are evaluated at the
gridpoints, source terms under the integrals are evaluated at midpoints
between two successive gridpoints since the corresponding integrals use
the midpoint rule. For integrations in $r$ and $\phi$ we use a
second-order midpoint rule.  For integrations in $\theta$ we use a
fourth-order midpoint rule. This was proven necessary to keep second
order convergence at the region of maximum field strength
\cite{Uryu:2012b}.  Derivatives at midpoints are calculated using
second-order rule for the angular variables $\theta,\phi$, and
third-order order rule for the radial variable $r$ (again for keeping
second-order convergence at same regions \cite{Uryu2012}). Derivatives
evaluated at gridpoints always use a fourth-order formula in all
variables.

\begin{table}
\begin{tabular}{lll}
\hline
$r_{a}$ &:& Radial coordinate where the grid $r_i$ starts.                \\
$r_{b}$ &:& Radial coordinate where the grid $r_i$ ends.              \\
$r_{c}$ &:& Radial coordinate between $r_{a}$ and $r_{b}$ where the      \\
&\phantom{:}& grid changes from equidistant to non-equidistant.        \\
$N_{r}$ &:& Total number of intervals $\Dl r_i$ between $r_{a}$ and $r_{b}$. \\
$\Nrm$ &: & Number of intervals $\Dl r_i$ in $[0,r_{c}]$. \\
$\Nrf$ &:& Number of intervals $\Dl r_i$ in $[0,R(\theta,\phi)]$. \\
$N_{\theta}$ &:& Total number of intervals $\Dl \theta_i$ for $\theta\in[0,\pi]$. \\
$N_{\phi}$ &:& Total number of intervals $\Dl \phi_i$ for $\phi\in[0,2\pi]$. \\
$L$ &:& Number of multipole in the Legendre expansion. \\
\hline
\end{tabular}  
\caption{Summary of parameters used for rotating star configurations.}
\label{tab:grid}
\end{table}

\begin{table}
\begin{tabular}{ccccccccccc}
\hline
Type & $r_a$ & $r_b$ & $r_c$ & $N_r$ & $\Nrm$ & $\Nrf$ & $N_\theta$ & $N_\phi$ & $L$ \\
\hline
H2.0 & 0 & $10^6$ & 1.25 & 192 &  80 &  64 &  48 &  48 & 12 \\
H2.5 & 0 & $10^6$ & 1.25 & 288 & 120 &  96 &  72 &  72 & 12 \\
H3.0 & 0 & $10^6$ & 1.25 & 384 & 160 & 128 &  96 &  96 & 12 \\
H3.5 & 0 & $10^6$ & 1.25 & 576 & 240 & 192 & 144 & 144 & 12 \\
H4.0 & 0 & $10^6$ & 1.25 & 768 & 320 & 256 & 192 & 192 & 12\\
\hline
\end{tabular}
\caption{Five different resolutions used for convergence tests.
  Parameters are shown in Table \ref{tab:grid}. The number of points that
  covers the largest star radius is $\Nrf$.}
\label{tab:reso}
\end{table}

It is worth noting that, since for QSs the relationship between the
specific enthalpy and the rest-mass density follows a general polynomial
function [\cf Eq.~\eqref{eq:enthalpy}], a root-finding method needs to be
employed when calculating the thermodynamical quantities from the
enthalpy. The regular polynomial expression of the specific enthalpy with
respect to rest-mass density allows us to use a Newton-Raphson method as
the derivative can also be expressed easily. In view of this, the
computational costs with a QS EOS are not significantly larger than those
with NS EOS. However, in order to guarantee a solution of rest-mass
density when the specific enthalpy is given, a bi-section root-finding
method needs to be employed if the Newton-Raphson method does not converge
sufficiently rapidly. In this case, the initial range of the bi-section
method is set to be the specific enthalpy corresponding to the rest-mass
densities at the stellar center and at the surface, respectively.

In previous works \cite{Huang08,Uryu:2012b,Uryu2016a,Tsokaros2016}, the
\cocal{} code has been extensively tested, both with respect to its
convergence properties, as well as with respect in actual
evolutions with other well established codes \cite{Tsokaros2016}. 

In what follows we report the convergence tests we have performed in
order to investigate the properties of the code under these new
conditions. Before doing that, we note that special care is needed when
using a root-finding method to calculate thermodynamical quantities for a
given specific enthalpy in the case of rotating QSs. In particular, it is
crucial to consider what is the accuracy set during the root-finding
step. Of course, it is possible to require that the accuracy in the
root-finding step is much higher than the other convergence criteria in
the code to guarantee an accurate result. In this way, however, the
computational costs will increase considerably, since the thermodynamical
quantities need to be calculated at every gridpoint and at every
iteration. We found that an accuracy of $10^{-10}$ for the thermodynamic
variables solutions neither compromises the accuracy of the solutions,
nor slows down the code significantly.

\subsection{Comparison with rotating NSs}

Although the newly developed code presented here is intended for QS EOSs,
it can be also used to produce rotating NSs if one restricts the EOS to a
single polytrope. This can be accomplished by setting the polynomial
terms to be only one, the surface rest-mass density $\rho_s=0$ and the
energy integral constant $C=0$. In this case,
Eq.~\eqref{eq:qsEOS} becomes
\begin{equation}
p=\kappa\rho^\Gamma=\kappa\rho^{1+1/n}\,,
\end{equation}
and the relationship between the energy density, the specific enthalpy
and the rest-mass density [Eqs. \eqref{eq:ene} and \ref{eq:enthalpy}]
will be exactly the same as that for a polytropic NS.

We choose a stiff EOS with $n=0.3$ and produce axisymmetric solution
sequences for small and high compactness $\compa:=M_{\rm ADM}/R=0.1,
0.2$, and $M_{\rm ADM}$ is the corresponding Arnowitt-Deser-Misner (ADM)
mass for a nonrotating model, both with the original rotating-NS and the
modified rotating-QS solver. The grid-structure parameters
used are $N_r=240$, $\Nrm=80$, $\Nrf=64$, $N_\theta=96$, $N_\phi=192$,
$L=12$, $r_b=10^4$, and $r_c=1.25$ (see Table \ref{tab:grid}). Overall,
we have found that the relative difference in all physical quantities is
of the order of $10^{-6}$, which is what is expected since the criteria
for convergence in \cocal{} are that the relative difference in metric
and fluid variables between two successive iterations is less than
$10^{-6}$.

\subsection{Convergence test for rotating QSs}

\begin{figure*}
\begin{center}
\includegraphics[width=0.95\columnwidth]{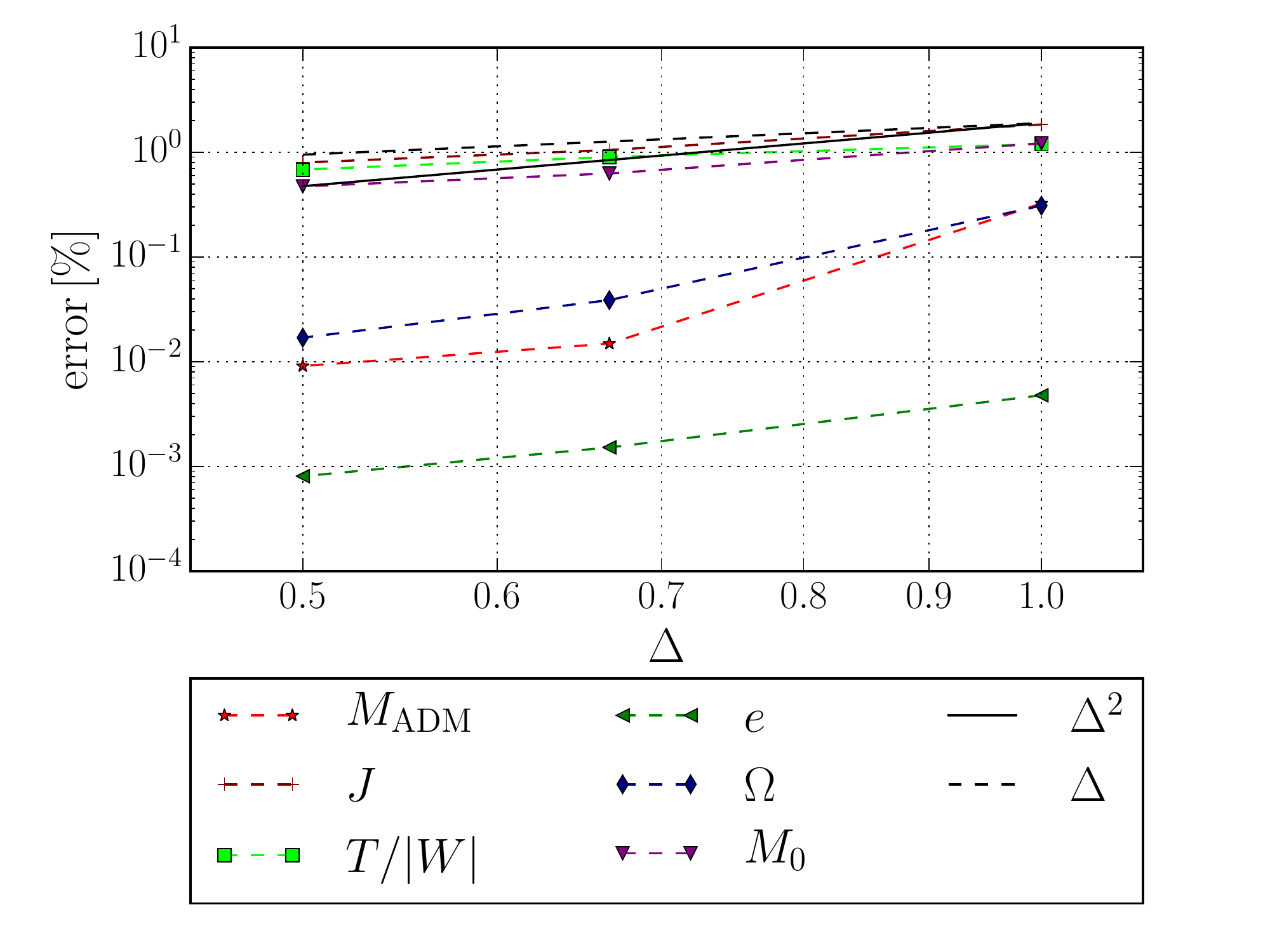}
\includegraphics[width=0.95\columnwidth]{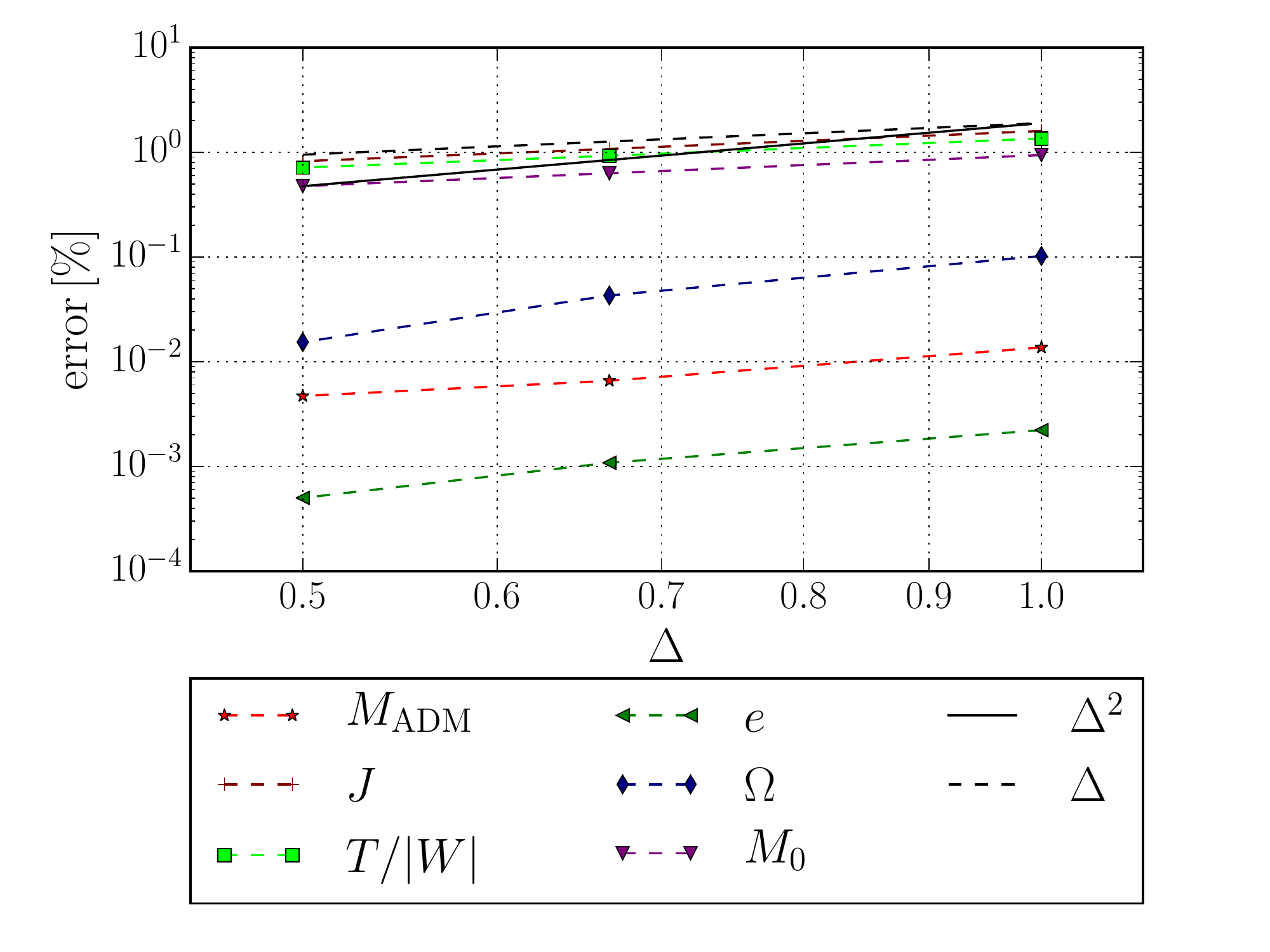}
\end{center}
\caption{Normalized differences $|(f_{{\rm H}3.0} - f_{{\rm
      H}2.0})/f_{{\rm H}4.0}|$, $|(f_{{\rm H}3.5} - f_{{\rm
      H}2.5})/f_{{\rm H}4.0}|$, and $|(f_{{\rm H}4.0} - f_{{\rm
      H}3.0})/f_{{\rm H}4.0}|$ are plotted for $\Omega$, $\Madm$, $J$,
  $T/|W|$, $M_0$ and $e:=\sqrt{1-(\bR_z/\bR_x)^2}$ against the
  resolutions $\Delta_{{\rm H}3.0}$, $\Delta_{{\rm H}2.5}$, and
  $\Delta_{{\rm H}2.0}$.  Black solid line is proportional to $\Delta^2$,
  while black dashed line is proportional to
  $\Delta$. The left panel refers to the LX EOS (Table
  \ref{tab:EOSparfile}) with central rest-mass density $\epsilon_c=1.301
  \times 10^{-3}$ and axis ratio in the coordinate length $R_z/R_x =
  0.75$.  The right panel is the same but for the MIT bag-model EOS with
  $\epsilon_c=7.361\times10^{-4}$ and the same deformation as the LX
  plot. }
\label{fig:convtest_RQS}
\end{figure*}

For the convergence analysis in this work we use the five resolutions
shown in Table \ref{tab:reso}. The outer boundary of the domain is placed
at $r_b=10^6$, while the surface of the star is always inside the sphere
$r=1$. The radius along the $x$-axis is exactly $r=1$ in the normalized
variables. There are exactly $\Nrf$ intervals along the radii in the $x$,
$y$, and $z$ directions. The number of Legendre terms used in the
expansions is kept constant ($L=12$) in all resolutions since convergence
with respect of those has been already investigated in
\cite{Tsokaros2007}. When going from the low-resolution setup H2.0 to the
high-resolution one H4.0, the spacings $\Delta r,\Delta\theta,
\Delta\phi$ decrease as $2/3,\ 3/4,\ 2/3,\ 3/4$.

As a result, if we denote as $f_{\mu}$, $f_{\nu}$ a quantity evaluated at
two different resolutions, then
\begin{equation}
f_{\mu}\,-\,f_{\nu} \, \approx \,
A\left[ \left(\frac{\Delta_\mu}{\Delta_\nu}\right)^n-1 \right] 
\Delta^n_\nu\,,
\label{eq:convtest}
\end{equation}
where $A$ is a constant and $\Delta_\mu$ is the grid separation at
resolution ${\rm H}\mu$. Choosing the combinations $f_{{\rm
    H}3.0}-f_{{\rm H}2.0}$, $f_{{\rm H}3.5}-f_{{\rm H}2.5}$, and $f_{{\rm
    H}4.0}-f_{{\rm H}3.0}$ so that we have $\Delta_{\mu} /
\Delta_{\nu}=1/2$ and normalizing by $f_{{\rm H}4.0}$ we plot in
Fig. \ref{fig:convtest_RQS} the relative error with respect to the grid
spacing for $\Omega$, $\Madm$, $J$, $T/|W|$, $M_0$ and the eccentricity
$e:=\sqrt{1-(\bR_z/\bR_x)^2}$, both for the LX EOS (left panel) and for
the MIT bag-model EOS (right plot). The deformation is kept at
$R_z/R_x=0.75$ for both EOSs, while the central densities are
$\epsilon_c=1.301\times10^{-3}$, and $\epsilon_c = 7.361\times10^{-4}$
for LX and MIT bag-model EOS respectively. The dashed black line reports
a reference first-order convergence, while the solid black line refers to
second-order convergence.

Note that quantities like the ADM mass, the angular velocity and the
eccentricity converge to second order, while quantities like the angular
momentum, the ratio $T/|W|$, and the rest-mass converge to an order that
is closer to first. Furthermore, Fig. \ref{fig:convtest_RQS} shows that
some quantities (\eg the ADM
mass of the LX EOS) shows a convergence order that is larger than 
second, 
but this is an artefact of the specific deformation. In
general, we found second-order convergence in $M_{\rm ADM},\ \Omega,\ e$
and at least first order for $J,\ T/|W|,\ M_0$. 

Note also that the two panels in Fig. \ref{fig:convtest_RQS} are very
similar, even though the EOSs are quite different, with the MIT bag-model
EOS being relatively soft (\ie $p\propto \rho^{4/3}$), while the LX EOS
is comparatively stiff (\ie $p\propto \rho^5$). Hence, the overall larger
error that is reported in Fig. \ref{fig:convtest_RQS} when compared to
the corresponding Fig. 1 in Ref. \cite{Huang08}, is mostly due to the
finite rest-mass density at the stellar surface. In the original
rotating-NS code, in fact, the surface was determined through a
first-order interpolation scheme. This approach, however, is not
sufficiently accurate for rotating QSs and would not lead to the desired
convergence order unless the surface finder scheme was upgraded to second
order.

\section{Triaxial solutions}
\label{sec:triasolution}

The onset of a secular instability to triaxial solutions for the MIT
bag-model EOS stars has been studied previously via a similar method in
Ref. \cite{rosinska2003}. Surface-fitted coordinates have been used to
accurately describe the discontinuous density at the surface of the star,
and a set of equations similar to the one of the conformal flat
approximation used here was solved. In order to find the secular bar-mode
instability point, the authors of Ref. \cite{rosinska2003} performed a
perturbation on the lapse function of an axisymmetric solution and build
a series of triaxial quasi-equilibrium configurations to see whether this
perturbation is damped or grows.

Here, we build quasi-equilibrium sequences with constant rest
mass (axisymmetric and triaxial) for both the MIT bag-model EOS and the LX
EOS. We begin with the axisymmetric sequence in which we calculate a series of
solutions with varying parameters, \ie the parameters that determine the
compactness (\eg the central rest-mass density $\rho_c$) and the rotation
($R_z/R_x$). In doing so, we impose axisymmetry as a separate condition
and manage to reach eccentricities as high as $e\simeq 0.96$ for
$R_z/R_x=0.2656$ and compactness $\compa=0.1$.  In order to access the
triaxial branch of solutions, we recompute the above sequence of
solutions but this time \textit{without} imposing axisymmetry. As the
rotation rate increases ($R_z/R_x$ decreases) the triaxial deformation
($R_y/R_x <1$) is \textit{spontaneously} triggered, since at large
rotation rate the triaxial configuration possesses lower total energy and
is therefore favoured over the axisymmetric solution. This approach is
different from the approach followed in Ref. \cite{rosinska2002}, where
the triaxial $m=2$ perturbation was triggered after a suitable
modification of a metric potential.

\begin{figure}
  \begin{center}
    \includegraphics[width=0.95\columnwidth]{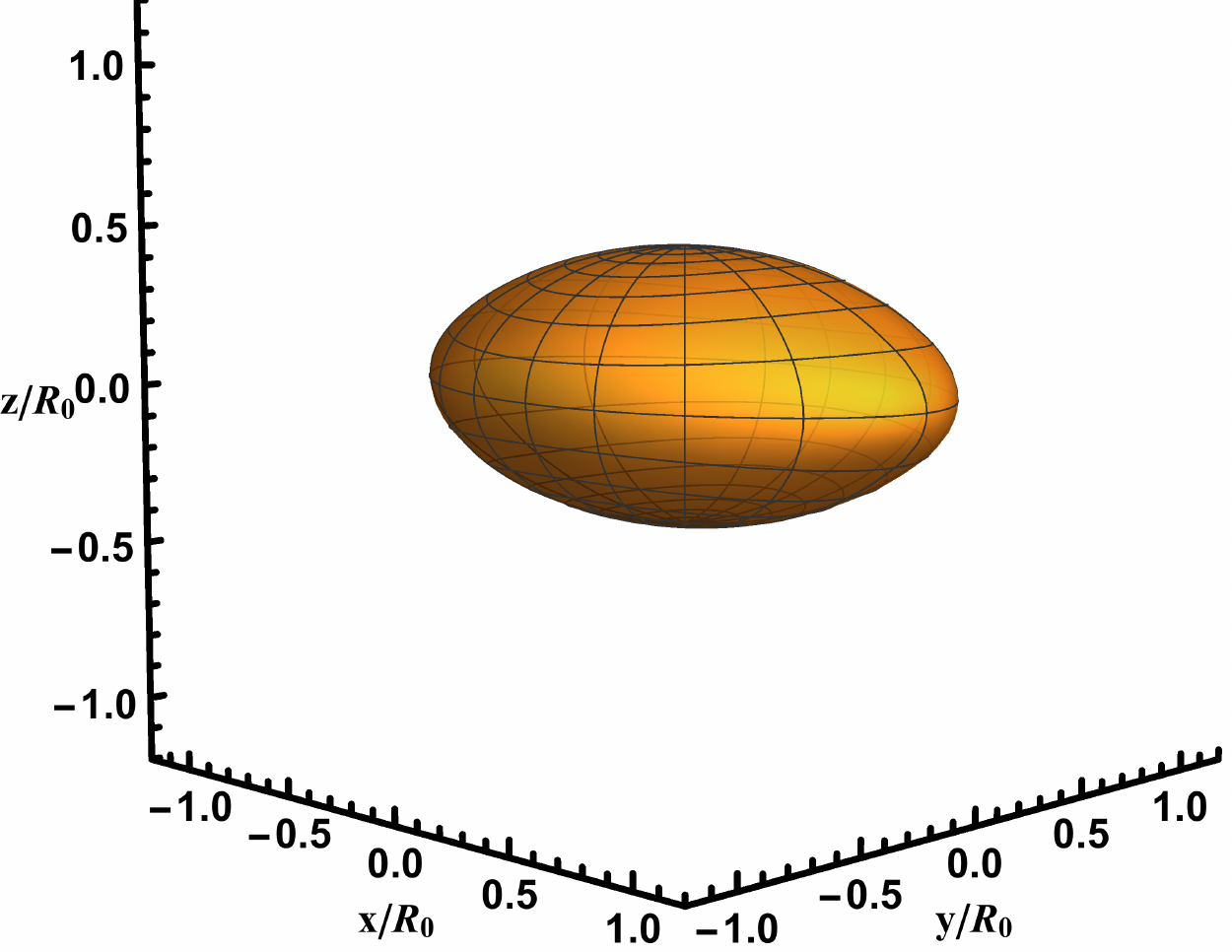}
  \end{center}
  \caption{Illustration of the three-dimensional surface of a QS solution
    with the largest triaxial deformation for the MIT bag-model with
    corresponding spherical compactness $\compa=0.2$. The axis ratio
    $R_y/R_x$ is 0.6757 and $R_z/R_x=0.4375$. The solid black lines on
    the surface corresponds to fixed values of the latitude angle and the
    fact that they are not parallel is a result of the triaxial
    deformation.}
	\label{fig:qs_surf}
\end{figure}

We keep decreasing $R_z/R_x$ to reach the mass shedding limit with the
triaxial configuration. We can then move along the triaxial solution
sequence by increasing $R_y/R_x$ which now acts as the new rotating
parameter. The sequence is then terminated close to the axisymmetric
sequence. The bifurcation point can be found by extrapolating this
triaxial sequence towards the axisymmetric solutions. The largest
triaxial deformation calculated in this work, for both the MIT bag-model
EOS and the LX EOS, is $R_y/R_x=0.5078$ for the $\compa=0.1$ case (a
three-dimensional of the surface for this solution is shown in
Fig.~\ref{fig:qs_surf}), $R_y/R_x=0.5234$ for $\compa=0.15$, and
$R_y/R_x=0.6757$ for $\compa=0.2$. Similar with NSs \cite{Huang2007}, the
endpoint of the triaxial sequence happens in lower eccentricities as the
compactness increases.

\begin{figure*}
\begin{center}
\includegraphics[width=0.95\columnwidth]{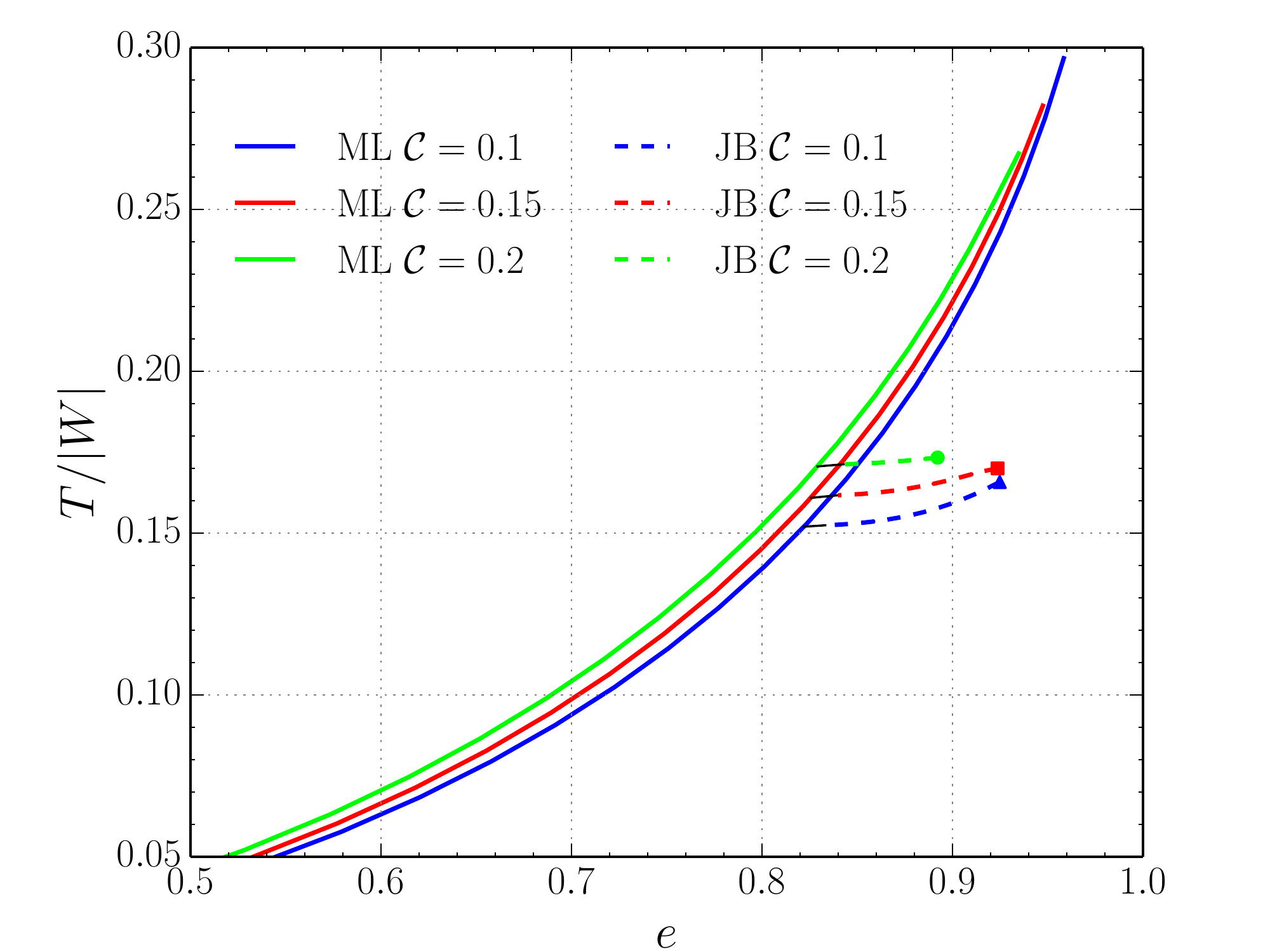}
\includegraphics[width=0.95\columnwidth]{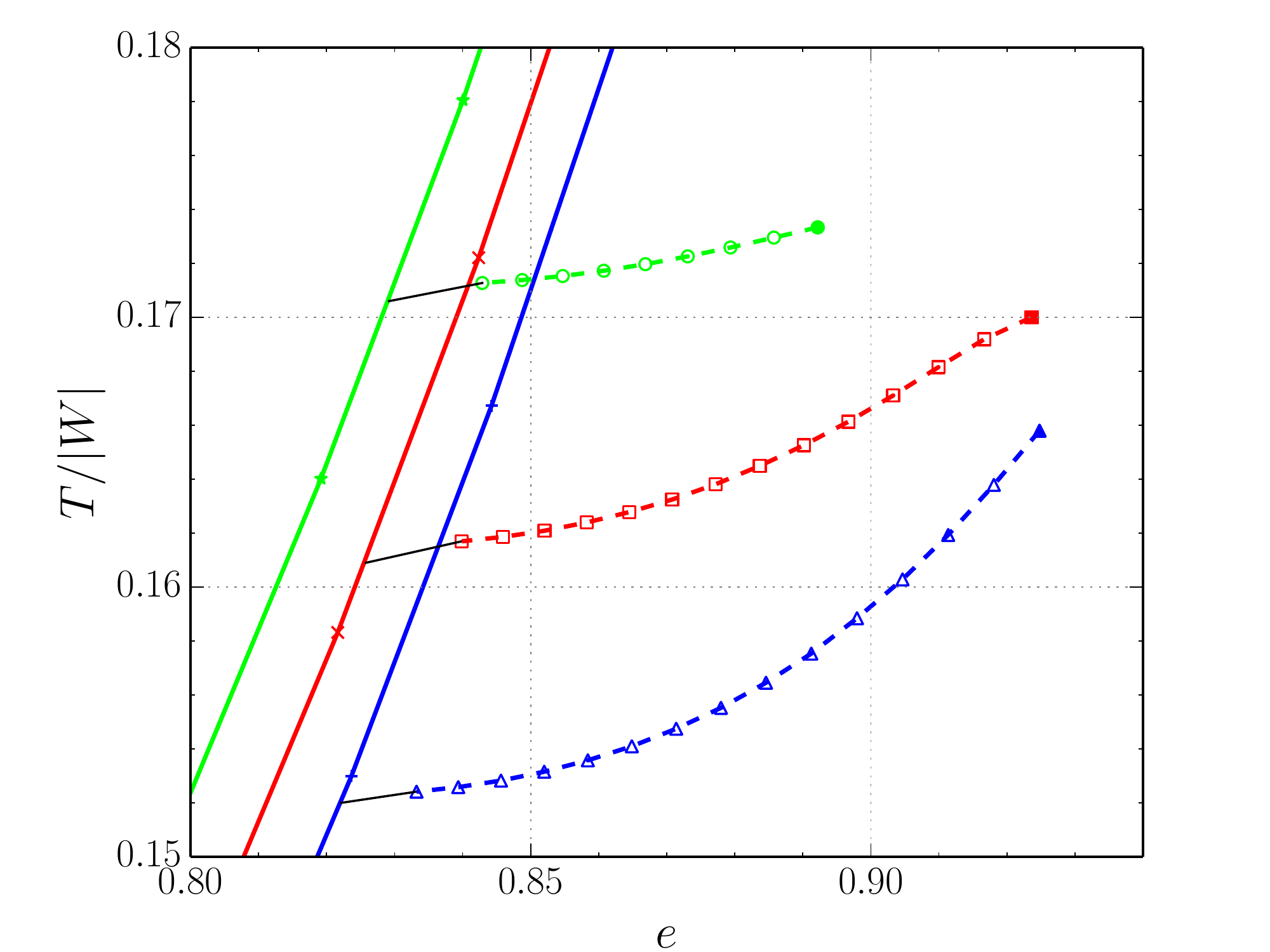}
\end{center}
\caption{\textit{Left panel:} $T/|W|$ versus eccentricity
  $e:=\sqrt{1-(\bR_z/\bR_x)^2}$ (in proper length) for MIT bag-model EOS
  sequences. Solid curves are axisymmetric solution sequences, and dashed
  curves are triaxial solution sequences, that correspond, to
  $\compa=M/R=0.2$ (green curves), $0.15$ (red curves) and $0.1$ (blue
  curves) respectively. Note that $M$ is the spherical ADM mass.
  \textit{Right panel:} magnification of the region near the onset of the
  triaxial solutions marked with empty symbols, while filled symbols mark
  the models at the mass-shedding limit. Solutions labelled with "ML" are
  axisymmetric solutions (Maclaurin spheroids), while those labeled
  ``JB'' are triaxial solutions (Jacobi ellipsoids). }
\label{fig:plot_mit}
\end{figure*}

\begin{figure*}
\begin{center}
\includegraphics[width=0.95\columnwidth]{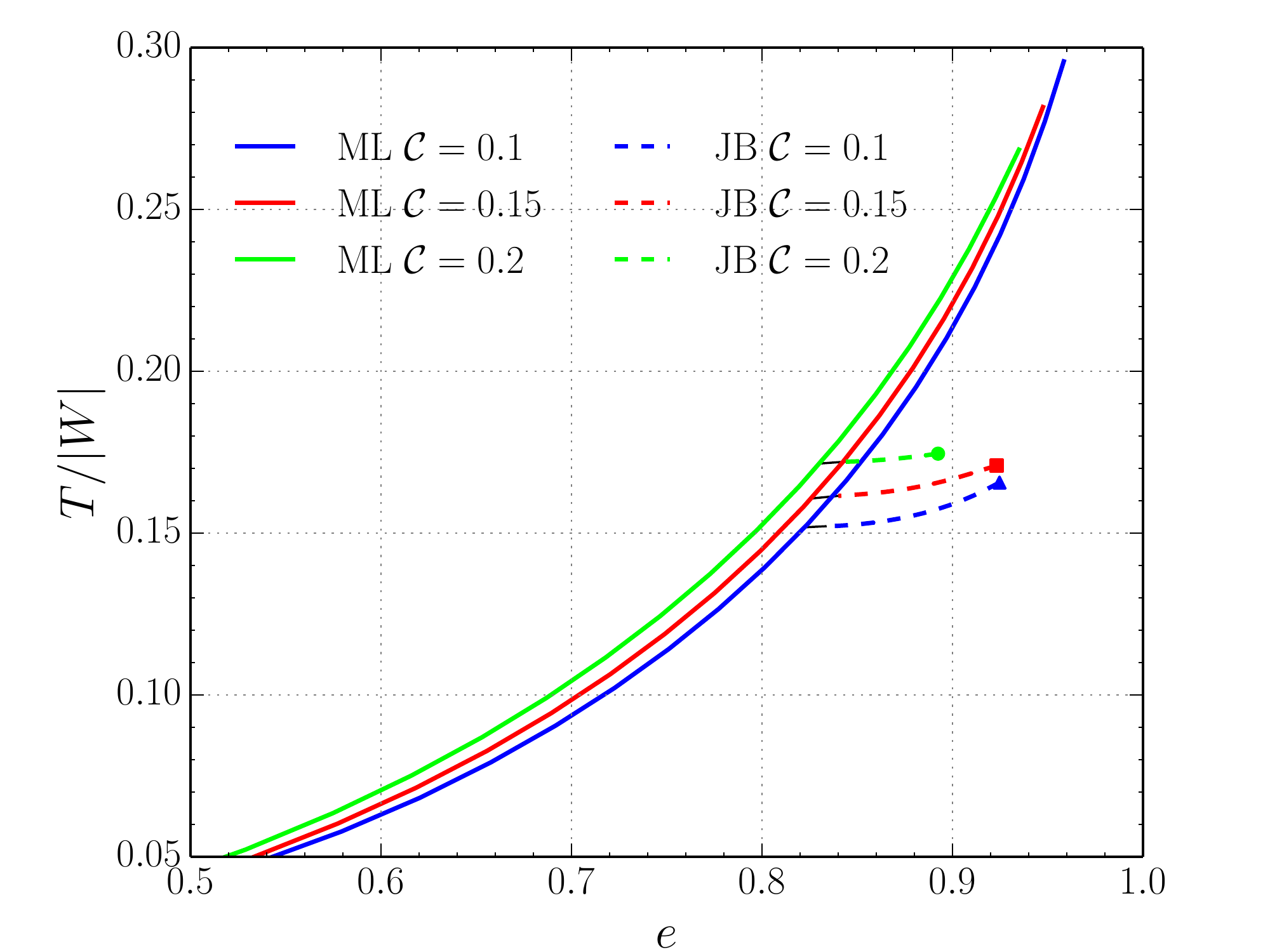}
\includegraphics[width=0.95\columnwidth]{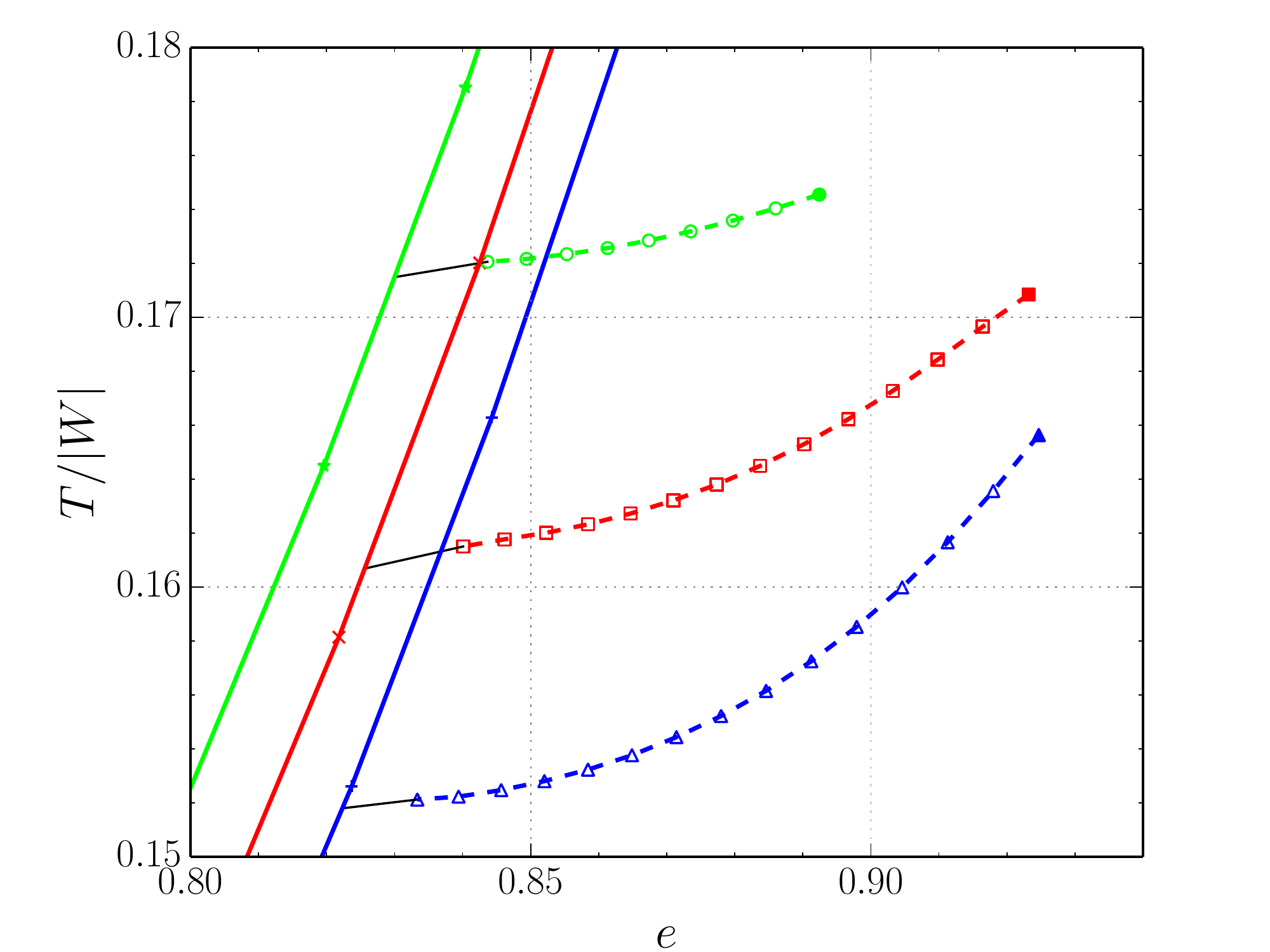}
\end{center}
\caption{The same as Fig.~\ref{fig:plot_mit} but for the LX EOS sequences.}
\label{fig:plot_lx}
\end{figure*}


In Figs. \ref{fig:plot_mit} and \ref{fig:plot_lx}, the relation between
the $T/|W|$ ratio versus the eccentricity of the star has been plotted
for three different compactnesses
($\compa=0.1,\ 0.15\ \mathrm{and}\ 0.2$) for both the MIT bag-model EOS
and the LX EOS.  Unlike in a Newtonian incompressible star, for which the
bifurcation to triaxial deformation happens at $(T/|W|)_{\rm crit,
 Newt}\simeq 0.1375$ for any compactness, in general relativity the
bifurcation point depends on the compactness. According to
\cite{rosinska2002},
\begin{equation}
\left(\frac{T}{|W|}\right)_\mathrm{\rm
  crit}=\left(\frac{T}{|W|}\right)_{\rm crit, Newt}
+0.126 \, \compa\left(1+\compa\right) \,.
\label{eq:twrelation}
\end{equation}
This relation holds true not only for NSs but also for QSs with the MIT
bag-model EOS (see Fig. 1 of \cite{rosinska2003}). The largest $T/|W|$
for the onset of secular instability is found to be $\simeq 0.17$ for
rotating QSs in the configurations that we considered for both the MIT
bag-model EOS and the LX EOS, and it will be even larger for higher
compactnesses. Both the LX EOS and the MIT bag-model EOS in our
calculations follow this relationship within a maximum error of
$3\%$. This implies that the secular instability to a "Jacobi type"
ellipsoidal figure in general relativity is not particularly affected by
the stiffness of the EOS for quark matter.

\begin{figure}
\begin{center}
\includegraphics[width=0.95\columnwidth]{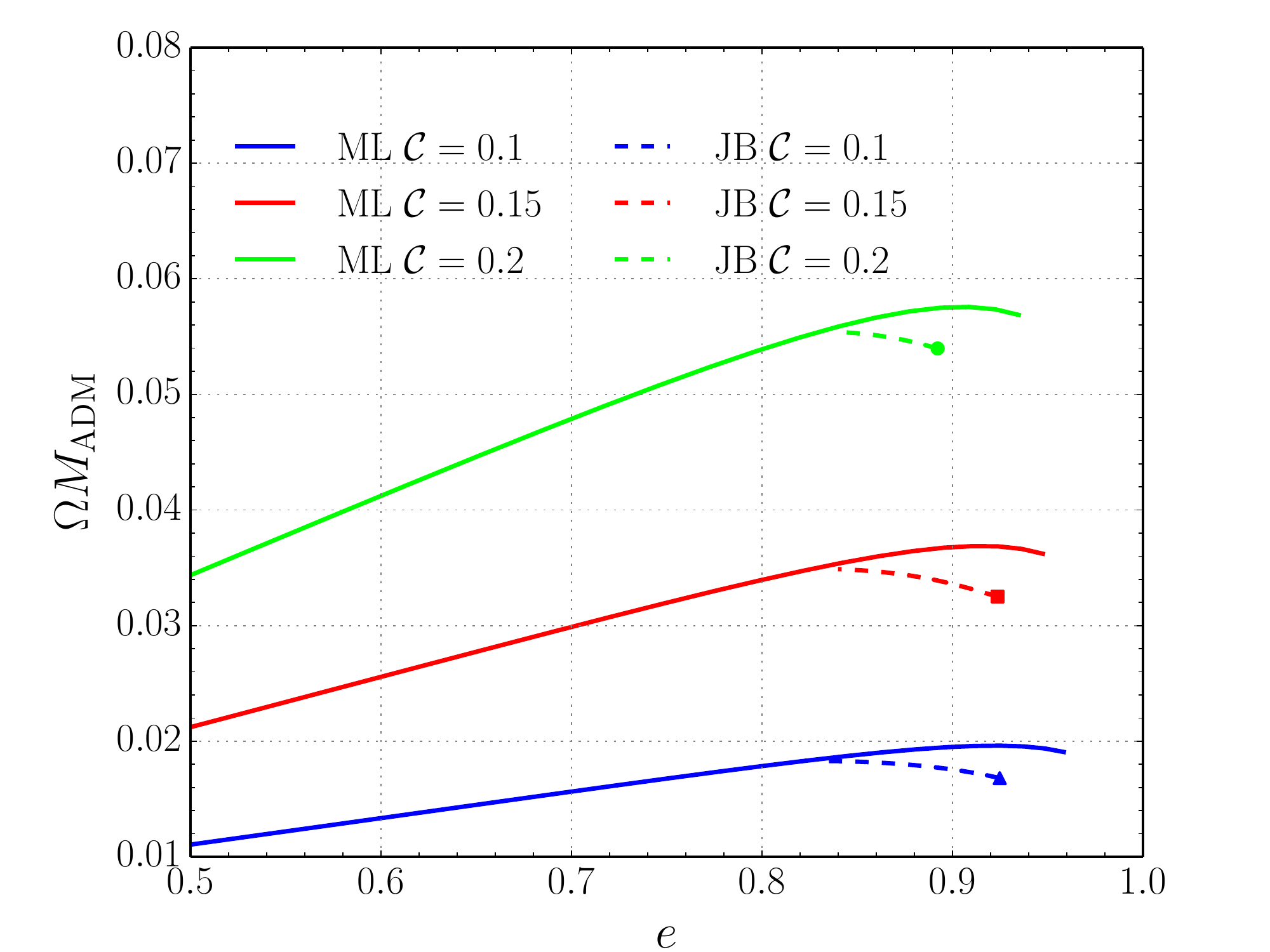}
\end{center}
\caption{Plots of $\Omega M_{\rm ADM}$ versus eccentricity for MIT bag
  model sequences. Solid curves are axisymmetric solution sequences, and
  dashed curves are triaxial solution sequences, that correspond to
  $\compa=M/R=0.2$ (top green curve), $0.15$ (middle red curve) and $0.1$
  (bottom blue curve) respectively. Note that $M$ is the spherical ADM
  mass. }
\label{fig:plot_mit_omeM}
\end{figure}

\begin{figure}
\begin{center}
\includegraphics[width=0.95\columnwidth]{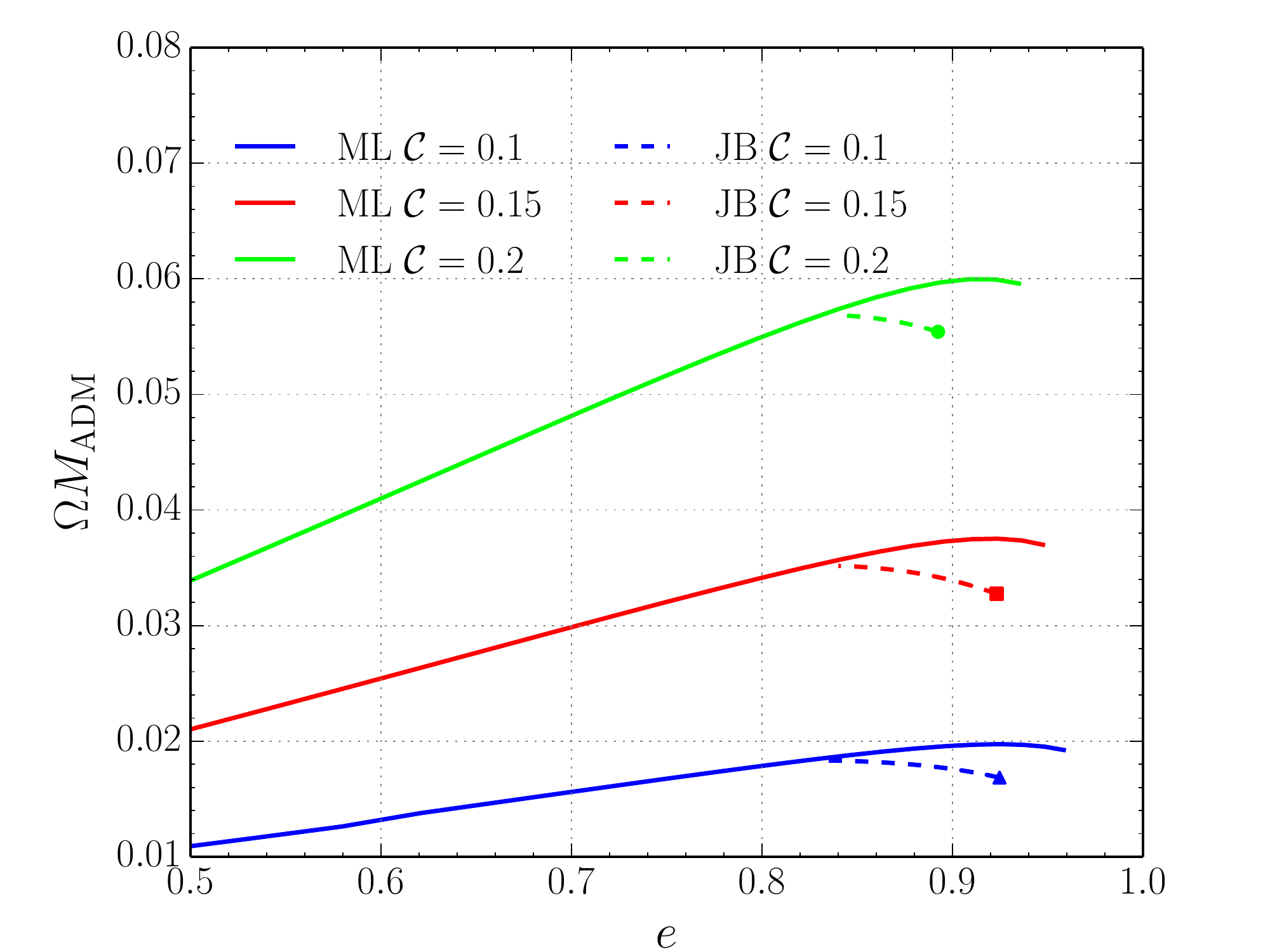}
\end{center}
\caption{The same as Fig.~\ref{fig:plot_mit_omeM} but
  for the LX EOS sequences. }
\label{fig:plot_lx_omeM}
\end{figure}

\begin{table*}
\begin{tabular}{ccccccccccccc}
\hline
EOS&$\compa$&$R_x$&$R_z/R_x$&$\epsilon_c$&$\Omega$&$\Madm$&$J$&$T/|W|$&$I$&$Z_{\rm p}$ \\
\hline
MIT& 0.1& 7.021 (8.077) & 	0.5647	(0.5693) & 	$ 6.200 \times 10^{-4} $	 & 	0.02808 & 	0.6515 & 	0.4580 & 	0.1520  & 	16.31  & 	0.1343 & 	 \\
MIT& 0.15& 7.962 (9.922) & 	0.5565 	(0.5640) & 	$ 6.811 \times 10^{-4} $ & 	0.02985 & 	1.169	 & 	1.293	 & 	0.1609	 & 	43.30	 & 	0.2231	  & \\
MIT& 0.2& 8.415 (11.43)	 & 	0.5478 	(0.5590)& 	$ 7.696	\times 10^{-4} $ & 	0.03199 & 	1.731 & 	2.649	 & 	0.1706 & 	82.83	 & 	0.3308	 & \\
LX& 0.1& 5.698	(6.557)	& 0.5644 (0.5689) & $9.144 \times 10^{-4} $	 & 	0.03451	 & 	0.5312 & 	0.3039 & 	0.1518	 & 	8.805	 & 	0.1343	 & 	 \\
LX& 0.15& 6.515 (8.130)& 0.5566 (0.5639)& 	$9.542 \times 10^{-4} $	 & 	0.03630 & 	0.9686& 	0.8850	 & 	0.1607	 & 	24.38	 & 	0.2251 & 	 	 \\
LX& 0.2& 6.972 (9.528)& 0.5469 (0.5574) & 	$9.977 \times 10^{-4} $	 & 	0.03838& 1.481& 	1.932  & 	0.1715 & 	50.34	 & 	0.3401	 & 	\\
\hline
$n=0.3$ & 0.1 & 6.624 (7.634) & 0.5634(0.5693) & $9.221 \times 10^{-4}$ & 0.03180 & 0.5841 & 0.3708 & 0.1507 & 11.66 & 0.1328 \\  
$n=0.3$ & 0.2 & 7.312 (9.979) & 0.5394(0.5535) & $1.243 \times 10^{-3}$ & 0.03926 & 1.435 & 1.835 & 0.1688 & 46.72 & 0.3311 \\  
$n=0.5$ & 0.1 & 10.30 (11.83) & 0.5461(0.5536) & $5.153 \times 10^{-4}$ & 0.02197 & 0.8416 & 0.7644 & 0.1493 & 34.80 & 0.1281 \\  
\hline
\end{tabular}
\caption{Quantities at the point of bifurcation of triaxial sequences
  from axisymmetric ones for the two EOSs considered. The compactness of
  the spherical star with the same rest mass $\compa$ are the model
  parameters. In the above, $R_x$ is the equatorial radius, and $R_z/R_x$
  is the ratio of polar to the equatorial radius. Each has two values;
  one is measured in the coordinate length, and the other in parenthesis
  is in proper length. $\epsilon_c$ is the energy density at the center
  of the compact star, $\Omega$ is the angular velocity. In the last
  three lines we report the bifurcation point of simple polytropes with
  polytropic index $n$ as computed in \cite{Huang08} for comparison.
  Note that we have chosen appropriate values for $\kappa$ such that the
  TOV maximum mass for those polytropic EOS reach 2.5 $M_\odot$.
  The definitions of $\Madm$, $J$, $T/|W|$, and $I$ can be found
    in the Appendix A of \cite{Tsokaros2015}. $Z_{\rm p}$ is the polar
  redshift.}
\label{tab:secular}
\end{table*}

It is worth noting that when compared with the rotating NSs calculated in
Ref. \cite{Uryu2016a}, rotating QSs have longer triaxial sequences. In
another word, the triaxial sequence of rotating QSs terminates at larger
eccentricity as well as larger triaxial deformation (in another word,
smaller $R_y/R_x$ ratio).  A rotating NS with $\Gamma=4$ and compactness
$\compa=0.1$ bifurcates from axisymmetry at $e\simeq 0.825$ and can
rotate as fast as to reach eccentricities $e< 0.9$ (see Fig. 6 in
\cite{Uryu2016a}). For the QS models considered here and both
EOSs, we have a bifurcation point at $e\simeq 0.825$
and the mass shedding limit at $e\simeq 0.93$. For more compact NSs with
$\compa=0.2$, the bifurcation point happens at $e\simeq 0.835$ and the
mass shedding limit at $e\simeq 0.88$. The corresponding compactness QS
models bifurcate at $e\simeq 0.83$ and rotate as fast as $e\simeq 0.89$.

A few remarks are useful to make at this point. First, we note that these
values of eccentricity are strictly valid under the
assumption of the conformal flatness approximation, which is however
accurate for smaller compactnesses. These estimates are
less accurate when the compactness increases
and are slightly different when adopting more accurate
formulations, such as the waveless approximation (see Fig. 6 in
\cite{Uryu2016a}). Second, another difference between triaxial NSs and
QSs is that for triaxial NSs, the ratio $T/|W|$ is essentially constant
along the triaxial sequence, especially for higher compactnesses (for
lower compactnesses there is an increase towards the mass shedding limit,
but this is very slight). For rotating triaxial QSs, on the other hand,
although this qualitative behaviour is still true, a greater curvature
towards higher $T/|W|$ ratios can be seen. For example, for the
$\compa=0.1$ models mentioned above, the difference between the critical
value of $T/|W|$ and the one at mass-shedding limit is $(T/|W|)_{\rm
  ms}-(T/|W|)_{\rm crit}\simeq 0.0015$ for NSs while it is $0.0137$ for
QSs. Third, in Ref. \cite{rosinska2003} it has been shown that at the
bifurcation point the relation between the scaled angular frequency,
$f/\bar{\epsilon}_s^{1/2}$ where $\bar{\epsilon}_s = \epsilon_s/(c^2\,
10^{14}\,\mathrm{g\,cm^{-3}}$), and the scaled gravitational mass
$M_\mathrm{ADM}\bar{\epsilon}_s^{1/2}$, depends only very weakly on the
bag constant. If we consider models with compactness $\compa=0.1$ model
(see top line in Table \ref{tab:secular}),
$M_\mathrm{ADM}\bar{\epsilon}_s^{1/2} = 1.193\,M_\odot$ and $f/
\bar{\epsilon}_s^{1/2} = 495.9\,\mathrm{Hz}$, while the scaled
bifurcation frequency for such a scaled mass model is roughly
$492-494\,\mathrm{Hz}$ as deduced from Fig. 7 in \cite{rosinska2003}.
Similarly, for the $\compa=0.15$ models, the renormalized ADM mass and
frequency are $2.140\,M_\odot$ and $527.2\,\mathrm{Hz}$, while the
corresponding range is $523-527\,\mathrm{Hz}$ in
Ref. \cite{rosinska2003}; finally, for the $\compa=0.2$ case, the values
are $3.168\,M_\odot$ and $565.0\,\mathrm{Hz}$, respectively, while the
range $558-566\,\mathrm{Hz}$ is found in \cite{rosinska2003}.  Overall,
the comparison of these three values shows a very good agreement with the
results presented in Fig. 7 of Ref. \cite{rosinska2003}.

\begin{figure*}
\begin{center}
\includegraphics[width=0.95\columnwidth]{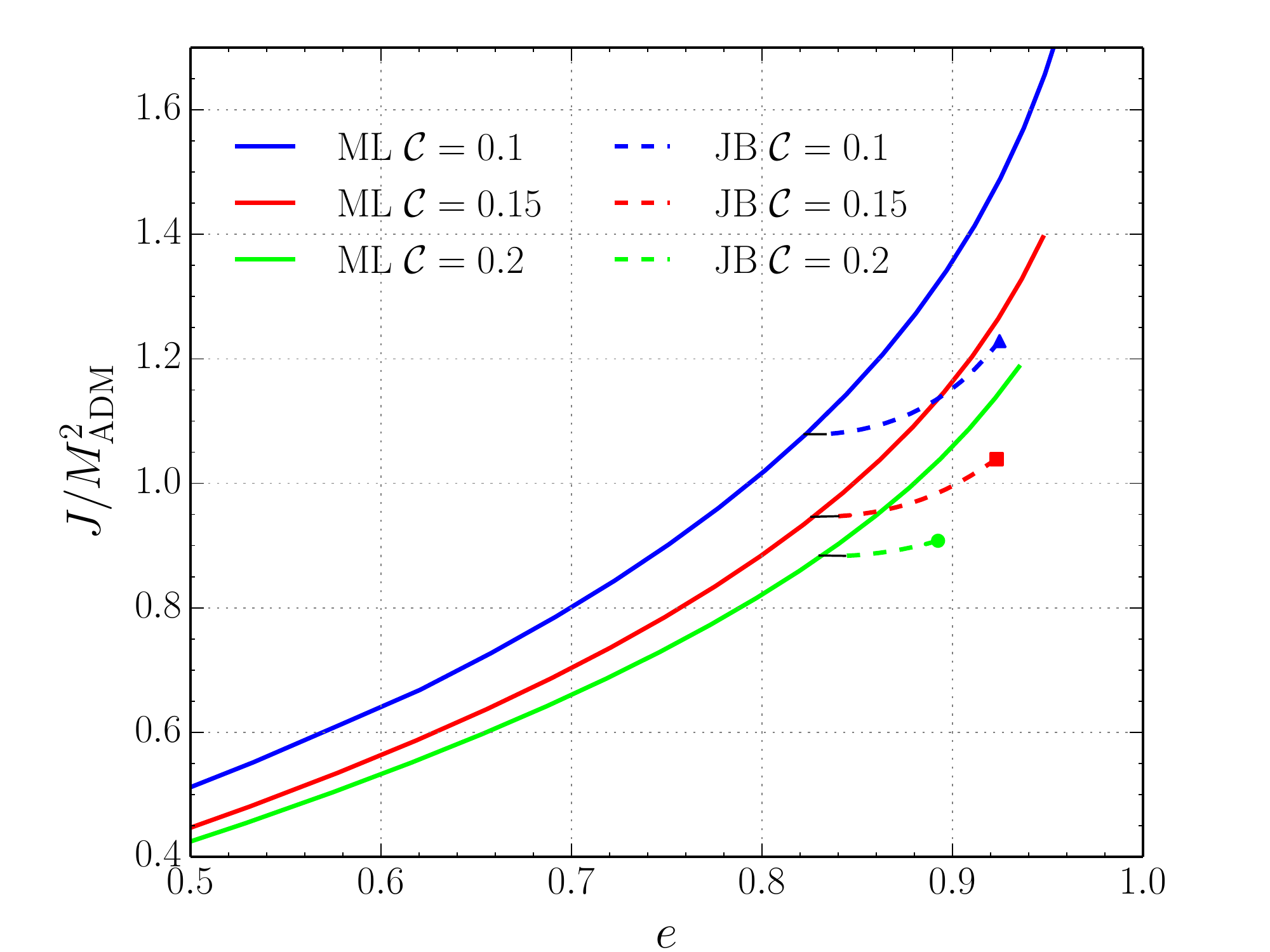}
\includegraphics[width=0.95\columnwidth]{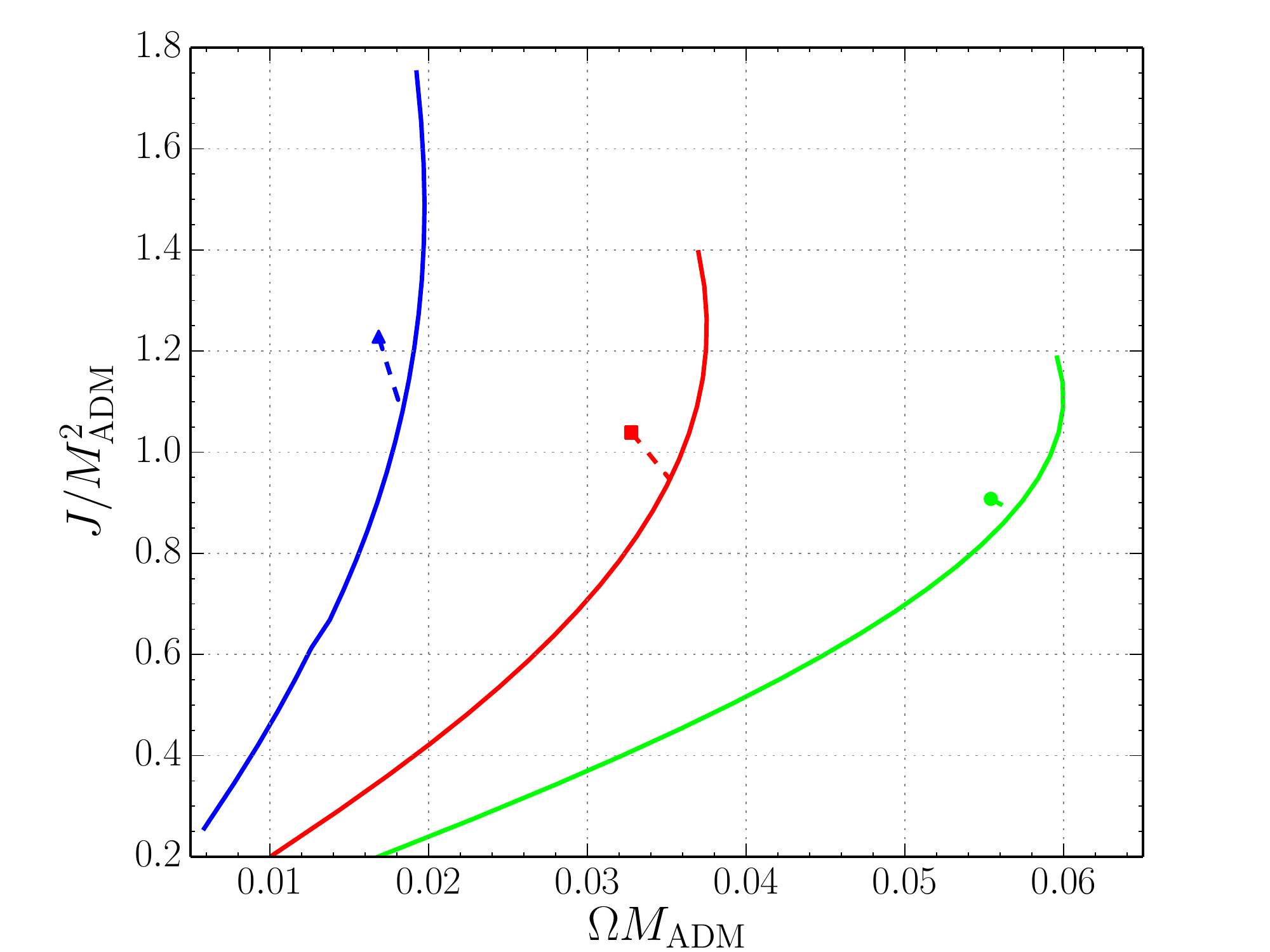}
\end{center}
\caption{Spin angular momentum versus the eccentricity and angular
  velocity for the LX EOS sequences. Dashed curves and solid curves from
  the top to the bottom in each panel correspond to $\compa=0.2$ (green),
  $0.15$ (red) and $0.1$ (blue) respectively.}
\label{fig:plot_lx_J}
\end{figure*}

\begin{figure}
  \begin{center}
    \includegraphics[width=0.95\columnwidth]{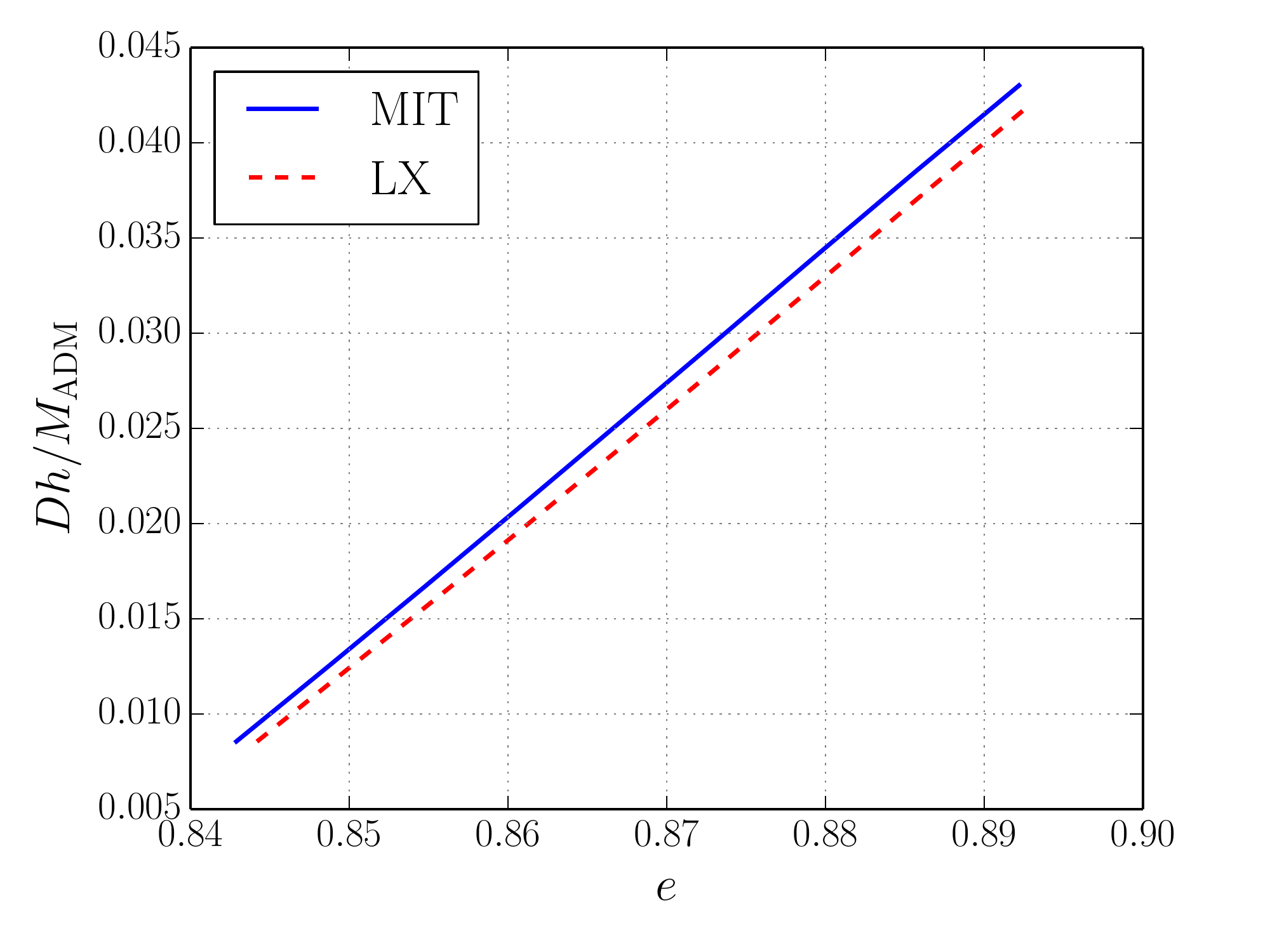}
  \end{center}
  \caption{Estimate of the GW strain amplitude for the $\compa=0.2$
    triaxial sequence for both the MIT bag-model EOS (blue solid curve)
    and the LX EOS (red dashed curve). The quantities are estimated
    according to the quadrupole formula. Shown is the GW strain for the
    $\ell=m=2$ mode normalized by the distance $D$ and the ADM mass
    $M_\mathrm{ADM}$ of the source.}
  \label{fig:plot_gw_quad}
\end{figure}

In order to understand the rotation properties of the triaxial solutions,
we also report quantities such as dimensionless spin and dimensionless
angular momentum in Fig. \ref{fig:plot_mit_omeM}- \ref{fig:plot_lx_J}.
The dimensionless spin as a function of the eccentricity for MIT
bag-model and LX model are shown in Fig. \ref{fig:plot_mit_omeM} and
\ref{fig:plot_lx_omeM}, respectively.  The top panel of
Fig. \ref{fig:plot_lx_J} reports the spin angular momentum as a function
of the eccentricity for the LX EOS. Similar as $T/|W|$, the angular
momentum increases with the eccentricity. The main difference is that the
relative positioning of the curves as a function of compactness is
reversed when compared with the $T/|W|$ plots. In other words for a given
eccentricity the greatest angular momentum is achieved for the smallest
compactness, while the greatest $T/|W|$ for the largest one. This is true
both for axisymmetric and triaxial solutions. Also as we can see from the
bottom panel of Fig. \ref{fig:plot_lx_J}, more compact objects can reach
greater rotational frequencies while less compact can reach larger
angular momenta, which can exceed unity. According to
Fig. \ref{fig:plot_mit_omeM} and \ref{fig:plot_lx_omeM} as well as the
bottom panel of Fig. \ref{fig:plot_lx_J}, triaxial sequences lose angular
velocity and gain spin angular momentum as one moves towards the mass
shedding limit.

An obviously interesting property of triaxially rotating compact stars is
that they can act as strong sources of GWs. A full general-relativistic
evolution needs to be employed in order to determine accurately the
details of the GW emission from such triaxially rotating stars and this
is beyond the scope of this paper (see however Ref. \cite{Tsokaros2017}
for the case of NSs). At the same time, we can apply the quadrupole
formula to make reasonable estimates
using the quasi-equilibrium initial data we have computed. The
relationship between the normalized GW strain and the eccentricity of the
star is shown in the top panel of Fig.~\ref{fig:plot_gw_quad}. Compared
with the results of triaxially rotating NSs calculated in
\cite{Tsokaros2017}, we find that the GW strain for QSs are several times
larger for similar values of the compactness. For example, model G4C025
in \cite{Tsokaros2017} with $e=0.8685$ radiates GW with normalized strain
0.007357, while the corresponding amplitude for both the MIT bag-model
EOS and the LX EOS is around 0.025 with same eccentricity (see
Fig.~\ref{fig:plot_gw_quad}). Also shown in Fig.~\ref{fig:plot_gw_quad}
for the two EOSs considered, are the relations between the strain and the
eccentricity, which are very similar and both essentially linear.

\section{Triaxial supramassive solutions}
\label{sec:triasupraQS}

Besides the constant rest-mass sequences mentioned above, we have also
built sequences with constant central rest-mass density for both the MIT
bag-model EOS and the LX EOS. We recall that when fixing the central
rest-mass density, the mass of the solutions will increase as the axis
ratio $R_z/R_x$ decreases. Furthermore, since we do not impose
axisymmetry, the triaxial deformation will be spontaneously triggered
when $T/|W|$ is large enough. Therefore, with such calculations we can
determine whether triaxial supramassive QSs, \ie triaxial solutions with
ADM mass larger than the TOV maximum mass ($M_\mathrm{TOV}$) exist and
the properties of such solutions\footnote{We recall that for NSs, a
  universal relation has been found between $M_\mathrm{TOV}$ and the
  maximum mass that can be sustained by axisymmetric solutions in uniform
  rotation, $M_{\rm max}$ (see also \cite{Weih2017} for the case of
  differentially rotating stars). More specifically, Breu and Rezzolla
  \cite{Breu2016} found that $M_{\rm max} \simeq (1.203 \pm 0.022)
  M_\mathrm{TOV}$ for a large class of EOSs; we expect a similar
  universal behaviour to be present also for QSs,
  although the scaling between $M_{\rm max}$ and
  $M_\mathrm{TOV}$ is likely to be different.} (Note that all models
shown in Figs. \ref{fig:plot_mit}--\ref{fig:plot_lx_J} are not
supramassive).

According to \cite{Uryu2016b}, triaxial supramassive NS does exist for
the case with polytropic EOS in the range $\Gamma \agt 4$.  Furthermore,
for the case with a two segments piecewise polytropic EOS, sequences of
triaxial supramassive NS solutions become longer, and hence the existence
of supramassive triaxial NS becomes evident, when the EOS of the lower
density region is stiff ($\Gamma=4$), and the higher density region is
soft ($\Gamma=2.5$)\cite{Uryu2016b}.  Therefore, it is likely that the
triaxial supramassive QS also exists because the QS EOS used in this
paper has an analogous property, namely, the effective $\Gamma$ is
smaller (softer) in the higher density region, and larger (stiffer) in
the lower density region (for the MIT model, see
\cite{gourgoulhon1999}). Besides having an interest of its own,
determining the existence of such solutions could be relevant to
establish whether a BNS merger could lead to the formation of such an
object. Based on current mass measurement constraints \cite{Demorest2010,
  Antoniadis2013} and on known BNS systems, the mass of the post-merger
product will very likely be larger than $M_\mathrm{TOV}$.

In order to study this, we have fixed the central rest-mass density close
to the value corresponding to $M_\mathrm{TOV}$ and built rotating
solution sequences for both the MIT bag-model EOS and the LX EOS,
respectively. In this way, we were indeed able to find triaxial
supramassive solutions for both EOSs, reporting in
Table.\ref{tab:triasupraqs} the solutions with largest triaxial
deformation, \ie the smallest ratio $R_y/R_x$. 

Finally, we note that although such models have large compactnesses and
we are aware that the IWM formalism becomes increasingly inaccurate for
large compactnesses (\ie with $\compa \gtrsim 0.3$), we also believe that
the associated $\sim3\%$ errors will not change the qualitative result,
namely, that triaxial supramassive QS models exist for the EOSs
considered here. At the same time, we plan to re-investigate this point
in the future, when more accurate methods, such as waveless formulation,
will be employed to compute QS solutions.

\begin{table*}
  \begin{tabular}{ccccccccccccc}
    \hline
    EOS&$R_z/R_x$&$R_y/R_x$&$R_x$&$\epsilon_c$&$\Omega$&$\Madm$&$J$&$T/|W|$&$I$&$M_\mathrm{TOV}$ \\
    \hline
    MIT& 0.4375 (0.4713) & 0.7657 (0.7938) & 9.978	(16.32) & 	$ 1.259 \times 10^{-3} $	 & 	0.03870 & 	2.862 & 	6.847 & 	0.1839  & 	173.1  & 	2.217 & 	 \\
    LX& 0.4375 (0.4912) & 0.7586 (0.8104) & 7.660 (16.49)& 	$1.348 \times 10^{-3} $	 & 	0.05001 & 	3.727 & 	11.30	 & 	0.1948	 & 	222.1	 & 3.325 & 	 	 \\
    \hline
  \end{tabular}
  \caption{Quantities of triaxial supramassive QS solutions with the
    largest triaxial deformation (smallest $R_y/R_x$ ratio) in our
    calculations. The above quantities are defined in the same way as in
    Table.\ref{tab:secular}. The TOV maximum mass of each EOS is also
    shown as a comparison. Due to the limitation of IWM formulation,
    there might be $\sim3\%$ errors on the quantities listed above (see
    related discussions in Sec.\ref{sec:triasupraQS}).
}
\label{tab:triasupraqs}
\end{table*}

\section{Conclusion}
\label{sec:disandconclu}

We have presented a new version of the \cocal{} code to compute
axisymmetric and triaxial solutions of uniformly rotating QSs in general
relativity with two EOSs, \ie the MIT bag-model and the LX EOS.
Comparisons have been made with NSs as well. Overall, three main
properties are found when comparing solution sequences of QSs with of
NSs. Firstly, QSs generally have a longer triaxial
sequences of solutions than NSs. In another words, QSs can reach a larger
triaxial deformation (or smaller $R_y/R_x$ ratio) before terminating the
sequence at the mass-shedding limit; this is mostly due to the larger
$T/|W|$ ratio that can be attained by QSs.  Secondly, when considering
similar triaxial configurations, QSs are (slightly) more efficient GW
sources; this is mostly due to the finite surface rest-mass density and
hence larger mass quadrupole for QSs. Thirdly, triaxial supramassive
solutions can be found for QSs as well; this is due again to the fact
that larger values of the $T/|W|$ ratio can be sustained before reaching
the mass-shedding limit.

Besides having an interest of its own within solutions of
self-gravitating objects in general relativity, triaxially rotating
compact stars are important sources for ground-based GW
observatories. Our calculations have shown that for rotating QSs with
different EOSs, the bifurcation point to triaxial
sequence happens at a spin period of $\sim 1\,{\rm ms}$, so that the
corresponding GW frequency is $\sim 2\,{\rm kHz}$ and hence within the
band of GW observatories such as Advanced LIGO or Virgo. Indeed,
exploiting the largest triaxial deformation solution obtained in our
calculations, the GW strain amplitude can be as large as $10^{-23}$ at a
distance of $\sim 30\,{\rm Mpc}$.

Although this is an interesting prospect, it is still unclear whether
such triaxial configurations can be produced in practice, since the
radiation-reaction timescales needed for the triggering of the secular
triaxial instability are still very uncertain, as are the other
mechanisms that could contrast the instability. For example, if the
triaxial deformation is induced in an isolated star, \eg a newly born
fast rotating star, GW radiation may take away the excess angular
momentum very rapidly so that the star would go back to the axisymmetric
sequence again after the $T/|W|$ ratio drops below the critical value.
Similarly, when considering stars in binary systems, there is the
prospect that an accreting system, such as the one spinning up pulsars,
could drive the accreting compact star to exceed the critical $T/|W|$
ratio, hence leading to a break of axisymmetry. In this process, which is
also known as forced GW emission, the triaxial deformation can be
maintained via the angular momentum supplied by the accreted
matter. Notwithstanding the large uncertainties involved with the details
of this picture, such as the presence or not of
bifurcation point or the realistic degree of deformation attained by the
unstable stars, the fact that these details depend sensitively on the EOS
\cite{Uryu2016b}, suggests that a detection of this type of signal could
serve as an important probe for distinguishing the EOS of compact stars.

Finally, we note that the triaxial configurations could also be invoked
to explain the spin-up limit for rotating compact stars, which is far
smaller than the mass-shedding limit.  The results presented here and in
Ref.  \cite{Huang08} suggest that when triaxial deformations are taken
into account, the rotational period of a compact star actually decreases
as it gains angular momentum, \eg by accretion, along the triaxial
sequence. As a result, the ``spin-up'' process provided by the accretion
of matter onto the pulsar can actually spin down the pulsar if the
bifurcation point is reached. In this case, no accreting pulsar could
spin up faster than the period at the bifurcation point. Of
  course, depending on the microphysical properties of the QS (\eg the
  magnitude of the shear viscosity or of the breaking strain in the
  crust) it is also possible that other mechanisms of emission of
  gravitational waves, (\eg other dynamical instabilities such as the
  barmode instability \cite{Watts:2003nn, Corvino:2010} or the $r$-mode
  instability \cite{Andersson1998, Friedman1998, Andersson99}, or nonzero
  ellipticities) could intervene at lower spinning frequencies and
  therefore before the onset of instability to a triaxial deformation is
  reached \cite{Bildsten98, Andersson:2009yt}. As a result, the search
  for fast spinning pulsars with more powerful radio telescopes, such as
  the Square Kilometre Array (SKA) and the Five-hundred-meter Aperture
  Spherical radio Telescope (FAST) \cite{SKAdoc, Nan2011} could provide
  important clues on the properties of pulsars and test the validity of
  the solid QS assumption \cite{Xu2003}.

\acknowledgments It is a pleasure to thank J. G. Lu, J. Papenfort,
Z. Younsi, and all the members of the Pulsar group in Peking University
and the Relastro group in Frankfurt for useful discussions.  We will
thank Dr. L. Shao and Dr. Y. Hu for their help with GW
experiments. E. Z. is grateful to the China Scholarship Council for
supporting the joint PhD training in Frankfurt. This research is
supported in part by the ERC synergy grant "BlackHoleCam: Imaging the
Event Horizon of Black Holes" (Grant No. 610058), by "NewCompStar", COST
Action MP1304, by the LOEWE-Program in the Helmholtz International Center
(HIC) for FAIR, by the European Union's Horizon 2020 Research and
Innovation Programme (Grant 671698) (call FETHPC-1-2014, project
ExaHyPE). This work is supported by National Key R\&D Program
(No.2017YFA0402600) and NNSF (11673002,U1531243). A. T. is supported by
NSF Grants PHY-1662211 and PHY-1602536, and NASA Grant 80NSSC17K0070.
K. U. is supported by JSPS Grant-in-Aid for Scientific Research(C)
15K05085.

\bibliographystyle{apsrev4-1}
\bibliography{aeireferences}

\end{document}

%% file: kigou.tex
\newcommand{\Madm}{M_{\rm ADM}}

\newcommand{\beq}{\begin{equation}} 
\newcommand{\eeq}{\end{equation}} 
\newcommand{\beqn}{\begin{eqnarray}} 
\newcommand{\eeqn}{\end{eqnarray}} 
\newcommand{\pa}{\partial}
\newcommand{\na}{\nabla}

\newcommand{\gmabd}{\gamma_{ab}}

\newcommand{\albe}{{\alpha\beta}}

\newcommand{\Tabd}{T_{\alpha\beta}}

\newcommand{\Gabd}{G_{\alpha\beta}}

\newcommand{\zD}{{\raise1.0ex\hbox{${}^{\ \circ}$}}\!\!\!\!\!D}
\newcommand{\alone}{{\raise0.5ex\hbox{${}^{\ 1}$}}\!\!\!\!\alpha}

\newcommand{\Dl}{\Delta}

\newcommand{\Lie}{\mbox{\pounds}}

\newcommand{\compa}{\mathcal{C}}

\newcommand{\nalam}{\mathrel{\raise0.9ex\hbox{$^\lambda$}\mkern-14mu
\lower0.0ex\hbox{$\nabla$}}}

\newcommand{\rhoH}{\rho_{\rm H}}

\newcommand{\Nrf}{{N_r^{\rm f}}}
\newcommand{\Nrm}{{N_r^{\rm m}}}

\newcommand{\Kabd}{K_{ab}}

\newcommand{\zeroD}{{\raise1.0ex\hbox{${}^{\ \circ}$}}\!\!\!\!\!D}
\newcommand{\zLap}{{\raise1.0ex\hbox{${}^{\ \circ}$}}\!\!\!\!\Delta}
\newcommand{\zna}{{\raise1.0ex\hbox{${}^{\ \circ}$}}\!\!\!\!\!\nabla}
\newcommand{\zS}{{\raise1.0ex\hbox{${}^{\ \circ}$}}\!\!\!\!\!S}

\newcommand{\bR}{{\bar R}}


%% file: main.bbl
\begin{thebibliography}{95}%
\makeatletter
\providecommand \@ifxundefined [1]{%
 \@ifx{#1\undefined}
}%
\providecommand \@ifnum [1]{%
 \ifnum #1\expandafter \@firstoftwo
 \else \expandafter \@secondoftwo
 \fi
}%
\providecommand \@ifx [1]{%
 \ifx #1\expandafter \@firstoftwo
 \else \expandafter \@secondoftwo
 \fi
}%
\providecommand \natexlab [1]{#1}%
\providecommand \enquote  [1]{``#1''}%
\providecommand \bibnamefont  [1]{#1}%
\providecommand \bibfnamefont [1]{#1}%
\providecommand \citenamefont [1]{#1}%
\providecommand \href@noop [0]{\@secondoftwo}%
\providecommand \href [0]{\begingroup \@sanitize@url \@href}%
\providecommand \@href[1]{\@@startlink{#1}\@@href}%
\providecommand \@@href[1]{\endgroup#1\@@endlink}%
\providecommand \@sanitize@url [0]{\catcode `\\12\catcode `\$12\catcode
  `\&12\catcode `\#12\catcode `\^12\catcode `\_12\catcode `\%12\relax}%
\providecommand \@@startlink[1]{}%
\providecommand \@@endlink[0]{}%
\providecommand \url  [0]{\begingroup\@sanitize@url \@url }%
\providecommand \@url [1]{\endgroup\@href {#1}{\urlprefix }}%
\providecommand \urlprefix  [0]{URL }%
\providecommand \Eprint [0]{\href }%
\providecommand \doibase [0]{http://dx.doi.org/}%
\providecommand \selectlanguage [0]{\@gobble}%
\providecommand \bibinfo  [0]{\@secondoftwo}%
\providecommand \bibfield  [0]{\@secondoftwo}%
\providecommand \translation [1]{[#1]}%
\providecommand \BibitemOpen [0]{}%
\providecommand \bibitemStop [0]{}%
\providecommand \bibitemNoStop [0]{.\EOS\space}%
\providecommand \EOS [0]{\spacefactor3000\relax}%
\providecommand \BibitemShut  [1]{\csname bibitem#1\endcsname}%
\let\auto@bib@innerbib\@empty
\bibitem [{\citenamefont {{The LIGO Scientific Collaboration}}\ and\
  \citenamefont {{The Virgo Collaboration}}(2017)}]{Abbott2017}%
  \BibitemOpen
  \bibfield  {author} {\bibinfo {author} {\bibnamefont {{The LIGO Scientific
  Collaboration}}}\ and\ \bibinfo {author} {\bibnamefont {{The Virgo
  Collaboration}}} (\bibinfo {collaboration} {LIGO Scientific Collaboration and
  Virgo Collaboration}),\ }\href {\doibase 10.1103/PhysRevLett.119.161101}
  {\bibfield  {journal} {\bibinfo  {journal} {Phys. Rev. Lett.}\ }\textbf
  {\bibinfo {volume} {119}},\ \bibinfo {pages} {161101} (\bibinfo {year}
  {2017})}\BibitemShut {NoStop}%
\bibitem [{\citenamefont {{The LIGO Scientific Collaboration}}\ \emph
  {et~al.}(2017{\natexlab{a}})\citenamefont {{The LIGO Scientific
  Collaboration}}, \citenamefont {{the Virgo Collaboration}}, \citenamefont
  {{Abbott}}, \citenamefont {{Abbott}}, \citenamefont {{Abbott}}, \citenamefont
  {{Acernese}}, \citenamefont {{Ackley}}, \citenamefont {a{Adams}},
  \citenamefont {{Adams}}, \citenamefont {{Addesso}},\ and\ \citenamefont
  {et~al.}}]{Abbott2017b}%
  \BibitemOpen
  \bibfield  {author} {\bibinfo {author} {\bibnamefont {{The LIGO Scientific
  Collaboration}}}, \bibinfo {author} {\bibnamefont {{the Virgo
  Collaboration}}}, \bibinfo {author} {\bibfnamefont {B.~P.}\ \bibnamefont
  {{Abbott}}}, \bibinfo {author} {\bibfnamefont {R.}~\bibnamefont {{Abbott}}},
  \bibinfo {author} {\bibfnamefont {T.~D.}\ \bibnamefont {{Abbott}}}, \bibinfo
  {author} {\bibfnamefont {F.}~\bibnamefont {{Acernese}}}, \bibinfo {author}
  {\bibfnamefont {K.}~\bibnamefont {{Ackley}}}, \bibinfo {author}
  {\bibfnamefont {C.}~\bibnamefont {a{Adams}}}, \bibinfo {author}
  {\bibfnamefont {T.}~\bibnamefont {{Adams}}}, \bibinfo {author} {\bibfnamefont
  {P.}~\bibnamefont {{Addesso}}}, \ and\ \bibinfo {author} {\bibnamefont
  {et~al.}} (\bibinfo {collaboration} {LIGO Scientific Collaboration and Virgo
  Collaboration}),\ }\href {http://stacks.iop.org/2041-8205/848/i=2/a=L12}
  {\bibfield  {journal} {\bibinfo  {journal} {Astrophys. J. Lett.}\ }\textbf
  {\bibinfo {volume} {848}},\ \bibinfo {pages} {L12} (\bibinfo {year}
  {2017}{\natexlab{a}})}\BibitemShut {NoStop}%
\bibitem [{\citenamefont {{LIGO Scientific Collaboration}}\ \emph
  {et~al.}(2017)\citenamefont {{LIGO Scientific Collaboration}}, \citenamefont
  {{Virgo Collaboration}}, \citenamefont {{Gamma-Ray Burst Monitor}},\ and\
  \citenamefont {{INTEGRAL}}}]{Abbott2017d}%
  \BibitemOpen
  \bibfield  {author} {\bibinfo {author} {\bibnamefont {{LIGO Scientific
  Collaboration}}}, \bibinfo {author} {\bibnamefont {{Virgo Collaboration}}},
  \bibinfo {author} {\bibfnamefont {F.}~\bibnamefont {{Gamma-Ray Burst
  Monitor}}}, \ and\ \bibinfo {author} {\bibnamefont {{INTEGRAL}}},\ }\href
  {http://stacks.iop.org/2041-8205/848/i=2/a=L13} {\bibfield  {journal}
  {\bibinfo  {journal} {Astrophys. J. Lett.}\ }\textbf {\bibinfo {volume}
  {848}},\ \bibinfo {pages} {L13} (\bibinfo {year} {2017})},\ \Eprint
  {http://arxiv.org/abs/1710.05834} {arXiv:1710.05834 [astro-ph.HE]}
  \BibitemShut {NoStop}%
\bibitem [{\citenamefont {{The LIGO Scientific Collaboration}}\ \emph
  {et~al.}(2017{\natexlab{b}})\citenamefont {{The LIGO Scientific
  Collaboration}}, \citenamefont {{the Virgo Collaboration}}, \citenamefont
  {{Abbott}}, \citenamefont {{Abbott}}, \citenamefont {{Abbott}}, \citenamefont
  {{Acernese}}, \citenamefont {{Ackley}}, \citenamefont {{Adams}},
  \citenamefont {{Adams}}, \citenamefont {{Addesso}},\ and\ \citenamefont
  {et~al.}}]{Abbott2017c}%
  \BibitemOpen
  \bibfield  {author} {\bibinfo {author} {\bibnamefont {{The LIGO Scientific
  Collaboration}}}, \bibinfo {author} {\bibnamefont {{the Virgo
  Collaboration}}}, \bibinfo {author} {\bibfnamefont {B.~P.}\ \bibnamefont
  {{Abbott}}}, \bibinfo {author} {\bibfnamefont {R.}~\bibnamefont {{Abbott}}},
  \bibinfo {author} {\bibfnamefont {T.~D.}\ \bibnamefont {{Abbott}}}, \bibinfo
  {author} {\bibfnamefont {F.}~\bibnamefont {{Acernese}}}, \bibinfo {author}
  {\bibfnamefont {K.}~\bibnamefont {{Ackley}}}, \bibinfo {author}
  {\bibfnamefont {C.}~\bibnamefont {{Adams}}}, \bibinfo {author} {\bibfnamefont
  {T.}~\bibnamefont {{Adams}}}, \bibinfo {author} {\bibfnamefont
  {P.}~\bibnamefont {{Addesso}}}, \ and\ \bibinfo {author} {\bibnamefont
  {et~al.}},\ }\href@noop {} {\bibfield  {journal} {\bibinfo  {journal} {ArXiv
  e-prints}\ } (\bibinfo {year} {2017}{\natexlab{b}})},\ \Eprint
  {http://arxiv.org/abs/1710.05836} {arXiv:1710.05836 [astro-ph.HE]}
  \BibitemShut {NoStop}%
\bibitem [{\citenamefont {Baiotti}\ and\ \citenamefont
  {Rezzolla}(2017)}]{Baiotti2016}%
  \BibitemOpen
  \bibfield  {author} {\bibinfo {author} {\bibfnamefont {L.}~\bibnamefont
  {Baiotti}}\ and\ \bibinfo {author} {\bibfnamefont {L.}~\bibnamefont
  {Rezzolla}},\ }\href {\doibase 10.1088/1361-6633/aa67bb} {\bibfield
  {journal} {\bibinfo  {journal} {Rept. Prog. Phys.}\ }\textbf {\bibinfo
  {volume} {80}},\ \bibinfo {pages} {096901} (\bibinfo {year} {2017})},\
  \Eprint {http://arxiv.org/abs/1607.03540} {arXiv:1607.03540 [gr-qc]}
  \BibitemShut {NoStop}%
\bibitem [{\citenamefont {{Andersson}}\ \emph {et~al.}(2011)\citenamefont
  {{Andersson}}, \citenamefont {{Ferrari}}, \citenamefont {{Jones}},
  \citenamefont {{Kokkotas}}, \citenamefont {{Krishnan}}, \citenamefont
  {{Read}}, \citenamefont {{Rezzolla}},\ and\ \citenamefont
  {{Zink}}}]{Andersson:2009yt}%
  \BibitemOpen
  \bibfield  {author} {\bibinfo {author} {\bibfnamefont {N.}~\bibnamefont
  {{Andersson}}}, \bibinfo {author} {\bibfnamefont {V.}~\bibnamefont
  {{Ferrari}}}, \bibinfo {author} {\bibfnamefont {D.~I.}\ \bibnamefont
  {{Jones}}}, \bibinfo {author} {\bibfnamefont {K.~D.}\ \bibnamefont
  {{Kokkotas}}}, \bibinfo {author} {\bibfnamefont {B.}~\bibnamefont
  {{Krishnan}}}, \bibinfo {author} {\bibfnamefont {J.~S.}\ \bibnamefont
  {{Read}}}, \bibinfo {author} {\bibfnamefont {L.}~\bibnamefont {{Rezzolla}}},
  \ and\ \bibinfo {author} {\bibfnamefont {B.}~\bibnamefont {{Zink}}},\ }\href
  {\doibase 10.1007/s10714-010-1059-4} {\bibfield  {journal} {\bibinfo
  {journal} {General Relativity and Gravitation}\ }\textbf {\bibinfo {volume}
  {43}},\ \bibinfo {pages} {409} (\bibinfo {year} {2011})},\ \Eprint
  {http://arxiv.org/abs/0912.0384} {arXiv:0912.0384 [astro-ph.SR]} \BibitemShut
  {NoStop}%
\bibitem [{\citenamefont {Abramovici}\ \emph {et~al.}(1992)\citenamefont
  {Abramovici}, \citenamefont {Althouse}, \citenamefont {Drever}, \citenamefont
  {Gursel}, \citenamefont {Kawamura}, \citenamefont {Raab}, \citenamefont
  {Shoemaker}, \citenamefont {Sievers}, \citenamefont {Spero}, \citenamefont
  {Thorne}, \citenamefont {Vogt}, \citenamefont {Weiss}, \citenamefont
  {Whitcomb},\ and\ \citenamefont {Zuker}}]{Abramovici92}%
  \BibitemOpen
  \bibfield  {author} {\bibinfo {author} {\bibfnamefont {A.~A.}\ \bibnamefont
  {Abramovici}}, \bibinfo {author} {\bibfnamefont {W.}~\bibnamefont
  {Althouse}}, \bibinfo {author} {\bibfnamefont {R.~P.}\ \bibnamefont
  {Drever}}, \bibinfo {author} {\bibfnamefont {Y.}~\bibnamefont {Gursel}},
  \bibinfo {author} {\bibfnamefont {S.}~\bibnamefont {Kawamura}}, \bibinfo
  {author} {\bibfnamefont {F.}~\bibnamefont {Raab}}, \bibinfo {author}
  {\bibfnamefont {D.}~\bibnamefont {Shoemaker}}, \bibinfo {author}
  {\bibfnamefont {L.}~\bibnamefont {Sievers}}, \bibinfo {author} {\bibfnamefont
  {R.}~\bibnamefont {Spero}}, \bibinfo {author} {\bibfnamefont {K.~S.}\
  \bibnamefont {Thorne}}, \bibinfo {author} {\bibfnamefont {R.}~\bibnamefont
  {Vogt}}, \bibinfo {author} {\bibfnamefont {R.}~\bibnamefont {Weiss}},
  \bibinfo {author} {\bibfnamefont {S.}~\bibnamefont {Whitcomb}}, \ and\
  \bibinfo {author} {\bibfnamefont {M.}~\bibnamefont {Zuker}},\ }\href
  {\doibase 10.1126/science.256.5055.325} {\bibfield  {journal} {\bibinfo
  {journal} {Science}\ }\textbf {\bibinfo {volume} {256}},\ \bibinfo {pages}
  {325} (\bibinfo {year} {1992})}\BibitemShut {NoStop}%
\bibitem [{\citenamefont {Punturo}\ \emph {et~al.}(2010)\citenamefont {Punturo}
  \emph {et~al.}}]{Punturo:2010}%
  \BibitemOpen
  \bibfield  {author} {\bibinfo {author} {\bibfnamefont {M.}~\bibnamefont
  {Punturo}} \emph {et~al.},\ }\href {\doibase 10.1088/0264-9381/27/8/084007}
  {\bibfield  {journal} {\bibinfo  {journal} {Class. Quantum Grav.}\ }\textbf
  {\bibinfo {volume} {27}},\ \bibinfo {pages} {084007} (\bibinfo {year}
  {2010})}\BibitemShut {NoStop}%
\bibitem [{\citenamefont {{Accadia}}\ \emph {et~al.}(2011)\citenamefont
  {{Accadia}} \emph {et~al.}}]{Accadia2011_etal}%
  \BibitemOpen
  \bibfield  {author} {\bibinfo {author} {\bibfnamefont {T.}~\bibnamefont
  {{Accadia}}} \emph {et~al.},\ }\href {\doibase
  10.1088/0264-9381/28/11/114002} {\bibfield  {journal} {\bibinfo  {journal}
  {Class. Quantum Grav.}\ }\textbf {\bibinfo {volume} {28}},\ \bibinfo {eid}
  {114002} (\bibinfo {year} {2011})}\BibitemShut {NoStop}%
\bibitem [{\citenamefont {{Kuroda}}\ and\ \citenamefont {{LCGT
  Collaboration}}(2010)}]{Kuroda2010}%
  \BibitemOpen
  \bibfield  {author} {\bibinfo {author} {\bibfnamefont {K.}~\bibnamefont
  {{Kuroda}}}\ and\ \bibinfo {author} {\bibnamefont {{LCGT Collaboration}}},\
  }\href {\doibase 10.1088/0264-9381/27/8/084004} {\bibfield  {journal}
  {\bibinfo  {journal} {Class. Quantum Grav.}\ }\textbf {\bibinfo {volume}
  {27}},\ \bibinfo {eid} {084004} (\bibinfo {year} {2010})}\BibitemShut
  {NoStop}%
\bibitem [{\citenamefont {{Aso}}\ \emph {et~al.}(2013)\citenamefont {{Aso}},
  \citenamefont {{Michimura}}, \citenamefont {{Somiya}}, \citenamefont
  {{Ando}}, \citenamefont {{Miyakawa}}, \citenamefont {{Sekiguchi}},
  \citenamefont {{Tatsumi}},\ and\ \citenamefont {{Yamamoto}}}]{Aso:2013}%
  \BibitemOpen
  \bibfield  {author} {\bibinfo {author} {\bibfnamefont {Y.}~\bibnamefont
  {{Aso}}}, \bibinfo {author} {\bibfnamefont {Y.}~\bibnamefont {{Michimura}}},
  \bibinfo {author} {\bibfnamefont {K.}~\bibnamefont {{Somiya}}}, \bibinfo
  {author} {\bibfnamefont {M.}~\bibnamefont {{Ando}}}, \bibinfo {author}
  {\bibfnamefont {O.}~\bibnamefont {{Miyakawa}}}, \bibinfo {author}
  {\bibfnamefont {T.}~\bibnamefont {{Sekiguchi}}}, \bibinfo {author}
  {\bibfnamefont {D.}~\bibnamefont {{Tatsumi}}}, \ and\ \bibinfo {author}
  {\bibfnamefont {H.}~\bibnamefont {{Yamamoto}}},\ }\href {\doibase
  10.1103/PhysRevD.88.043007} {\bibfield  {journal} {\bibinfo  {journal} {Phys.
  Rev. D}\ }\textbf {\bibinfo {volume} {88}},\ \bibinfo {eid} {043007}
  (\bibinfo {year} {2013})},\ \Eprint {http://arxiv.org/abs/1306.6747}
  {arXiv:1306.6747 [gr-qc]} \BibitemShut {NoStop}%
\bibitem [{\citenamefont {{Chandrasekhar}}(1969)}]{Chandrasekhar1969book}%
  \BibitemOpen
  \bibfield  {author} {\bibinfo {author} {\bibfnamefont {S.}~\bibnamefont
  {{Chandrasekhar}}},\ }\href@noop {} {\emph {\bibinfo {title} {The Silliman
  Foundation Lectures, New Haven: Yale University Press, 1969}}}\ (\bibinfo
  {year} {1969})\BibitemShut {NoStop}%
\bibitem [{\citenamefont {{Meinel}}\ \emph {et~al.}(2008)\citenamefont
  {{Meinel}}, \citenamefont {{Ansorg}}, \citenamefont {{Kleinw{\"a}chter}},
  \citenamefont {{Neugebauer}},\ and\ \citenamefont {{Petroff}}}]{Meinel:2008}%
  \BibitemOpen
  \bibfield  {author} {\bibinfo {author} {\bibfnamefont {R.}~\bibnamefont
  {{Meinel}}}, \bibinfo {author} {\bibfnamefont {M.}~\bibnamefont {{Ansorg}}},
  \bibinfo {author} {\bibfnamefont {A.}~\bibnamefont {{Kleinw{\"a}chter}}},
  \bibinfo {author} {\bibfnamefont {G.}~\bibnamefont {{Neugebauer}}}, \ and\
  \bibinfo {author} {\bibfnamefont {D.}~\bibnamefont {{Petroff}}},\ }\href
  {\doibase 10.1017/CBO9780511535154} {\emph {\bibinfo {title} {{Relativistic
  Figures of Equilibrium}}}}\ (\bibinfo  {publisher} {Cambridge University
  Press},\ \bibinfo {year} {2008})\BibitemShut {NoStop}%
\bibitem [{\citenamefont {{Friedman}}\ and\ \citenamefont
  {{Stergioulas}}(2013)}]{Friedman2012}%
  \BibitemOpen
  \bibfield  {author} {\bibinfo {author} {\bibfnamefont {J.~L.}\ \bibnamefont
  {{Friedman}}}\ and\ \bibinfo {author} {\bibfnamefont {N.}~\bibnamefont
  {{Stergioulas}}},\ }\href@noop {} {\emph {\bibinfo {title} {Rotating
  Relativistic Stars, by John L.~Friedman , Nikolaos Stergioulas, Cambridge,
  UK: Cambridge University Press, 2013}}}\ (\bibinfo {year} {2013})\BibitemShut
  {NoStop}%
\bibitem [{\citenamefont {{Shapiro}}\ and\ \citenamefont
  {{Zane}}(1998)}]{shapiro98b}%
  \BibitemOpen
  \bibfield  {author} {\bibinfo {author} {\bibfnamefont {S.~L.}\ \bibnamefont
  {{Shapiro}}}\ and\ \bibinfo {author} {\bibfnamefont {S.}~\bibnamefont
  {{Zane}}},\ }\href {\doibase 10.1086/313124} {\bibfield  {journal} {\bibinfo
  {journal} {Astrop. J. Supp.}\ }\textbf {\bibinfo {volume} {117}},\ \bibinfo
  {pages} {531} (\bibinfo {year} {1998})}\BibitemShut {NoStop}%
\bibitem [{\citenamefont {{Bonazzola}}\ \emph {et~al.}(1996)\citenamefont
  {{Bonazzola}}, \citenamefont {{Frieben}},\ and\ \citenamefont
  {{Gourgoulhon}}}]{Bonazzola1996b}%
  \BibitemOpen
  \bibfield  {author} {\bibinfo {author} {\bibfnamefont {S.}~\bibnamefont
  {{Bonazzola}}}, \bibinfo {author} {\bibfnamefont {J.}~\bibnamefont
  {{Frieben}}}, \ and\ \bibinfo {author} {\bibfnamefont {E.}~\bibnamefont
  {{Gourgoulhon}}},\ }\href {\doibase 10.1086/176977} {\bibfield  {journal}
  {\bibinfo  {journal} {Astrophys. J.}\ }\textbf {\bibinfo {volume} {460}},\
  \bibinfo {pages} {379} (\bibinfo {year} {1996})},\ \Eprint
  {http://arxiv.org/abs/gr-qc/9509023} {gr-qc/9509023} \BibitemShut {NoStop}%
\bibitem [{\citenamefont {{Gondek-Rosi{\'n}ska}}\ and\ \citenamefont
  {{Gourgoulhon}}(2002)}]{rosinska2002}%
  \BibitemOpen
  \bibfield  {author} {\bibinfo {author} {\bibfnamefont {D.}~\bibnamefont
  {{Gondek-Rosi{\'n}ska}}}\ and\ \bibinfo {author} {\bibfnamefont
  {E.}~\bibnamefont {{Gourgoulhon}}},\ }\href {\doibase
  10.1103/PhysRevD.66.044021} {\bibfield  {journal} {\bibinfo  {journal} {Phys.
  Rev. D}\ }\textbf {\bibinfo {volume} {66}},\ \bibinfo {eid} {044021}
  (\bibinfo {year} {2002})},\ \Eprint {http://arxiv.org/abs/gr-qc/0205102}
  {gr-qc/0205102} \BibitemShut {NoStop}%
\bibitem [{\citenamefont {{Bonazzola}}\ \emph {et~al.}(1998)\citenamefont
  {{Bonazzola}}, \citenamefont {{Frieben}},\ and\ \citenamefont
  {{Gourgoulhon}}}]{Bonazzola1998c}%
  \BibitemOpen
  \bibfield  {author} {\bibinfo {author} {\bibfnamefont {S.}~\bibnamefont
  {{Bonazzola}}}, \bibinfo {author} {\bibfnamefont {J.}~\bibnamefont
  {{Frieben}}}, \ and\ \bibinfo {author} {\bibfnamefont {E.}~\bibnamefont
  {{Gourgoulhon}}},\ }\href@noop {} {\bibfield  {journal} {\bibinfo  {journal}
  {Astron. Astrophys.}\ }\textbf {\bibinfo {volume} {331}},\ \bibinfo {pages}
  {280} (\bibinfo {year} {1998})},\ \Eprint
  {http://arxiv.org/abs/gr-qc/9710121} {gr-qc/9710121} \BibitemShut {NoStop}%
\bibitem [{\citenamefont {{Houser}}\ and\ \citenamefont
  {{Centrella}}(1996)}]{Houser96}%
  \BibitemOpen
  \bibfield  {author} {\bibinfo {author} {\bibfnamefont {J.~L.}\ \bibnamefont
  {{Houser}}}\ and\ \bibinfo {author} {\bibfnamefont {J.~M.}\ \bibnamefont
  {{Centrella}}},\ }\href@noop {} {\bibfield  {journal} {\bibinfo  {journal}
  {Phys. Rev. D}\ }\textbf {\bibinfo {volume} {54}},\ \bibinfo {pages} {7278}
  (\bibinfo {year} {1996})},\ \Eprint {http://arxiv.org/abs/gr-qc/9611033}
  {gr-qc/9611033} \BibitemShut {NoStop}%
\bibitem [{\citenamefont {{Pickett}}\ \emph {et~al.}(1996)\citenamefont
  {{Pickett}}, \citenamefont {{Durisen}},\ and\ \citenamefont
  {{Davis}}}]{Pickett96}%
  \BibitemOpen
  \bibfield  {author} {\bibinfo {author} {\bibfnamefont {B.~K.}\ \bibnamefont
  {{Pickett}}}, \bibinfo {author} {\bibfnamefont {R.~H.}\ \bibnamefont
  {{Durisen}}}, \ and\ \bibinfo {author} {\bibfnamefont {G.~A.}\ \bibnamefont
  {{Davis}}},\ }\href {\doibase 10.1086/176852} {\bibfield  {journal} {\bibinfo
   {journal} {Astrophys. J.}\ }\textbf {\bibinfo {volume} {458}},\ \bibinfo
  {pages} {714} (\bibinfo {year} {1996})}\BibitemShut {NoStop}%
\bibitem [{\citenamefont {{Brown}}(2000)}]{brown2000}%
  \BibitemOpen
  \bibfield  {author} {\bibinfo {author} {\bibfnamefont {J.~D.}\ \bibnamefont
  {{Brown}}},\ }\href@noop {} {\bibfield  {journal} {\bibinfo  {journal} {Phys.
  Rev. D}\ }\textbf {\bibinfo {volume} {62}},\ \bibinfo {pages} {084024}
  (\bibinfo {year} {2000})},\ \Eprint {http://arxiv.org/abs/gr-qc/0004002}
  {gr-qc/0004002} \BibitemShut {NoStop}%
\bibitem [{\citenamefont {{New}}\ \emph {et~al.}(2000)\citenamefont {{New}},
  \citenamefont {{Centrella}},\ and\ \citenamefont {{Tohline}}}]{New2000}%
  \BibitemOpen
  \bibfield  {author} {\bibinfo {author} {\bibfnamefont {K.~C.~B.}\
  \bibnamefont {{New}}}, \bibinfo {author} {\bibfnamefont {J.~M.}\ \bibnamefont
  {{Centrella}}}, \ and\ \bibinfo {author} {\bibfnamefont {J.~E.}\ \bibnamefont
  {{Tohline}}},\ }\href {\doibase 10.1103/PhysRevD.62.064019} {\bibfield
  {journal} {\bibinfo  {journal} {Phys. Rev. D}\ }\textbf {\bibinfo {volume}
  {62}},\ \bibinfo {pages} {064019} (\bibinfo {year} {2000})},\ \Eprint
  {http://arxiv.org/abs/arXiv:astro-ph/9911525} {arXiv:astro-ph/9911525}
  \BibitemShut {NoStop}%
\bibitem [{\citenamefont {{Liu}}(2002)}]{Liu02}%
  \BibitemOpen
  \bibfield  {author} {\bibinfo {author} {\bibfnamefont {Y.~T.}\ \bibnamefont
  {{Liu}}},\ }\href {\doibase 10.1103/PhysRevD.65.124003} {\bibfield  {journal}
  {\bibinfo  {journal} {Phys. Rev. D}\ }\textbf {\bibinfo {volume} {65}},\
  \bibinfo {pages} {124003} (\bibinfo {year} {2002})},\ \Eprint
  {http://arxiv.org/abs/arXiv:gr-qc/0109078} {arXiv:gr-qc/0109078} \BibitemShut
  {NoStop}%
\bibitem [{\citenamefont {Watts}\ \emph {et~al.}(2005)\citenamefont {Watts},
  \citenamefont {Andersson},\ and\ \citenamefont {Jones}}]{Watts:2003nn}%
  \BibitemOpen
  \bibfield  {author} {\bibinfo {author} {\bibfnamefont {A.~L.}\ \bibnamefont
  {Watts}}, \bibinfo {author} {\bibfnamefont {N.}~\bibnamefont {Andersson}}, \
  and\ \bibinfo {author} {\bibfnamefont {D.~I.}\ \bibnamefont {Jones}},\
  }\href@noop {} {\bibfield  {journal} {\bibinfo  {journal} {Astrophys. J.}\
  }\textbf {\bibinfo {volume} {618}},\ \bibinfo {pages} {L37} (\bibinfo {year}
  {2005})},\ \Eprint {http://arxiv.org/abs/astro-ph/0309554} {astro-ph/0309554}
  \BibitemShut {NoStop}%
\bibitem [{\citenamefont {{Baiotti}}\ \emph {et~al.}(2007)\citenamefont
  {{Baiotti}}, \citenamefont {{De Pietri}}, \citenamefont {{Manca}},\ and\
  \citenamefont {{Rezzolla}}}]{Baiotti06b}%
  \BibitemOpen
  \bibfield  {author} {\bibinfo {author} {\bibfnamefont {L.}~\bibnamefont
  {{Baiotti}}}, \bibinfo {author} {\bibfnamefont {R.}~\bibnamefont {{De
  Pietri}}}, \bibinfo {author} {\bibfnamefont {G.~M.}\ \bibnamefont {{Manca}}},
  \ and\ \bibinfo {author} {\bibfnamefont {L.}~\bibnamefont {{Rezzolla}}},\
  }\href {\doibase 10.1103/PhysRevD.75.044023} {\bibfield  {journal} {\bibinfo
  {journal} {Phys. Rev. D}\ }\textbf {\bibinfo {volume} {75}},\ \bibinfo {eid}
  {044023} (\bibinfo {year} {2007})},\ \Eprint
  {http://arxiv.org/abs/astro-ph/0609473} {astro-ph/0609473} \BibitemShut
  {NoStop}%
\bibitem [{\citenamefont {{Manca}}\ \emph {et~al.}(2007)\citenamefont
  {{Manca}}, \citenamefont {{Baiotti}}, \citenamefont {{DePietri}},\ and\
  \citenamefont {{Rezzolla}}}]{Manca07}%
  \BibitemOpen
  \bibfield  {author} {\bibinfo {author} {\bibfnamefont {G.~M.}\ \bibnamefont
  {{Manca}}}, \bibinfo {author} {\bibfnamefont {L.}~\bibnamefont {{Baiotti}}},
  \bibinfo {author} {\bibfnamefont {R.}~\bibnamefont {{DePietri}}}, \ and\
  \bibinfo {author} {\bibfnamefont {L.}~\bibnamefont {{Rezzolla}}},\ }\href
  {\doibase 10.1088/0264-9381/24/12/S12} {\bibfield  {journal} {\bibinfo
  {journal} {Class. Quantum Grav.}\ }\textbf {\bibinfo {volume} {24}},\
  \bibinfo {pages} {S171} (\bibinfo {year} {2007})},\ \Eprint
  {http://arxiv.org/abs/0705.1826} {arXiv:0705.1826} \BibitemShut {NoStop}%
\bibitem [{\citenamefont {{Corvino}}\ \emph {et~al.}(2010)\citenamefont
  {{Corvino}}, \citenamefont {{Rezzolla}}, \citenamefont {{Bernuzzi}},
  \citenamefont {{De Pietri}},\ and\ \citenamefont
  {{Giacomazzo}}}]{Corvino:2010}%
  \BibitemOpen
  \bibfield  {author} {\bibinfo {author} {\bibfnamefont {G.}~\bibnamefont
  {{Corvino}}}, \bibinfo {author} {\bibfnamefont {L.}~\bibnamefont
  {{Rezzolla}}}, \bibinfo {author} {\bibfnamefont {S.}~\bibnamefont
  {{Bernuzzi}}}, \bibinfo {author} {\bibfnamefont {R.}~\bibnamefont {{De
  Pietri}}}, \ and\ \bibinfo {author} {\bibfnamefont {B.}~\bibnamefont
  {{Giacomazzo}}},\ }\href {\doibase 10.1088/0264-9381/27/11/114104} {\bibfield
   {journal} {\bibinfo  {journal} {Class. Quantum Grav.}\ }\textbf {\bibinfo
  {volume} {27}},\ \bibinfo {pages} {114104} (\bibinfo {year} {2010})},\
  \Eprint {http://arxiv.org/abs/1001.5281} {arXiv:1001.5281 [gr-qc]}
  \BibitemShut {NoStop}%
\bibitem [{\citenamefont {Camarda}\ \emph {et~al.}(2009)\citenamefont
  {Camarda}, \citenamefont {Anninos}, \citenamefont {Fragile},\ and\
  \citenamefont {Font}}]{Camarda:2009mk}%
  \BibitemOpen
  \bibfield  {author} {\bibinfo {author} {\bibfnamefont {K.~D.}\ \bibnamefont
  {Camarda}}, \bibinfo {author} {\bibfnamefont {P.}~\bibnamefont {Anninos}},
  \bibinfo {author} {\bibfnamefont {P.~C.}\ \bibnamefont {Fragile}}, \ and\
  \bibinfo {author} {\bibfnamefont {J.~A.}\ \bibnamefont {Font}},\ }\href
  {\doibase 10.1088/0004-637X/707/2/1610} {\bibfield  {journal} {\bibinfo
  {journal} {Astrophys. J.}\ }\textbf {\bibinfo {volume} {707}},\ \bibinfo
  {pages} {1610} (\bibinfo {year} {2009})},\ \Eprint
  {http://arxiv.org/abs/0911.0670} {arXiv:0911.0670 [astro-ph.SR]} \BibitemShut
  {NoStop}%
\bibitem [{\citenamefont {{Franci}}\ \emph {et~al.}(2013)\citenamefont
  {{Franci}}, \citenamefont {{De Pietri}}, \citenamefont {{Dionysopoulou}},\
  and\ \citenamefont {{Rezzolla}}}]{Franci2013}%
  \BibitemOpen
  \bibfield  {author} {\bibinfo {author} {\bibfnamefont {L.}~\bibnamefont
  {{Franci}}}, \bibinfo {author} {\bibfnamefont {R.}~\bibnamefont {{De
  Pietri}}}, \bibinfo {author} {\bibfnamefont {K.}~\bibnamefont
  {{Dionysopoulou}}}, \ and\ \bibinfo {author} {\bibfnamefont {L.}~\bibnamefont
  {{Rezzolla}}},\ }\href {\doibase 10.1088/1742-6596/470/1/012008} {\bibfield
  {journal} {\bibinfo  {journal} {Journal of Physics Conference Series}\
  }\textbf {\bibinfo {volume} {470}},\ \bibinfo {eid} {012008} (\bibinfo {year}
  {2013})},\ \Eprint {http://arxiv.org/abs/1309.6549} {arXiv:1309.6549 [gr-qc]}
  \BibitemShut {NoStop}%
\bibitem [{\citenamefont {{Muhlberger}}\ \emph {et~al.}(2014)\citenamefont
  {{Muhlberger}}, \citenamefont {{Nouri}}, \citenamefont {{Duez}},
  \citenamefont {{Foucart}}, \citenamefont {{Kidder}}, \citenamefont {{Ott}},
  \citenamefont {{Scheel}}, \citenamefont {{Szil{\'a}gyi}},\ and\ \citenamefont
  {{Teukolsky}}}]{Muhlberger2014}%
  \BibitemOpen
  \bibfield  {author} {\bibinfo {author} {\bibfnamefont {C.~D.}\ \bibnamefont
  {{Muhlberger}}}, \bibinfo {author} {\bibfnamefont {F.~H.}\ \bibnamefont
  {{Nouri}}}, \bibinfo {author} {\bibfnamefont {M.~D.}\ \bibnamefont {{Duez}}},
  \bibinfo {author} {\bibfnamefont {F.}~\bibnamefont {{Foucart}}}, \bibinfo
  {author} {\bibfnamefont {L.~E.}\ \bibnamefont {{Kidder}}}, \bibinfo {author}
  {\bibfnamefont {C.~D.}\ \bibnamefont {{Ott}}}, \bibinfo {author}
  {\bibfnamefont {M.~A.}\ \bibnamefont {{Scheel}}}, \bibinfo {author}
  {\bibfnamefont {B.}~\bibnamefont {{Szil{\'a}gyi}}}, \ and\ \bibinfo {author}
  {\bibfnamefont {S.~A.}\ \bibnamefont {{Teukolsky}}},\ }\href {\doibase
  10.1103/PhysRevD.90.104014} {\bibfield  {journal} {\bibinfo  {journal} {Phys.
  Rev. D}\ }\textbf {\bibinfo {volume} {90}},\ \bibinfo {eid} {104014}
  (\bibinfo {year} {2014})},\ \Eprint {http://arxiv.org/abs/1405.2144}
  {arXiv:1405.2144 [astro-ph.HE]} \BibitemShut {NoStop}%
\bibitem [{\citenamefont {{Andersson}}(2003)}]{Andersson03}%
  \BibitemOpen
  \bibfield  {author} {\bibinfo {author} {\bibfnamefont {N.}~\bibnamefont
  {{Andersson}}},\ }\href@noop {} {\bibfield  {journal} {\bibinfo  {journal}
  {Class. Quantum Grav.}\ }\textbf {\bibinfo {volume} {20}},\ \bibinfo {pages}
  {105} (\bibinfo {year} {2003})},\ \Eprint
  {http://arxiv.org/abs/arXiv:astro-ph/0211057} {arXiv:astro-ph/0211057}
  \BibitemShut {NoStop}%
\bibitem [{\citenamefont {{Lai}}\ and\ \citenamefont
  {{Shapiro}}(1995)}]{Lai95}%
  \BibitemOpen
  \bibfield  {author} {\bibinfo {author} {\bibfnamefont {D.}~\bibnamefont
  {{Lai}}}\ and\ \bibinfo {author} {\bibfnamefont {S.~L.}\ \bibnamefont
  {{Shapiro}}},\ }\href {\doibase 10.1086/175438} {\bibfield  {journal}
  {\bibinfo  {journal} {Astrophys. J.}\ }\textbf {\bibinfo {volume} {442}},\
  \bibinfo {pages} {259} (\bibinfo {year} {1995})},\ \Eprint
  {http://arxiv.org/abs/astro-ph/9408053} {astro-ph/9408053} \BibitemShut
  {NoStop}%
\bibitem [{\citenamefont {{Bildsten}}(1998)}]{Bildsten98}%
  \BibitemOpen
  \bibfield  {author} {\bibinfo {author} {\bibfnamefont {L.}~\bibnamefont
  {{Bildsten}}},\ }\href {\doibase 10.1086/311440} {\bibfield  {journal}
  {\bibinfo  {journal} {Astrophys. J. L}\ }\textbf {\bibinfo {volume} {501}},\
  \bibinfo {pages} {L89} (\bibinfo {year} {1998})},\ \Eprint
  {http://arxiv.org/abs/astro-ph/9804325} {astro-ph/9804325} \BibitemShut
  {NoStop}%
\bibitem [{\citenamefont {{Woosley}}\ and\ \citenamefont
  {{Janka}}(2005)}]{woosley2005}%
  \BibitemOpen
  \bibfield  {author} {\bibinfo {author} {\bibfnamefont {S.}~\bibnamefont
  {{Woosley}}}\ and\ \bibinfo {author} {\bibfnamefont {T.}~\bibnamefont
  {{Janka}}},\ }\href {\doibase 10.1038/nphys172} {\bibfield  {journal}
  {\bibinfo  {journal} {Nature Physics}\ }\textbf {\bibinfo {volume} {1}},\
  \bibinfo {pages} {147} (\bibinfo {year} {2005})},\ \Eprint
  {http://arxiv.org/abs/astro-ph/0601261} {astro-ph/0601261} \BibitemShut
  {NoStop}%
\bibitem [{\citenamefont {{Watts}}\ \emph {et~al.}(2008)\citenamefont
  {{Watts}}, \citenamefont {{Krishnan}}, \citenamefont {{Bildsten}},\ and\
  \citenamefont {{Schutz}}}]{watts2008}%
  \BibitemOpen
  \bibfield  {author} {\bibinfo {author} {\bibfnamefont {A.~L.}\ \bibnamefont
  {{Watts}}}, \bibinfo {author} {\bibfnamefont {B.}~\bibnamefont {{Krishnan}}},
  \bibinfo {author} {\bibfnamefont {L.}~\bibnamefont {{Bildsten}}}, \ and\
  \bibinfo {author} {\bibfnamefont {B.~F.}\ \bibnamefont {{Schutz}}},\ }\href
  {\doibase 10.1111/j.1365-2966.2008.13594.x} {\bibfield  {journal} {\bibinfo
  {journal} {Mon. Not. R. Astron. Soc.}\ }\textbf {\bibinfo {volume} {389}},\
  \bibinfo {pages} {839} (\bibinfo {year} {2008})},\ \Eprint
  {http://arxiv.org/abs/0803.4097} {arXiv:0803.4097} \BibitemShut {NoStop}%
\bibitem [{\citenamefont {{Piro}}\ and\ \citenamefont
  {{Thrane}}(2012)}]{piro2012b}%
  \BibitemOpen
  \bibfield  {author} {\bibinfo {author} {\bibfnamefont {A.~L.}\ \bibnamefont
  {{Piro}}}\ and\ \bibinfo {author} {\bibfnamefont {E.}~\bibnamefont
  {{Thrane}}},\ }\href {\doibase 10.1088/0004-637X/761/1/63} {\bibfield
  {journal} {\bibinfo  {journal} {Astrophys. J.}\ }\textbf {\bibinfo {volume}
  {761}},\ \bibinfo {eid} {63} (\bibinfo {year} {2012})},\ \Eprint
  {http://arxiv.org/abs/1207.3805} {arXiv:1207.3805 [astro-ph.HE]} \BibitemShut
  {NoStop}%
\bibitem [{\citenamefont {Huang}\ \emph {et~al.}(2008)\citenamefont {Huang},
  \citenamefont {Markakis}, \citenamefont {Sugiyama},\ and\ \citenamefont
  {Ury{\={u}}}}]{Huang08}%
  \BibitemOpen
  \bibfield  {author} {\bibinfo {author} {\bibfnamefont {X.}~\bibnamefont
  {Huang}}, \bibinfo {author} {\bibfnamefont {C.}~\bibnamefont {Markakis}},
  \bibinfo {author} {\bibfnamefont {N.}~\bibnamefont {Sugiyama}}, \ and\
  \bibinfo {author} {\bibfnamefont {K.}~\bibnamefont {Ury{\={u}}}},\ }\href
  {\doibase 10.1103/PhysRevD.78.124023} {\bibfield  {journal} {\bibinfo
  {journal} {Phys. Rev.}\ }\textbf {\bibinfo {volume} {D78}},\ \bibinfo {pages}
  {124023} (\bibinfo {year} {2008})},\ \Eprint {http://arxiv.org/abs/0809.0673}
  {arXiv:0809.0673 [astro-ph]} \BibitemShut {NoStop}%
\bibitem [{\citenamefont {Ury{\={u}}}\ \emph {et~al.}(2016)\citenamefont
  {Ury{\={u}}}, \citenamefont {Tsokaros}, \citenamefont {Galeazzi},
  \citenamefont {Hotta}, \citenamefont {Sugimura}, \citenamefont {Taniguchi},\
  and\ \citenamefont {Yoshida}}]{Uryu2016a}%
  \BibitemOpen
  \bibfield  {author} {\bibinfo {author} {\bibfnamefont {K.}~\bibnamefont
  {Ury{\={u}}}}, \bibinfo {author} {\bibfnamefont {A.}~\bibnamefont
  {Tsokaros}}, \bibinfo {author} {\bibfnamefont {F.}~\bibnamefont {Galeazzi}},
  \bibinfo {author} {\bibfnamefont {H.}~\bibnamefont {Hotta}}, \bibinfo
  {author} {\bibfnamefont {M.}~\bibnamefont {Sugimura}}, \bibinfo {author}
  {\bibfnamefont {K.}~\bibnamefont {Taniguchi}}, \ and\ \bibinfo {author}
  {\bibfnamefont {S.}~\bibnamefont {Yoshida}},\ }\href {\doibase
  10.1103/PhysRevD.93.044056} {\bibfield  {journal} {\bibinfo  {journal} {Phys.
  Rev.}\ }\textbf {\bibinfo {volume} {D93}},\ \bibinfo {pages} {044056}
  (\bibinfo {year} {2016})}\BibitemShut {NoStop}%
\bibitem [{\citenamefont {{James}}(1964)}]{Jame64}%
  \BibitemOpen
  \bibfield  {author} {\bibinfo {author} {\bibfnamefont {R.~A.}\ \bibnamefont
  {{James}}},\ }\href {\doibase 10.1086/147949} {\bibfield  {journal} {\bibinfo
   {journal} {Astrophys. J.}\ }\textbf {\bibinfo {volume} {140}},\ \bibinfo
  {pages} {552} (\bibinfo {year} {1964})}\BibitemShut {NoStop}%
\bibitem [{\citenamefont {{Hachisu}}\ and\ \citenamefont
  {{Eriguchi}}(1982)}]{hachisu1982}%
  \BibitemOpen
  \bibfield  {author} {\bibinfo {author} {\bibfnamefont {I.}~\bibnamefont
  {{Hachisu}}}\ and\ \bibinfo {author} {\bibfnamefont {Y.}~\bibnamefont
  {{Eriguchi}}},\ }\href {\doibase 10.1143/PTP.68.206} {\bibfield  {journal}
  {\bibinfo  {journal} {Progress of Theoretical Physics}\ }\textbf {\bibinfo
  {volume} {68}},\ \bibinfo {pages} {206} (\bibinfo {year} {1982})}\BibitemShut
  {NoStop}%
\bibitem [{\citenamefont {{Lai}}\ \emph {et~al.}(1993)\citenamefont {{Lai}},
  \citenamefont {{Rasio}},\ and\ \citenamefont {{Shapiro}}}]{Lai93}%
  \BibitemOpen
  \bibfield  {author} {\bibinfo {author} {\bibfnamefont {D.}~\bibnamefont
  {{Lai}}}, \bibinfo {author} {\bibfnamefont {F.~A.}\ \bibnamefont {{Rasio}}},
  \ and\ \bibinfo {author} {\bibfnamefont {S.~L.}\ \bibnamefont {{Shapiro}}},\
  }\href {\doibase 10.1086/191822} {\bibfield  {journal} {\bibinfo  {journal}
  {Astrophys. J., Suppl. Ser.}\ }\textbf {\bibinfo {volume} {88}},\ \bibinfo
  {pages} {205} (\bibinfo {year} {1993})}\BibitemShut {NoStop}%
\bibitem [{\citenamefont {{Bodmer}}(1971)}]{Bodmer1971}%
  \BibitemOpen
  \bibfield  {author} {\bibinfo {author} {\bibfnamefont {A.~R.}\ \bibnamefont
  {{Bodmer}}},\ }\href {\doibase 10.1103/PhysRevD.4.1601} {\bibfield  {journal}
  {\bibinfo  {journal} {Phys. Rev. D}\ }\textbf {\bibinfo {volume} {4}},\
  \bibinfo {pages} {1601} (\bibinfo {year} {1971})}\BibitemShut {NoStop}%
\bibitem [{\citenamefont {Witten}(1984)}]{Witten84}%
  \BibitemOpen
  \bibfield  {author} {\bibinfo {author} {\bibfnamefont {E.}~\bibnamefont
  {Witten}},\ }\href {\doibase 10.1103/physrevd.30.272} {\bibfield  {journal}
  {\bibinfo  {journal} {Phys. Rev. D}\ }\textbf {\bibinfo {volume} {30}},\
  \bibinfo {pages} {272} (\bibinfo {year} {1984})}\BibitemShut {NoStop}%
\bibitem [{\citenamefont {{Dai}}\ \emph {et~al.}(2016)\citenamefont {{Dai}},
  \citenamefont {{Wang}}, \citenamefont {{Wang}}, \citenamefont {{Wang}},\ and\
  \citenamefont {{Yu}}}]{Dai_ZG:2016}%
  \BibitemOpen
  \bibfield  {author} {\bibinfo {author} {\bibfnamefont {Z.~G.}\ \bibnamefont
  {{Dai}}}, \bibinfo {author} {\bibfnamefont {S.~Q.}\ \bibnamefont {{Wang}}},
  \bibinfo {author} {\bibfnamefont {J.~S.}\ \bibnamefont {{Wang}}}, \bibinfo
  {author} {\bibfnamefont {L.~J.}\ \bibnamefont {{Wang}}}, \ and\ \bibinfo
  {author} {\bibfnamefont {Y.~W.}\ \bibnamefont {{Yu}}},\ }\href {\doibase
  10.3847/0004-637X/817/2/132} {\bibfield  {journal} {\bibinfo  {journal}
  {Astrophys. J.}\ }\textbf {\bibinfo {volume} {817}},\ \bibinfo {eid} {132}
  (\bibinfo {year} {2016})},\ \Eprint {http://arxiv.org/abs/1508.07745}
  {arXiv:1508.07745 [astro-ph.HE]} \BibitemShut {NoStop}%
\bibitem [{\citenamefont {{Lai}}\ \emph {et~al.}(2017)\citenamefont {{Lai}},
  \citenamefont {{Yu}}, \citenamefont {{Zhou}}, \citenamefont {{Li}},\ and\
  \citenamefont {{Xu}}}]{Lai2017b}%
  \BibitemOpen
  \bibfield  {author} {\bibinfo {author} {\bibfnamefont {X.~Y.}\ \bibnamefont
  {{Lai}}}, \bibinfo {author} {\bibfnamefont {Y.~W.}\ \bibnamefont {{Yu}}},
  \bibinfo {author} {\bibfnamefont {E.~P.}\ \bibnamefont {{Zhou}}}, \bibinfo
  {author} {\bibfnamefont {Y.~Y.}\ \bibnamefont {{Li}}}, \ and\ \bibinfo
  {author} {\bibfnamefont {R.~X.}\ \bibnamefont {{Xu}}},\ }\href@noop {}
  {\bibfield  {journal} {\bibinfo  {journal} {Research in Astronomy and
  Astrophysics}\ ,\ \bibinfo {pages} {in press}} (\bibinfo {year} {2017})},\
  \Eprint {http://arxiv.org/abs/1710.04964} {arXiv:1710.04964 [astro-ph.HE]}
  \BibitemShut {NoStop}%
\bibitem [{\citenamefont {{Li}}\ \emph {et~al.}(2016)\citenamefont {{Li}},
  \citenamefont {{Zhang}}, \citenamefont {{Zhang}}, \citenamefont {{Gao}},
  \citenamefont {{Qi}},\ and\ \citenamefont {{Liu}}}]{Li2016}%
  \BibitemOpen
  \bibfield  {author} {\bibinfo {author} {\bibfnamefont {A.}~\bibnamefont
  {{Li}}}, \bibinfo {author} {\bibfnamefont {B.}~\bibnamefont {{Zhang}}},
  \bibinfo {author} {\bibfnamefont {N.-B.}\ \bibnamefont {{Zhang}}}, \bibinfo
  {author} {\bibfnamefont {H.}~\bibnamefont {{Gao}}}, \bibinfo {author}
  {\bibfnamefont {B.}~\bibnamefont {{Qi}}}, \ and\ \bibinfo {author}
  {\bibfnamefont {T.}~\bibnamefont {{Liu}}},\ }\href {\doibase
  10.1103/PhysRevD.94.083010} {\bibfield  {journal} {\bibinfo  {journal} {Phys.
  Rev. D}\ }\textbf {\bibinfo {volume} {94}},\ \bibinfo {eid} {083010}
  (\bibinfo {year} {2016})},\ \Eprint {http://arxiv.org/abs/1606.02934}
  {arXiv:1606.02934 [astro-ph.HE]} \BibitemShut {NoStop}%
\bibitem [{\citenamefont {{Itoh}}(1970)}]{Itoh70}%
  \BibitemOpen
  \bibfield  {author} {\bibinfo {author} {\bibfnamefont {N.}~\bibnamefont
  {{Itoh}}},\ }\href@noop {} {\bibfield  {journal} {\bibinfo  {journal}
  {Progress of Theoretical Physics}\ }\textbf {\bibinfo {volume} {44}},\
  \bibinfo {pages} {291} (\bibinfo {year} {1970})}\BibitemShut {NoStop}%
\bibitem [{\citenamefont {{Alcock}}\ \emph {et~al.}(1986)\citenamefont
  {{Alcock}}, \citenamefont {{Farhi}},\ and\ \citenamefont
  {{Olinto}}}]{Alcock86}%
  \BibitemOpen
  \bibfield  {author} {\bibinfo {author} {\bibfnamefont {C.}~\bibnamefont
  {{Alcock}}}, \bibinfo {author} {\bibfnamefont {E.}~\bibnamefont {{Farhi}}}, \
  and\ \bibinfo {author} {\bibfnamefont {A.}~\bibnamefont {{Olinto}}},\ }\href
  {\doibase 10.1086/164679} {\bibfield  {journal} {\bibinfo  {journal}
  {Astrophys. J.}\ }\textbf {\bibinfo {volume} {310}},\ \bibinfo {pages} {261}
  (\bibinfo {year} {1986})}\BibitemShut {NoStop}%
\bibitem [{\citenamefont {{Haensel}}\ \emph {et~al.}(1986)\citenamefont
  {{Haensel}}, \citenamefont {{Zdunik}},\ and\ \citenamefont
  {{Schaefer}}}]{Haensel1986}%
  \BibitemOpen
  \bibfield  {author} {\bibinfo {author} {\bibfnamefont {P.}~\bibnamefont
  {{Haensel}}}, \bibinfo {author} {\bibfnamefont {J.~L.}\ \bibnamefont
  {{Zdunik}}}, \ and\ \bibinfo {author} {\bibfnamefont {R.}~\bibnamefont
  {{Schaefer}}},\ }\href
  {http://articles.adsabs.harvard.edu/full/1986A%26A...160..121H} {\bibfield
  {journal} {\bibinfo  {journal} {Astron. Astrophys.}\ }\textbf {\bibinfo
  {volume} {160}},\ \bibinfo {pages} {121} (\bibinfo {year}
  {1986})}\BibitemShut {NoStop}%
\bibitem [{\citenamefont {{Gondek-Rosi{\'n}ska}}\ \emph
  {et~al.}(2000)\citenamefont {{Gondek-Rosi{\'n}ska}}, \citenamefont {{Bulik}},
  \citenamefont {{Zdunik}}, \citenamefont {{Gourgoulhon}}, \citenamefont
  {{Ray}}, \citenamefont {{Dey}},\ and\ \citenamefont {{Dey}}}]{rosinska2000b}%
  \BibitemOpen
  \bibfield  {author} {\bibinfo {author} {\bibfnamefont {D.}~\bibnamefont
  {{Gondek-Rosi{\'n}ska}}}, \bibinfo {author} {\bibfnamefont {T.}~\bibnamefont
  {{Bulik}}}, \bibinfo {author} {\bibfnamefont {L.}~\bibnamefont {{Zdunik}}},
  \bibinfo {author} {\bibfnamefont {E.}~\bibnamefont {{Gourgoulhon}}}, \bibinfo
  {author} {\bibfnamefont {S.}~\bibnamefont {{Ray}}}, \bibinfo {author}
  {\bibfnamefont {J.}~\bibnamefont {{Dey}}}, \ and\ \bibinfo {author}
  {\bibfnamefont {M.}~\bibnamefont {{Dey}}},\ }\href@noop {} {\bibfield
  {journal} {\bibinfo  {journal} {Astron. Astrophys.}\ }\textbf {\bibinfo
  {volume} {363}},\ \bibinfo {pages} {1005} (\bibinfo {year} {2000})},\ \Eprint
  {http://arxiv.org/abs/astro-ph/0007004} {astro-ph/0007004} \BibitemShut
  {NoStop}%
\bibitem [{\citenamefont {{Gourgoulhon}}\ \emph {et~al.}(1999)\citenamefont
  {{Gourgoulhon}}, \citenamefont {{Haensel}}, \citenamefont {{Livine}},
  \citenamefont {{Paluch}}, \citenamefont {{Bonazzola}},\ and\ \citenamefont
  {{Marck}}}]{gourgoulhon1999}%
  \BibitemOpen
  \bibfield  {author} {\bibinfo {author} {\bibfnamefont {E.}~\bibnamefont
  {{Gourgoulhon}}}, \bibinfo {author} {\bibfnamefont {P.}~\bibnamefont
  {{Haensel}}}, \bibinfo {author} {\bibfnamefont {R.}~\bibnamefont {{Livine}}},
  \bibinfo {author} {\bibfnamefont {E.}~\bibnamefont {{Paluch}}}, \bibinfo
  {author} {\bibfnamefont {S.}~\bibnamefont {{Bonazzola}}}, \ and\ \bibinfo
  {author} {\bibfnamefont {J.-A.}\ \bibnamefont {{Marck}}},\ }\href@noop {}
  {\bibfield  {journal} {\bibinfo  {journal} {Astron. Astrophys.}\ }\textbf
  {\bibinfo {volume} {349}},\ \bibinfo {pages} {851} (\bibinfo {year}
  {1999})},\ \Eprint {http://arxiv.org/abs/astro-ph/9907225} {astro-ph/9907225}
  \BibitemShut {NoStop}%
\bibitem [{\citenamefont {{Stergioulas}}\ \emph {et~al.}(1999)\citenamefont
  {{Stergioulas}}, \citenamefont {{Klu{\'z}niak}},\ and\ \citenamefont
  {{Bulik}}}]{Stergioulas99a}%
  \BibitemOpen
  \bibfield  {author} {\bibinfo {author} {\bibfnamefont {N.}~\bibnamefont
  {{Stergioulas}}}, \bibinfo {author} {\bibfnamefont {W.}~\bibnamefont
  {{Klu{\'z}niak}}}, \ and\ \bibinfo {author} {\bibfnamefont {T.}~\bibnamefont
  {{Bulik}}},\ }\href@noop {} {\bibfield  {journal} {\bibinfo  {journal}
  {Astron. Astrophys.}\ }\textbf {\bibinfo {volume} {352}},\ \bibinfo {pages}
  {L116} (\bibinfo {year} {1999})},\ \Eprint
  {http://arxiv.org/abs/astro-ph/9909152} {astro-ph/9909152} \BibitemShut
  {NoStop}%
\bibitem [{\citenamefont {{Szkudlarek}}\ \emph {et~al.}(2012)\citenamefont
  {{Szkudlarek}}, \citenamefont {{Gondek-Rosi{\'n}}}, \citenamefont {{ska}},
  \citenamefont {{Villain}},\ and\ \citenamefont {{Ansorg}}}]{szkudlarek2012}%
  \BibitemOpen
  \bibfield  {author} {\bibinfo {author} {\bibfnamefont {M.}~\bibnamefont
  {{Szkudlarek}}}, \bibinfo {author} {\bibnamefont {{Gondek-Rosi{\'n}}}},
  \bibinfo {author} {\bibfnamefont {D.}~\bibnamefont {{ska}}}, \bibinfo
  {author} {\bibfnamefont {L.}~\bibnamefont {{Villain}}}, \ and\ \bibinfo
  {author} {\bibfnamefont {M.}~\bibnamefont {{Ansorg}}},\ }in\ \href
  {http://aspbooks.org/custom/publications/paper/466-0231.html} {\emph
  {\bibinfo {booktitle} {Electromagnetic Radiation from Pulsars and
  Magnetars}}},\ \bibinfo {series} {Astronomical Society of the Pacific
  Conference Series}, Vol.\ \bibinfo {volume} {466},\ \bibinfo {editor} {edited
  by\ \bibinfo {editor} {\bibfnamefont {W.}~\bibnamefont {{Lewandowski}}},
  \bibinfo {editor} {\bibfnamefont {O.}~\bibnamefont {{Maron}}}, \ and\
  \bibinfo {editor} {\bibfnamefont {J.}~\bibnamefont {{Kijak}}}}\ (\bibinfo
  {year} {2012})\ p.\ \bibinfo {pages} {231}\BibitemShut {NoStop}%
\bibitem [{\citenamefont {{Gondek-Rosinska}}\ \emph {et~al.}(2000)\citenamefont
  {{Gondek-Rosinska}}, \citenamefont {{Haensel}}, \citenamefont {{Zdunik}},\
  and\ \citenamefont {{Gourgoulhon}}}]{rosinska2000a}%
  \BibitemOpen
  \bibfield  {author} {\bibinfo {author} {\bibfnamefont {D.}~\bibnamefont
  {{Gondek-Rosinska}}}, \bibinfo {author} {\bibfnamefont {P.}~\bibnamefont
  {{Haensel}}}, \bibinfo {author} {\bibfnamefont {J.~L.}\ \bibnamefont
  {{Zdunik}}}, \ and\ \bibinfo {author} {\bibfnamefont {E.}~\bibnamefont
  {{Gourgoulhon}}},\ }in\ \href@noop {} {\emph {\bibinfo {booktitle} {IAU
  Colloq. 177: Pulsar Astronomy - 2000 and Beyond}}},\ \bibinfo {series}
  {Astronomical Society of the Pacific Conference Series}, Vol.\ \bibinfo
  {volume} {202},\ \bibinfo {editor} {edited by\ \bibinfo {editor}
  {\bibfnamefont {M.}~\bibnamefont {{Kramer}}}, \bibinfo {editor}
  {\bibfnamefont {N.}~\bibnamefont {{Wex}}}, \ and\ \bibinfo {editor}
  {\bibfnamefont {R.}~\bibnamefont {{Wielebinski}}}}\ (\bibinfo {year} {2000})\
  p.\ \bibinfo {pages} {661},\ \Eprint {http://arxiv.org/abs/astro-ph/0009282}
  {astro-ph/0009282} \BibitemShut {NoStop}%
\bibitem [{\citenamefont {{Gondek-Rosi{\'n}ska}}\ \emph
  {et~al.}(2001)\citenamefont {{Gondek-Rosi{\'n}ska}}, \citenamefont
  {{Stergioulas}}, \citenamefont {{Bulik}}, \citenamefont {{Klu{\'z}niak}},\
  and\ \citenamefont {{Gourgoulhon}}}]{rosinska2001}%
  \BibitemOpen
  \bibfield  {author} {\bibinfo {author} {\bibfnamefont {D.}~\bibnamefont
  {{Gondek-Rosi{\'n}ska}}}, \bibinfo {author} {\bibfnamefont {N.}~\bibnamefont
  {{Stergioulas}}}, \bibinfo {author} {\bibfnamefont {T.}~\bibnamefont
  {{Bulik}}}, \bibinfo {author} {\bibfnamefont {W.}~\bibnamefont
  {{Klu{\'z}niak}}}, \ and\ \bibinfo {author} {\bibfnamefont {E.}~\bibnamefont
  {{Gourgoulhon}}},\ }\href {\doibase 10.1051/0004-6361:20011328} {\bibfield
  {journal} {\bibinfo  {journal} {Astron. Astrophys.}\ }\textbf {\bibinfo
  {volume} {380}},\ \bibinfo {pages} {190} (\bibinfo {year} {2001})},\ \Eprint
  {http://arxiv.org/abs/astro-ph/0110209} {astro-ph/0110209} \BibitemShut
  {NoStop}%
\bibitem [{\citenamefont {{Gondek-Rosi{\'n}ska}}\ \emph
  {et~al.}(2003)\citenamefont {{Gondek-Rosi{\'n}ska}}, \citenamefont
  {{Gourgoulhon}},\ and\ \citenamefont {{Haensel}}}]{rosinska2003}%
  \BibitemOpen
  \bibfield  {author} {\bibinfo {author} {\bibfnamefont {D.}~\bibnamefont
  {{Gondek-Rosi{\'n}ska}}}, \bibinfo {author} {\bibfnamefont {E.}~\bibnamefont
  {{Gourgoulhon}}}, \ and\ \bibinfo {author} {\bibfnamefont {P.}~\bibnamefont
  {{Haensel}}},\ }\href {\doibase 10.1051/0004-6361:20031431} {\bibfield
  {journal} {\bibinfo  {journal} {Astron. Astrophys.}\ }\textbf {\bibinfo
  {volume} {412}},\ \bibinfo {pages} {777} (\bibinfo {year} {2003})},\ \Eprint
  {http://arxiv.org/abs/astro-ph/0311128} {astro-ph/0311128} \BibitemShut
  {NoStop}%
\bibitem [{\citenamefont {{Ury{\={u}}}}\ and\ \citenamefont
  {{Tsokaros}}(2012)}]{Uryu2012}%
  \BibitemOpen
  \bibfield  {author} {\bibinfo {author} {\bibfnamefont {K.}~\bibnamefont
  {{Ury{\={u}}}}}\ and\ \bibinfo {author} {\bibfnamefont {A.}~\bibnamefont
  {{Tsokaros}}},\ }\href {\doibase 10.1103/PhysRevD.85.064014} {\bibfield
  {journal} {\bibinfo  {journal} {Phys. Rev. D}\ }\textbf {\bibinfo {volume}
  {85}},\ \bibinfo {eid} {064014} (\bibinfo {year} {2012})},\ \Eprint
  {http://arxiv.org/abs/1108.3065} {arXiv:1108.3065 [gr-qc]} \BibitemShut
  {NoStop}%
\bibitem [{\citenamefont {Ury{\={u}}}\ \emph {et~al.}(2012)\citenamefont
  {Ury{\={u}}}, \citenamefont {Tsokaros},\ and\ \citenamefont
  {Grandclement}}]{Uryu:2012b}%
  \BibitemOpen
  \bibfield  {author} {\bibinfo {author} {\bibfnamefont {K.}~\bibnamefont
  {Ury{\={u}}}}, \bibinfo {author} {\bibfnamefont {A.}~\bibnamefont
  {Tsokaros}}, \ and\ \bibinfo {author} {\bibfnamefont {P.}~\bibnamefont
  {Grandclement}},\ }\href {\doibase 10.1103/PhysRevD.86.104001} {\bibfield
  {journal} {\bibinfo  {journal} {Phys. Rev.}\ }\textbf {\bibinfo {volume}
  {D86}},\ \bibinfo {pages} {104001} (\bibinfo {year} {2012})},\ \Eprint
  {http://arxiv.org/abs/1210.5811} {arXiv:1210.5811 [gr-qc]} \BibitemShut
  {NoStop}%
\bibitem [{\citenamefont {Tsokaros}\ and\ \citenamefont
  {Ury{\={u}}}(2012)}]{Tsokaros2012}%
  \BibitemOpen
  \bibfield  {author} {\bibinfo {author} {\bibfnamefont {A.}~\bibnamefont
  {Tsokaros}}\ and\ \bibinfo {author} {\bibfnamefont {K.}~\bibnamefont
  {Ury{\={u}}}},\ }\href {\doibase 10.1007/s10665-012-9585-6} {\bibfield
  {journal} {\bibinfo  {journal} {Journal of Engineering Mathematics}\ }\textbf
  {\bibinfo {volume} {82}},\ \bibinfo {pages} {133} (\bibinfo {year}
  {2012})}\BibitemShut {NoStop}%
\bibitem [{\citenamefont {{Tsokaros}}\ \emph {et~al.}(2015)\citenamefont
  {{Tsokaros}}, \citenamefont {{Ury{\={u}}}},\ and\ \citenamefont
  {{Rezzolla}}}]{Tsokaros2015}%
  \BibitemOpen
  \bibfield  {author} {\bibinfo {author} {\bibfnamefont {A.}~\bibnamefont
  {{Tsokaros}}}, \bibinfo {author} {\bibfnamefont {K.}~\bibnamefont
  {{Ury{\={u}}}}}, \ and\ \bibinfo {author} {\bibfnamefont {L.}~\bibnamefont
  {{Rezzolla}}},\ }\href {\doibase 10.1103/PhysRevD.91.104030} {\bibfield
  {journal} {\bibinfo  {journal} {Phys. Rev. D}\ }\textbf {\bibinfo {volume}
  {91}},\ \bibinfo {eid} {104030} (\bibinfo {year} {2015})},\ \Eprint
  {http://arxiv.org/abs/1502.05674} {arXiv:1502.05674 [gr-qc]} \BibitemShut
  {NoStop}%
\bibitem [{\citenamefont {{Tsokaros}}\ \emph {et~al.}(2017)\citenamefont
  {{Tsokaros}}, \citenamefont {{Ruiz}}, \citenamefont {{Paschalidis}},
  \citenamefont {{Shapiro}}, \citenamefont {{Baiotti}},\ and\ \citenamefont
  {{Ury{\= u}}}}]{Tsokaros2017}%
  \BibitemOpen
  \bibfield  {author} {\bibinfo {author} {\bibfnamefont {A.}~\bibnamefont
  {{Tsokaros}}}, \bibinfo {author} {\bibfnamefont {M.}~\bibnamefont {{Ruiz}}},
  \bibinfo {author} {\bibfnamefont {V.}~\bibnamefont {{Paschalidis}}}, \bibinfo
  {author} {\bibfnamefont {S.~L.}\ \bibnamefont {{Shapiro}}}, \bibinfo {author}
  {\bibfnamefont {L.}~\bibnamefont {{Baiotti}}}, \ and\ \bibinfo {author}
  {\bibfnamefont {K.}~\bibnamefont {{Ury{\= u}}}},\ }\href {\doibase
  10.1103/PhysRevD.95.124057} {\bibfield  {journal} {\bibinfo  {journal}
  {\prd}\ }\textbf {\bibinfo {volume} {95}},\ \bibinfo {eid} {124057} (\bibinfo
  {year} {2017})},\ \Eprint {http://arxiv.org/abs/1704.00038} {arXiv:1704.00038
  [gr-qc]} \BibitemShut {NoStop}%
\bibitem [{\citenamefont {{Rezzolla}}\ and\ \citenamefont
  {{Zanotti}}(2013)}]{Rezzolla_book:2013}%
  \BibitemOpen
  \bibfield  {author} {\bibinfo {author} {\bibfnamefont {L.}~\bibnamefont
  {{Rezzolla}}}\ and\ \bibinfo {author} {\bibfnamefont {O.}~\bibnamefont
  {{Zanotti}}},\ }\href {\doibase 10.1093/acprof:oso/9780198528906.001.0001}
  {\emph {\bibinfo {title} {Relativistic Hydrodynamics}}}\ (\bibinfo
  {publisher} {Oxford University Press},\ \bibinfo {address} {Oxford, UK},\
  \bibinfo {year} {2013})\BibitemShut {NoStop}%
\bibitem [{\citenamefont {{Isenberg}}\ and\ \citenamefont
  {{Nester}}(1980)}]{Isenberg1980}%
  \BibitemOpen
  \bibfield  {author} {\bibinfo {author} {\bibfnamefont {J.}~\bibnamefont
  {{Isenberg}}}\ and\ \bibinfo {author} {\bibfnamefont {J.}~\bibnamefont
  {{Nester}}},\ }in\ \href@noop {} {\emph {\bibinfo {booktitle} {General
  Relativity and Gravitation. Vol. 1. One hundred years after the birth of
  Albert Einstein. Edited by A. Held. New York, NY: Plenum Press, p. 23,
  1980}}},\ Vol.~\bibinfo {volume} {1},\ \bibinfo {editor} {edited by\ \bibinfo
  {editor} {\bibfnamefont {A.}~\bibnamefont {{Held}}}}\ (\bibinfo {year}
  {1980})\ p.~\bibinfo {pages} {23}\BibitemShut {NoStop}%
\bibitem [{\citenamefont {{Isenberg}}(2008)}]{Isenberg08}%
  \BibitemOpen
  \bibfield  {author} {\bibinfo {author} {\bibfnamefont {J.~A.}\ \bibnamefont
  {{Isenberg}}},\ }\href {\doibase 10.1142/S0218271808011997} {\bibfield
  {journal} {\bibinfo  {journal} {International Journal of Modern Physics D}\
  }\textbf {\bibinfo {volume} {17}},\ \bibinfo {pages} {265} (\bibinfo {year}
  {2008})},\ \Eprint {http://arxiv.org/abs/arXiv:gr-qc/0702113}
  {arXiv:gr-qc/0702113} \BibitemShut {NoStop}%
\bibitem [{\citenamefont {{Wilson}}\ and\ \citenamefont
  {{Mathews}}(1989)}]{Wilson89}%
  \BibitemOpen
  \bibfield  {author} {\bibinfo {author} {\bibfnamefont {J.~R.}\ \bibnamefont
  {{Wilson}}}\ and\ \bibinfo {author} {\bibfnamefont {G.~J.}\ \bibnamefont
  {{Mathews}}},\ }\enquote {\bibinfo {title} {{Relativistic hydrodynamics.}}}\
  in\ \href@noop {} {\emph {\bibinfo {booktitle} {Frontiers in Numerical
  Relativity}}},\ \bibinfo {editor} {edited by\ \bibinfo {editor}
  {\bibfnamefont {C.~R.}\ \bibnamefont {{Evans}}}, \bibinfo {editor}
  {\bibfnamefont {L.~S.}\ \bibnamefont {{Finn}}}, \ and\ \bibinfo {editor}
  {\bibfnamefont {D.~W.}\ \bibnamefont {{Hobill}}}}\ (\bibinfo {year} {1989})\
  pp.\ \bibinfo {pages} {306--314}\BibitemShut {NoStop}%
\bibitem [{\citenamefont {{Chodos}}\ \emph {et~al.}(1974)\citenamefont
  {{Chodos}}, \citenamefont {{Jaffe}}, \citenamefont {{Johnson}}, \citenamefont
  {{Thorn}},\ and\ \citenamefont {{Weisskopf}}}]{chodos1974}%
  \BibitemOpen
  \bibfield  {author} {\bibinfo {author} {\bibfnamefont {A.}~\bibnamefont
  {{Chodos}}}, \bibinfo {author} {\bibfnamefont {R.~L.}\ \bibnamefont
  {{Jaffe}}}, \bibinfo {author} {\bibfnamefont {K.}~\bibnamefont {{Johnson}}},
  \bibinfo {author} {\bibfnamefont {C.~B.}\ \bibnamefont {{Thorn}}}, \ and\
  \bibinfo {author} {\bibfnamefont {V.~F.}\ \bibnamefont {{Weisskopf}}},\
  }\href {\doibase 10.1103/PhysRevD.9.3471} {\bibfield  {journal} {\bibinfo
  {journal} {Phys. Rev. D}\ }\textbf {\bibinfo {volume} {9}},\ \bibinfo {pages}
  {3471} (\bibinfo {year} {1974})}\BibitemShut {NoStop}%
\bibitem [{\citenamefont {{Limousin}}\ \emph {et~al.}(2005)\citenamefont
  {{Limousin}}, \citenamefont {{Gondek-Rosi{\'n}ska}},\ and\ \citenamefont
  {{Gourgoulhon}}}]{limousin2005}%
  \BibitemOpen
  \bibfield  {author} {\bibinfo {author} {\bibfnamefont {F.}~\bibnamefont
  {{Limousin}}}, \bibinfo {author} {\bibfnamefont {D.}~\bibnamefont
  {{Gondek-Rosi{\'n}ska}}}, \ and\ \bibinfo {author} {\bibfnamefont
  {E.}~\bibnamefont {{Gourgoulhon}}},\ }\href {\doibase
  10.1103/PhysRevD.71.064012} {\bibfield  {journal} {\bibinfo  {journal} {Phys.
  Rev. D}\ }\textbf {\bibinfo {volume} {71}},\ \bibinfo {eid} {064012}
  (\bibinfo {year} {2005})},\ \Eprint {http://arxiv.org/abs/gr-qc/0411127}
  {gr-qc/0411127} \BibitemShut {NoStop}%
\bibitem [{\citenamefont {{Lai}}\ and\ \citenamefont {{Xu}}(2009)}]{lai2009}%
  \BibitemOpen
  \bibfield  {author} {\bibinfo {author} {\bibfnamefont {X.~Y.}\ \bibnamefont
  {{Lai}}}\ and\ \bibinfo {author} {\bibfnamefont {R.~X.}\ \bibnamefont
  {{Xu}}},\ }\href {\doibase 10.1111/j.1745-3933.2009.00701.x} {\bibfield
  {journal} {\bibinfo  {journal} {Mon. Not. R. Astron. Soc.}\ }\textbf
  {\bibinfo {volume} {398}},\ \bibinfo {pages} {L31} (\bibinfo {year}
  {2009})},\ \Eprint {http://arxiv.org/abs/0905.2839} {arXiv:0905.2839
  [astro-ph.HE]} \BibitemShut {NoStop}%
\bibitem [{\citenamefont {{Lai}}\ and\ \citenamefont {{Xu}}(2017)}]{lai2017}%
  \BibitemOpen
  \bibfield  {author} {\bibinfo {author} {\bibfnamefont {X.~Y.}\ \bibnamefont
  {{Lai}}}\ and\ \bibinfo {author} {\bibfnamefont {R.~X.}\ \bibnamefont
  {{Xu}}},\ }in\ \href {\doibase 10.1088/1742-6596/861/1/012027} {\emph
  {\bibinfo {booktitle} {Journal of Physics Conference Series}}},\ \bibinfo
  {series} {Journal of Physics Conference Series}, Vol.\ \bibinfo {volume}
  {861}\ (\bibinfo {year} {2017})\ p.\ \bibinfo {pages} {012027},\ \Eprint
  {http://arxiv.org/abs/1701.08463} {arXiv:1701.08463 [astro-ph.HE]}
  \BibitemShut {NoStop}%
\bibitem [{\citenamefont {{Guo}}\ \emph {et~al.}(2014)\citenamefont {{Guo}},
  \citenamefont {{Lai}},\ and\ \citenamefont {{Xu}}}]{Guo2014}%
  \BibitemOpen
  \bibfield  {author} {\bibinfo {author} {\bibfnamefont {Y.-J.}\ \bibnamefont
  {{Guo}}}, \bibinfo {author} {\bibfnamefont {X.-Y.}\ \bibnamefont {{Lai}}}, \
  and\ \bibinfo {author} {\bibfnamefont {R.-X.}\ \bibnamefont {{Xu}}},\ }\href
  {\doibase 10.1088/1674-1137/38/5/055101} {\bibfield  {journal} {\bibinfo
  {journal} {Chinese Physics C}\ }\textbf {\bibinfo {volume} {38}},\ \bibinfo
  {eid} {055101} (\bibinfo {year} {2014})}\BibitemShut {NoStop}%
\bibitem [{\citenamefont {{Lai}}\ and\ \citenamefont {{Xu}}(2011)}]{lai2011}%
  \BibitemOpen
  \bibfield  {author} {\bibinfo {author} {\bibfnamefont {X.-Y.}\ \bibnamefont
  {{Lai}}}\ and\ \bibinfo {author} {\bibfnamefont {R.-X.}\ \bibnamefont
  {{Xu}}},\ }\href {\doibase 10.1088/1674-4527/11/6/008} {\bibfield  {journal}
  {\bibinfo  {journal} {Research in Astronomy and Astrophysics}\ }\textbf
  {\bibinfo {volume} {11}},\ \bibinfo {pages} {687} (\bibinfo {year} {2011})},\
  \Eprint {http://arxiv.org/abs/1011.0526} {arXiv:1011.0526 [astro-ph.SR]}
  \BibitemShut {NoStop}%
\bibitem [{\citenamefont {{Demorest}}\ \emph {et~al.}(2010)\citenamefont
  {{Demorest}}, \citenamefont {{Pennucci}}, \citenamefont {{Ransom}},
  \citenamefont {{Roberts}},\ and\ \citenamefont {{Hessels}}}]{Demorest2010}%
  \BibitemOpen
  \bibfield  {author} {\bibinfo {author} {\bibfnamefont {P.~B.}\ \bibnamefont
  {{Demorest}}}, \bibinfo {author} {\bibfnamefont {T.}~\bibnamefont
  {{Pennucci}}}, \bibinfo {author} {\bibfnamefont {S.~M.}\ \bibnamefont
  {{Ransom}}}, \bibinfo {author} {\bibfnamefont {M.~S.~E.}\ \bibnamefont
  {{Roberts}}}, \ and\ \bibinfo {author} {\bibfnamefont {J.~W.~T.}\
  \bibnamefont {{Hessels}}},\ }\href {\doibase 10.1038/nature09466} {\bibfield
  {journal} {\bibinfo  {journal} {Nature}\ }\textbf {\bibinfo {volume} {467}},\
  \bibinfo {pages} {1081} (\bibinfo {year} {2010})},\ \Eprint
  {http://arxiv.org/abs/1010.5788} {arXiv:1010.5788 [astro-ph.HE]} \BibitemShut
  {NoStop}%
\bibitem [{\citenamefont {{Antoniadis}}\ \emph {et~al.}(2013)\citenamefont
  {{Antoniadis}}, \citenamefont {{Freire}}, \citenamefont {{Wex}},
  \citenamefont {{Tauris}}, \citenamefont {{Lynch}}, \citenamefont {{van
  Kerkwijk}}, \citenamefont {{Kramer}}, \citenamefont {{Bassa}}, \citenamefont
  {{Dhillon}}, \citenamefont {{Driebe}}, \citenamefont {{Hessels}},
  \citenamefont {{Kaspi}}, \citenamefont {{Kondratiev}}, \citenamefont
  {{Langer}}, \citenamefont {{Marsh}}, \citenamefont {{McLaughlin}},
  \citenamefont {{Pennucci}}, \citenamefont {{Ransom}}, \citenamefont
  {{Stairs}}, \citenamefont {{van Leeuwen}}, \citenamefont {{Verbiest}},\ and\
  \citenamefont {{Whelan}}}]{Antoniadis2013}%
  \BibitemOpen
  \bibfield  {author} {\bibinfo {author} {\bibfnamefont {J.}~\bibnamefont
  {{Antoniadis}}}, \bibinfo {author} {\bibfnamefont {P.~C.~C.}\ \bibnamefont
  {{Freire}}}, \bibinfo {author} {\bibfnamefont {N.}~\bibnamefont {{Wex}}},
  \bibinfo {author} {\bibfnamefont {T.~M.}\ \bibnamefont {{Tauris}}}, \bibinfo
  {author} {\bibfnamefont {R.~S.}\ \bibnamefont {{Lynch}}}, \bibinfo {author}
  {\bibfnamefont {M.~H.}\ \bibnamefont {{van Kerkwijk}}}, \bibinfo {author}
  {\bibfnamefont {M.}~\bibnamefont {{Kramer}}}, \bibinfo {author}
  {\bibfnamefont {C.}~\bibnamefont {{Bassa}}}, \bibinfo {author} {\bibfnamefont
  {V.~S.}\ \bibnamefont {{Dhillon}}}, \bibinfo {author} {\bibfnamefont
  {T.}~\bibnamefont {{Driebe}}}, \bibinfo {author} {\bibfnamefont {J.~W.~T.}\
  \bibnamefont {{Hessels}}}, \bibinfo {author} {\bibfnamefont {V.~M.}\
  \bibnamefont {{Kaspi}}}, \bibinfo {author} {\bibfnamefont {V.~I.}\
  \bibnamefont {{Kondratiev}}}, \bibinfo {author} {\bibfnamefont
  {N.}~\bibnamefont {{Langer}}}, \bibinfo {author} {\bibfnamefont {T.~R.}\
  \bibnamefont {{Marsh}}}, \bibinfo {author} {\bibfnamefont {M.~A.}\
  \bibnamefont {{McLaughlin}}}, \bibinfo {author} {\bibfnamefont {T.~T.}\
  \bibnamefont {{Pennucci}}}, \bibinfo {author} {\bibfnamefont {S.~M.}\
  \bibnamefont {{Ransom}}}, \bibinfo {author} {\bibfnamefont {I.~H.}\
  \bibnamefont {{Stairs}}}, \bibinfo {author} {\bibfnamefont {J.}~\bibnamefont
  {{van Leeuwen}}}, \bibinfo {author} {\bibfnamefont {J.~P.~W.}\ \bibnamefont
  {{Verbiest}}}, \ and\ \bibinfo {author} {\bibfnamefont {D.~G.}\ \bibnamefont
  {{Whelan}}},\ }\href {\doibase 10.1126/science.1233232} {\bibfield  {journal}
  {\bibinfo  {journal} {Science}\ }\textbf {\bibinfo {volume} {340}},\ \bibinfo
  {pages} {448} (\bibinfo {year} {2013})},\ \Eprint
  {http://arxiv.org/abs/1304.6875} {arXiv:1304.6875 [astro-ph.HE]} \BibitemShut
  {NoStop}%
\bibitem [{\citenamefont {{Zhou}}\ \emph {et~al.}(2014)\citenamefont {{Zhou}},
  \citenamefont {{Lu}}, \citenamefont {{Tong}},\ and\ \citenamefont
  {{Xu}}}]{zhou2014}%
  \BibitemOpen
  \bibfield  {author} {\bibinfo {author} {\bibfnamefont {E.~P.}\ \bibnamefont
  {{Zhou}}}, \bibinfo {author} {\bibfnamefont {J.~G.}\ \bibnamefont {{Lu}}},
  \bibinfo {author} {\bibfnamefont {H.}~\bibnamefont {{Tong}}}, \ and\ \bibinfo
  {author} {\bibfnamefont {R.~X.}\ \bibnamefont {{Xu}}},\ }\href {\doibase
  10.1093/mnras/stu1370} {\bibfield  {journal} {\bibinfo  {journal} {Mon. Not.
  R. Astron. Soc.}\ }\textbf {\bibinfo {volume} {443}},\ \bibinfo {pages}
  {2705} (\bibinfo {year} {2014})},\ \Eprint {http://arxiv.org/abs/1404.2793}
  {arXiv:1404.2793 [astro-ph.HE]} \BibitemShut {NoStop}%
\bibitem [{\citenamefont {{Xu}}\ and\ \citenamefont {{Liang}}(2009)}]{xu2009}%
  \BibitemOpen
  \bibfield  {author} {\bibinfo {author} {\bibfnamefont {R.}~\bibnamefont
  {{Xu}}}\ and\ \bibinfo {author} {\bibfnamefont {E.}~\bibnamefont {{Liang}}},\
  }\href {\doibase 10.1007/s11433-009-0045-x} {\bibfield  {journal} {\bibinfo
  {journal} {Science in China: Physics, Mechanics and Astronomy}\ }\textbf
  {\bibinfo {volume} {52}},\ \bibinfo {pages} {315} (\bibinfo {year} {2009})},\
  \Eprint {http://arxiv.org/abs/0811.0234} {arXiv:0811.0234} \BibitemShut
  {NoStop}%
\bibitem [{\citenamefont {{Wang}}\ \emph {et~al.}(2017)\citenamefont {{Wang}},
  \citenamefont {{Lu}}, \citenamefont {{Tong}}, \citenamefont {{Ge}},
  \citenamefont {{Li}}, \citenamefont {{Men}},\ and\ \citenamefont
  {{Xu}}}]{wang2017}%
  \BibitemOpen
  \bibfield  {author} {\bibinfo {author} {\bibfnamefont {W.}~\bibnamefont
  {{Wang}}}, \bibinfo {author} {\bibfnamefont {J.}~\bibnamefont {{Lu}}},
  \bibinfo {author} {\bibfnamefont {H.}~\bibnamefont {{Tong}}}, \bibinfo
  {author} {\bibfnamefont {M.}~\bibnamefont {{Ge}}}, \bibinfo {author}
  {\bibfnamefont {Z.}~\bibnamefont {{Li}}}, \bibinfo {author} {\bibfnamefont
  {Y.}~\bibnamefont {{Men}}}, \ and\ \bibinfo {author} {\bibfnamefont
  {R.}~\bibnamefont {{Xu}}},\ }\href {\doibase 10.3847/1538-4357/aa5e52}
  {\bibfield  {journal} {\bibinfo  {journal} {Astrophys. J.}\ }\textbf
  {\bibinfo {volume} {837}},\ \bibinfo {eid} {81} (\bibinfo {year} {2017})},\
  \Eprint {http://arxiv.org/abs/1603.08288} {arXiv:1603.08288 [astro-ph.HE]}
  \BibitemShut {NoStop}%
\bibitem [{\citenamefont {{Xu}}(2003)}]{Xu2003}%
  \BibitemOpen
  \bibfield  {author} {\bibinfo {author} {\bibfnamefont {R.~X.}\ \bibnamefont
  {{Xu}}},\ }\href {\doibase 10.1086/379209} {\bibfield  {journal} {\bibinfo
  {journal} {Astrophys. J. Lett.}\ }\textbf {\bibinfo {volume} {596}},\
  \bibinfo {pages} {L59} (\bibinfo {year} {2003})},\ \Eprint
  {http://arxiv.org/abs/astro-ph/0302165} {astro-ph/0302165} \BibitemShut
  {NoStop}%
\bibitem [{\citenamefont {{Read}}\ \emph {et~al.}(2009)\citenamefont {{Read}},
  \citenamefont {{Lackey}}, \citenamefont {{Owen}},\ and\ \citenamefont
  {{Friedman}}}]{Read:2009a}%
  \BibitemOpen
  \bibfield  {author} {\bibinfo {author} {\bibfnamefont {J.~S.}\ \bibnamefont
  {{Read}}}, \bibinfo {author} {\bibfnamefont {B.~D.}\ \bibnamefont
  {{Lackey}}}, \bibinfo {author} {\bibfnamefont {B.~J.}\ \bibnamefont
  {{Owen}}}, \ and\ \bibinfo {author} {\bibfnamefont {J.~L.}\ \bibnamefont
  {{Friedman}}},\ }\href {\doibase 10.1103/PhysRevD.79.124032} {\bibfield
  {journal} {\bibinfo  {journal} {Phys. Rev. D}\ }\textbf {\bibinfo {volume}
  {79}},\ \bibinfo {eid} {124032} (\bibinfo {year} {2009})},\ \Eprint
  {http://arxiv.org/abs/0812.2163} {arXiv:0812.2163} \BibitemShut {NoStop}%
\bibitem [{\citenamefont {{Li}}\ \emph {et~al.}(2017)\citenamefont {{Li}},
  \citenamefont {{Zhu}},\ and\ \citenamefont {{Zhou}}}]{Li2017}%
  \BibitemOpen
  \bibfield  {author} {\bibinfo {author} {\bibfnamefont {A.}~\bibnamefont
  {{Li}}}, \bibinfo {author} {\bibfnamefont {Z.-Y.}\ \bibnamefont {{Zhu}}}, \
  and\ \bibinfo {author} {\bibfnamefont {X.}~\bibnamefont {{Zhou}}},\ }\href
  {\doibase 10.3847/1538-4357/aa7a00} {\bibfield  {journal} {\bibinfo
  {journal} {\apj}\ }\textbf {\bibinfo {volume} {844}},\ \bibinfo {eid} {41}
  (\bibinfo {year} {2017})}\BibitemShut {NoStop}%
\bibitem [{\citenamefont {{Bhattacharyya}}\ \emph {et~al.}(2016)\citenamefont
  {{Bhattacharyya}}, \citenamefont {{Bombaci}}, \citenamefont {{Logoteta}},\
  and\ \citenamefont {{Thampan}}}]{Bhattacharyya2016}%
  \BibitemOpen
  \bibfield  {author} {\bibinfo {author} {\bibfnamefont {S.}~\bibnamefont
  {{Bhattacharyya}}}, \bibinfo {author} {\bibfnamefont {I.}~\bibnamefont
  {{Bombaci}}}, \bibinfo {author} {\bibfnamefont {D.}~\bibnamefont
  {{Logoteta}}}, \ and\ \bibinfo {author} {\bibfnamefont {A.~V.}\ \bibnamefont
  {{Thampan}}},\ }\href {\doibase 10.1093/mnras/stw206} {\bibfield  {journal}
  {\bibinfo  {journal} {Mon. Not. R. Astron. Soc.}\ }\textbf {\bibinfo {volume}
  {457}},\ \bibinfo {pages} {3101} (\bibinfo {year} {2016})},\ \Eprint
  {http://arxiv.org/abs/1601.06120} {arXiv:1601.06120 [astro-ph.HE]}
  \BibitemShut {NoStop}%
\bibitem [{\citenamefont {{Alford}}\ \emph {et~al.}(2005)\citenamefont
  {{Alford}}, \citenamefont {{Braby}}, \citenamefont {{Paris}},\ and\
  \citenamefont {{Reddy}}}]{Alford2005}%
  \BibitemOpen
  \bibfield  {author} {\bibinfo {author} {\bibfnamefont {M.}~\bibnamefont
  {{Alford}}}, \bibinfo {author} {\bibfnamefont {M.}~\bibnamefont {{Braby}}},
  \bibinfo {author} {\bibfnamefont {M.}~\bibnamefont {{Paris}}}, \ and\
  \bibinfo {author} {\bibfnamefont {S.}~\bibnamefont {{Reddy}}},\ }\href
  {\doibase 10.1086/430902} {\bibfield  {journal} {\bibinfo  {journal}
  {Astrophys. J.}\ }\textbf {\bibinfo {volume} {629}},\ \bibinfo {pages} {969}
  (\bibinfo {year} {2005})},\ \Eprint {http://arxiv.org/abs/nucl-th/0411016}
  {nucl-th/0411016} \BibitemShut {NoStop}%
\bibitem [{\citenamefont {{Baym}}\ \emph {et~al.}(2017)\citenamefont {{Baym}},
  \citenamefont {{Hatsuda}}, \citenamefont {{Kojo}}, \citenamefont {{Powell}},
  \citenamefont {{Song}},\ and\ \citenamefont {{Takatsuka}}}]{Baym2017}%
  \BibitemOpen
  \bibfield  {author} {\bibinfo {author} {\bibfnamefont {G.}~\bibnamefont
  {{Baym}}}, \bibinfo {author} {\bibfnamefont {T.}~\bibnamefont {{Hatsuda}}},
  \bibinfo {author} {\bibfnamefont {T.}~\bibnamefont {{Kojo}}}, \bibinfo
  {author} {\bibfnamefont {P.~D.}\ \bibnamefont {{Powell}}}, \bibinfo {author}
  {\bibfnamefont {Y.}~\bibnamefont {{Song}}}, \ and\ \bibinfo {author}
  {\bibfnamefont {T.}~\bibnamefont {{Takatsuka}}},\ }\href@noop {} {\bibfield
  {journal} {\bibinfo  {journal} {ArXiv e-prints}\ } (\bibinfo {year}
  {2017})},\ \Eprint {http://arxiv.org/abs/1707.04966} {arXiv:1707.04966
  [astro-ph.HE]} \BibitemShut {NoStop}%
\bibitem [{\citenamefont {{Ostriker}}\ and\ \citenamefont
  {{Mark}}(1968)}]{Ostriker1968}%
  \BibitemOpen
  \bibfield  {author} {\bibinfo {author} {\bibfnamefont {J.~P.}\ \bibnamefont
  {{Ostriker}}}\ and\ \bibinfo {author} {\bibfnamefont {J.~W.-K.}\ \bibnamefont
  {{Mark}}},\ }\href {\doibase 10.1086/149506} {\bibfield  {journal} {\bibinfo
  {journal} {Astrophys. J.}\ }\textbf {\bibinfo {volume} {151}},\ \bibinfo
  {pages} {1075} (\bibinfo {year} {1968})}\BibitemShut {NoStop}%
\bibitem [{\citenamefont {{Komatsu}}\ \emph {et~al.}(1989)\citenamefont
  {{Komatsu}}, \citenamefont {{Eriguchi}},\ and\ \citenamefont
  {{Hachisu}}}]{Komatsu89}%
  \BibitemOpen
  \bibfield  {author} {\bibinfo {author} {\bibfnamefont {H.}~\bibnamefont
  {{Komatsu}}}, \bibinfo {author} {\bibfnamefont {Y.}~\bibnamefont
  {{Eriguchi}}}, \ and\ \bibinfo {author} {\bibfnamefont {I.}~\bibnamefont
  {{Hachisu}}},\ }\href {\doibase 10.1093/mnras/237.2.355} {\bibfield
  {journal} {\bibinfo  {journal} {Mon. Not. R. Astron. Soc.}\ }\textbf
  {\bibinfo {volume} {237}},\ \bibinfo {pages} {355} (\bibinfo {year}
  {1989})}\BibitemShut {NoStop}%
\bibitem [{\citenamefont {{Tsokaros}}\ \emph {et~al.}(2016)\citenamefont
  {{Tsokaros}}, \citenamefont {{Mundim}}, \citenamefont {{Galeazzi}},
  \citenamefont {{Rezzolla}},\ and\ \citenamefont {{Ury{\=u}}}}]{Tsokaros2016}%
  \BibitemOpen
  \bibfield  {author} {\bibinfo {author} {\bibfnamefont {A.}~\bibnamefont
  {{Tsokaros}}}, \bibinfo {author} {\bibfnamefont {B.~C.}\ \bibnamefont
  {{Mundim}}}, \bibinfo {author} {\bibfnamefont {F.}~\bibnamefont
  {{Galeazzi}}}, \bibinfo {author} {\bibfnamefont {L.}~\bibnamefont
  {{Rezzolla}}}, \ and\ \bibinfo {author} {\bibfnamefont {K.}~\bibnamefont
  {{Ury{\=u}}}},\ }\href@noop {} {\bibfield  {journal} {\bibinfo  {journal}
  {arXiv:1605.07205}\ } (\bibinfo {year} {2016})},\ \Eprint
  {http://arxiv.org/abs/1605.07205} {arXiv:1605.07205 [gr-qc]} \BibitemShut
  {NoStop}%
\bibitem [{\citenamefont {Tsokaros}\ and\ \citenamefont
  {Ury{\={u}}}(2007)}]{Tsokaros2007}%
  \BibitemOpen
  \bibfield  {author} {\bibinfo {author} {\bibfnamefont {A.~A.}\ \bibnamefont
  {Tsokaros}}\ and\ \bibinfo {author} {\bibfnamefont {K.}~\bibnamefont
  {Ury{\={u}}}},\ }\href {\doibase 10.1103/PhysRevD.75.044026} {\bibfield
  {journal} {\bibinfo  {journal} {Phys. Rev. D}\ }\textbf {\bibinfo {volume}
  {75}},\ \bibinfo {pages} {044026} (\bibinfo {year} {2007})}\BibitemShut
  {NoStop}%
\bibitem [{\citenamefont {{Huang}}\ \emph {et~al.}(2007)\citenamefont
  {{Huang}}, \citenamefont {{Cai}}, \citenamefont {{Shen}},\ and\ \citenamefont
  {{Yuan}}}]{Huang2007}%
  \BibitemOpen
  \bibfield  {author} {\bibinfo {author} {\bibfnamefont {L.}~\bibnamefont
  {{Huang}}}, \bibinfo {author} {\bibfnamefont {M.}~\bibnamefont {{Cai}}},
  \bibinfo {author} {\bibfnamefont {Z.-Q.}\ \bibnamefont {{Shen}}}, \ and\
  \bibinfo {author} {\bibfnamefont {F.}~\bibnamefont {{Yuan}}},\ }\href
  {\doibase 10.1111/j.1365-2966.2007.11713.x} {\bibfield  {journal} {\bibinfo
  {journal} {Mon. Not. R. Astron. Soc.}\ }\textbf {\bibinfo {volume} {379}},\
  \bibinfo {pages} {833} (\bibinfo {year} {2007})},\ \Eprint
  {http://arxiv.org/abs/astro-ph/0703254} {astro-ph/0703254} \BibitemShut
  {NoStop}%
\bibitem [{\citenamefont {{Weih}}\ \emph {et~al.}(2018)\citenamefont {{Weih}},
  \citenamefont {{Most}},\ and\ \citenamefont {{Rezzolla}}}]{Weih2017}%
  \BibitemOpen
  \bibfield  {author} {\bibinfo {author} {\bibfnamefont {L.~R.}\ \bibnamefont
  {{Weih}}}, \bibinfo {author} {\bibfnamefont {E.~R.}\ \bibnamefont {{Most}}},
  \ and\ \bibinfo {author} {\bibfnamefont {L.}~\bibnamefont {{Rezzolla}}},\
  }\href {\doibase 10.1093/mnrasl/slx178} {\bibfield  {journal} {\bibinfo
  {journal} {Mon. Not. R. Astron. Soc.}\ }\textbf {\bibinfo {volume} {473}},\
  \bibinfo {pages} {L126} (\bibinfo {year} {2018})},\ \Eprint
  {http://arxiv.org/abs/1709.06058} {arXiv:1709.06058 [gr-qc]} \BibitemShut
  {NoStop}%
\bibitem [{\citenamefont {{Breu}}\ and\ \citenamefont
  {{Rezzolla}}(2016)}]{Breu2016}%
  \BibitemOpen
  \bibfield  {author} {\bibinfo {author} {\bibfnamefont {C.}~\bibnamefont
  {{Breu}}}\ and\ \bibinfo {author} {\bibfnamefont {L.}~\bibnamefont
  {{Rezzolla}}},\ }\href {\doibase 10.1093/mnras/stw575} {\bibfield  {journal}
  {\bibinfo  {journal} {Mon. Not. R. Astron. Soc.}\ }\textbf {\bibinfo {volume}
  {459}},\ \bibinfo {pages} {646} (\bibinfo {year} {2016})},\ \Eprint
  {http://arxiv.org/abs/1601.06083} {arXiv:1601.06083 [gr-qc]} \BibitemShut
  {NoStop}%
\bibitem [{\citenamefont {{Ury{\=u}}}\ \emph {et~al.}(2016)\citenamefont
  {{Ury{\=u}}}, \citenamefont {{Tsokaros}}, \citenamefont {{Baiotti}},
  \citenamefont {{Galeazzi}}, \citenamefont {{Sugiyama}}, \citenamefont
  {{Taniguchi}},\ and\ \citenamefont {{Yoshida}}}]{Uryu2016b}%
  \BibitemOpen
  \bibfield  {author} {\bibinfo {author} {\bibfnamefont {K.}~\bibnamefont
  {{Ury{\=u}}}}, \bibinfo {author} {\bibfnamefont {A.}~\bibnamefont
  {{Tsokaros}}}, \bibinfo {author} {\bibfnamefont {L.}~\bibnamefont
  {{Baiotti}}}, \bibinfo {author} {\bibfnamefont {F.}~\bibnamefont
  {{Galeazzi}}}, \bibinfo {author} {\bibfnamefont {N.}~\bibnamefont
  {{Sugiyama}}}, \bibinfo {author} {\bibfnamefont {K.}~\bibnamefont
  {{Taniguchi}}}, \ and\ \bibinfo {author} {\bibfnamefont {S.}~\bibnamefont
  {{Yoshida}}},\ }\href {\doibase 10.1103/PhysRevD.94.101302} {\bibfield
  {journal} {\bibinfo  {journal} {Phys. Rev. D}\ }\textbf {\bibinfo {volume}
  {94}},\ \bibinfo {eid} {101302} (\bibinfo {year} {2016})},\ \Eprint
  {http://arxiv.org/abs/1606.04604} {arXiv:1606.04604 [astro-ph.HE]}
  \BibitemShut {NoStop}%
\bibitem [{\citenamefont {{Andersson}}(1998)}]{Andersson1998}%
  \BibitemOpen
  \bibfield  {author} {\bibinfo {author} {\bibfnamefont {N.}~\bibnamefont
  {{Andersson}}},\ }\href {\doibase 10.1086/305919} {\bibfield  {journal}
  {\bibinfo  {journal} {Astrophys. J.}\ }\textbf {\bibinfo {volume} {502}},\
  \bibinfo {pages} {708} (\bibinfo {year} {1998})},\ \Eprint
  {http://arxiv.org/abs/arXiv:gr-qc/9706075} {arXiv:gr-qc/9706075} \BibitemShut
  {NoStop}%
\bibitem [{\citenamefont {{Friedman}}\ and\ \citenamefont
  {{Morsink}}(1998)}]{Friedman1998}%
  \BibitemOpen
  \bibfield  {author} {\bibinfo {author} {\bibfnamefont {J.~L.}\ \bibnamefont
  {{Friedman}}}\ and\ \bibinfo {author} {\bibfnamefont {S.~M.}\ \bibnamefont
  {{Morsink}}},\ }\href {\doibase 10.1086/305920} {\bibfield  {journal}
  {\bibinfo  {journal} {Astrophys. J.}\ }\textbf {\bibinfo {volume} {502}},\
  \bibinfo {pages} {714} (\bibinfo {year} {1998})},\ \Eprint
  {http://arxiv.org/abs/arXiv:gr-qc/9706073} {arXiv:gr-qc/9706073} \BibitemShut
  {NoStop}%
\bibitem [{\citenamefont {Andersson}\ \emph {et~al.}(1999)\citenamefont
  {Andersson}, \citenamefont {Kokkotas},\ and\ \citenamefont
  {Stergioulas}}]{Andersson99}%
  \BibitemOpen
  \bibfield  {author} {\bibinfo {author} {\bibfnamefont {N.}~\bibnamefont
  {Andersson}}, \bibinfo {author} {\bibfnamefont {K.~D.}\ \bibnamefont
  {Kokkotas}}, \ and\ \bibinfo {author} {\bibfnamefont {N.}~\bibnamefont
  {Stergioulas}},\ }\href@noop {} {\bibfield  {journal} {\bibinfo  {journal}
  {Astrophys. J.}\ }\textbf {\bibinfo {volume} {510}},\ \bibinfo {pages} {854}
  (\bibinfo {year} {1999})}\BibitemShut {NoStop}%
\bibitem [{\citenamefont {{Kramer}}\ and\ \citenamefont
  {{Stappers}}(2015)}]{SKAdoc}%
  \BibitemOpen
  \bibfield  {author} {\bibinfo {author} {\bibfnamefont {M.}~\bibnamefont
  {{Kramer}}}\ and\ \bibinfo {author} {\bibfnamefont {B.}~\bibnamefont
  {{Stappers}}},\ }\href@noop {} {\bibfield  {journal} {\bibinfo  {journal}
  {Advancing Astrophysics with the Square Kilometre Array (AASKA14)}\ ,\
  \bibinfo {eid} {36}} (\bibinfo {year} {2015})},\ \Eprint
  {http://arxiv.org/abs/1507.04423} {arXiv:1507.04423 [astro-ph.IM]}
  \BibitemShut {NoStop}%
\bibitem [{\citenamefont {{Nan}}\ \emph {et~al.}(2011)\citenamefont {{Nan}},
  \citenamefont {{Li}}, \citenamefont {{Jin}}, \citenamefont {{Wang}},
  \citenamefont {{Zhu}}, \citenamefont {{Zhu}}, \citenamefont {{Zhang}},
  \citenamefont {{Yue}},\ and\ \citenamefont {{Qian}}}]{Nan2011}%
  \BibitemOpen
  \bibfield  {author} {\bibinfo {author} {\bibfnamefont {R.}~\bibnamefont
  {{Nan}}}, \bibinfo {author} {\bibfnamefont {D.}~\bibnamefont {{Li}}},
  \bibinfo {author} {\bibfnamefont {C.}~\bibnamefont {{Jin}}}, \bibinfo
  {author} {\bibfnamefont {Q.}~\bibnamefont {{Wang}}}, \bibinfo {author}
  {\bibfnamefont {L.}~\bibnamefont {{Zhu}}}, \bibinfo {author} {\bibfnamefont
  {W.}~\bibnamefont {{Zhu}}}, \bibinfo {author} {\bibfnamefont
  {H.}~\bibnamefont {{Zhang}}}, \bibinfo {author} {\bibfnamefont
  {Y.}~\bibnamefont {{Yue}}}, \ and\ \bibinfo {author} {\bibfnamefont
  {L.}~\bibnamefont {{Qian}}},\ }\href {\doibase 10.1142/S0218271811019335}
  {\bibfield  {journal} {\bibinfo  {journal} {International Journal of Modern
  Physics D}\ }\textbf {\bibinfo {volume} {20}},\ \bibinfo {pages} {989}
  (\bibinfo {year} {2011})},\ \Eprint {http://arxiv.org/abs/1105.3794}
  {arXiv:1105.3794 [astro-ph.IM]} \BibitemShut {NoStop}%
\end{thebibliography}%
